%% file: kblas.tex
\documentclass[prodmode,acmtoms]{acmsmall}

\usepackage[ruled]{algorithm2e}

\SetAlFnt{\small}
\SetAlCapFnt{\small}
\SetAlCapNameFnt{\small}
\SetAlCapHSkip{0pt}
\IncMargin{-\parindent}

\usepackage{graphicx}
\usepackage{amsmath}
\usepackage{amsfonts}
\usepackage{algorithmic}
\usepackage{subfigure}
\usepackage{url}
\usepackage{array}
\usepackage{caption}

\newcolumntype{M}[1]{>{\centering\arraybackslash}m{#1}}
\newcolumntype{N}{@{}m{0pt}@{}}

\begin{document}

% Page heads
\markboth{A. Abdelfattah et al.}{KBLAS: An Optimized Library for Dense Matrix-Vector Multiplication on GPU Accelerators}

%Computing Classification Systems
%G.4 MATHEMATICAL SOFTWARE
%Parallel and vector implementations
%
% Date submitted: May 2, 2011

% Title portion
\title{KBLAS: An Optimized Library for Dense Matrix-Vector Multiplication on GPU Accelerators}

\author{Ahmad Abdelfattah\affil{Extreme Computing Research Center, KAUST}
David Keyes\affil{Extreme Computing Research Center, KAUST}
Hatem Ltaief\affil{Extreme Computing Research Center, KAUST}}

\begin{abstract}
\input{abstract}
\end{abstract}




\acmformat{Abdelfattah, A., Keyes, D., and Ltaief, H. 2014. KBLAS: An Optimized Library for Dense Matrix-Vector Multiplication on GPU Accelerators.}

\begin{bottomstuff}
% This research reported here was partially supported by the
% National Science Foundation, the Department of Energy and Microsoft Research.

Author's addresses: A. Abdelfattah (Ahmad.Ahmad@kaust.edu.sa), D. Keyes (David.Keyes@kaust.edu.sa) {and} H. Ltaief (Hatem.Ltaief@kaust.edu.sa),
Extreme Computing Research Center, King Abdullah University of Science and Technology, Thuwal, Saudi Arabia,
\end{bottomstuff}

\maketitle

\section{Introduction}
\label{sec:intro}
\input{intro.tex}

\section{Related Work}
\label{sec:related_work}
\input{rw.tex}

\section{Contributions}
\label{sec:contrib}
\input{contrib.tex}

\section{Overview Of GPUs}
\label{sec:gpu}
\input{gpu.tex}

\section{The KBLAS Library}
\label{sec:kblas_lib}
This section highlights all the functionalities supported by KBLAS, 
as well as the data layout required by its routines. 
\subsection{Main Structure}
\label{subsec:structure}
\input{structure.tex}

\subsection{Routines and APIs}
\label{subsec:routines}
\input{routines.tex}

\subsection{Data Layout}
\label{subsec:layout}
\input{layout.tex}

\section{High Performance Kernel Implementations}
\label{sec:impl}
\input{impl.tex}

\section{Performance Model}
\label{sec:model}
\input{model.tex}

\section{Performance Results and Analysis}
\label{sec:perf}
\input{perf.tex}

\section{Performance Tuning}
\label{sec:tuning}
\input{tuning.tex}

\section{Accelerating Existing Numerical Linear Algebra Libraries}
\label{sec:app}
\input{app.tex}

\section{Conclusion and Future Work}
\label{sec:summary}
\input{summary.tex}

\section*{Acknowledgment}
\label{sec:ack}
\input{ack.tex}

% Bibliography
\bibliographystyle{acmsmall}
\bibliography{kblas}

% History dates: received, revised, accepted
% \received{May 2011}{2011}{2011}

\end{document}

%% file: abstract.tex
%!TEX root = kblas.tex
%
% Hatem2Ahmad: will be moving this paragraph into intro
%
% Heterogeneity is a confirmed property of future HPC systems. Traditional multicore CPUs 
% are equipped with hardware accelerators, with Graphical Processing Units (GPUs) being 
% a common choice as high throughput coprocessors. GPUs are well fit in applications that 
% have a lot of parallelism. Applications that use dense linear algebra are a good example, 
% and depend on standard kernels that are the scope of this work. GPUs can achieve orders of 
% magnitude speedups over traditional 
% CPUs for such kernels. While good speedups are achievable through compiler generated 
% codes, hand tailored kernels usually achieve better performance. 
% However, in order for hand coded kernels to survive the evolving GPU architecture, 
% they have to be tunable through design parameters that can maintain the same performance 
% level on future GPUs. These kernels can then be shipped into a library which scientists can 
% use in their codes. This paper introduces an example of such libraries.

KBLAS is a new open source high performance library that provides 
optimized kernels for a subset of Level 2 BLAS functionalities
on CUDA-enabled GPUs. Since performance of dense matrix-vector multiplication  
is hindered by the overhead of memory accesses, a double-buffering
optimization technique is employed to overlap data motion with computation.
After identifying a proper set of tuning parameters, KBLAS is able to 
efficiently run on various GPU architectures across different generations, 
avoiding the time-consuming step of code rewriting, while still
being compliant with the standard BLAS API. Another advanced optimization technique 
allows to ensure coalesced memory access when dealing with submatrices, 
especially in the context of high level dense linear algebra algorithms.
All four precisions KBLAS kernels have been leveraged to multi-GPUs environment, which requires
the introduction of new APIs to ease users' experiences on these
challenging systems. The KBLAS performance
outperforms existing state-of-the-art implementations on all matrix sizes,
achieves asymptotically up to $50$\% and 
$60$\% speedup on single GPU and multi-GPUs systems, respectively,
and validates our performance model. 
A subset of KBLAS high performance kernels has been integrated 
into NVIDIA's standard BLAS implementation 
(cuBLAS) for larger dissemination, starting version $6.0$.

%% file: intro.tex
%!TEX root = kblas.tex
The increasing compute power of modern hardware accelerators, 
such as GPUs, has drawn interest for general purpose scientific 
computing, especially compute-intensive workloads, which expose
data parallelism at the forefront. 
%For instance, 
Dense Linear Algebra (DLA) is one area 
where GPUs can achieve orders of magnitude better performance than traditional 
multi-core architectures. This is because most DLA algorithms, as implemented 
in the standard LAPACK library~\cite{LAPACK}, are embarrassingly parallel with 
regular memory accesses and essentially rely on Level 3 BLAS~\citeA{BLAS}
(i.e., matrix-matrix multiplication) for high performance.
% These types of kernels fit well into modern GPU architectures. 
% However, it is non-trivial for these kernels to perform close to their sustained peak performance. 

Matrix-Vector Multiplication (MVM) kernels are also widely 
used in DLA algorithms. In fact, these Level 2 BLAS operations
are the building blocks of the panel factorization phase
for one-sided and two-sided transformations, while solving 
linear systems of equations and eigenvalue problems 
or singular value decomposition, respectively.
% the building blocks for the standard LAPACK algorithms~\cite{LAPACK}. 
% For example, matrix reduction algorithms, such as bidiagonal 
% or tridiagonal reductions, use MV kernels excessively 
% at their cores.
However, such kernels have low floating point operations per byte ratio (flops/byte), 
and so are bounded by the sustained memory bandwidth of the hardware. 
As part of the critical path, MVM kernels usually represent 
serious performance bottlenecks in the aforementioned algorithms.
In fact, current state-of-the-art MVM implementations on GPUs are capable of 
extracting only a small percentage of the bus bandwidth peak.
Therefore, optimization techniques should maximize the memory bus utilization, 
in order to push MVM's performance 
close to the STREAM benchmark~[\citeANP{stream1995} \citeyearNP{stream1995}; \citeyearNP{stream2007}],
which represents a tight upper-bound for applications performance 
hindered by the bus bandwidth.
% as they are launched among other compute bound 
% kernels which normally run much faster than memory bound routines. 
% Therefore, MV kernels should run as fast 
% as possible. 

This paper introduces KBLAS\footnote{KBLAS: KAUST Basic Linear Algebra Subprograms. Available at 
\url{http://cec.kaust.edu.sa/Pages/kblas.aspx}}, 
an optimized library for dense MVM kernels on GPUs. 
The KBLAS library supports all four standard precisions. It can also 
run on shared-memory compute nodes equipped with multiple GPUs. 
For easy integration, the single GPU kernels 
from KBLAS are fully compliant with the standard BLAS interface. KBLAS also 
provides new interfaces for advanced users 
to ensure coalesced memory access, 
when dealing with operations on submatrices. 
Moreover, new APIs for multi-GPUs systems 
are proposed to facilitate user code developments,
thanks to a transparent memory management.
The authors build over previous 
work~[\citeANP{ahmad_vecpar12} 2013a][\citeANP{ahmad_heteropar12} 2013b] and 
introduce more functionalities and performance tuning knobs
to maintain decent throughput across previous and current 
GPUs hardware generations, without code rewriting.
The KBLAS performance outperforms existing state-of-the-art 
open-source and commercial implementations 
(i.e., NVIDIA's standard BLAS implementation cuBLAS~\cite{CUBLAS}, MAGMABLAS~\cite{MAGMA} and 
CULA~\cite{humphrey2010cula}) on all matrix sizes.
KBLAS achieves asymptotically up to $50$\% and 
$60$\% speedup against the best implementations 
on single GPU and multi-GPUs systems, respectively. 
%validating by the same token our performance model (within $95$\%).
Most of KBLAS kernels run within at least 80\% of the sustained 
peak performance determined by our performance model, which is based 
on experimentally measuring the memory bandwidth using STREAM benchmark. 
The paper also shows up to $20$\% improvement of 
high performance DLA libraries after KBLAS integration.
Last, but not least, a subset of KBLAS high performance kernels has been integrated 
into cuBLAS for larger dissemination, starting version 
$6.0$\footnote{\url{http://docs.nvidia.com/cuda/cublas/#appendix-acknowledgements}}.
% Hatem2Ahmad: move this to genetal kblas section
% One routine from KBLAS has been integrated 
% into NVIDIA's BLAS implementation, cuBLAS \cite{CUBLAS}, 
% starting version $6.0$. KBLAS is written 
% using CUDA C/C++ extensions. It supports compute capability 2.0 (Fermi GPUs) or higher. 
% An end user only needs CUDA Toolkit in order to build 
% and use KBLAS. Sample testing codes are provided 
% to test every single kernel in KBLAS, and compare it 
% against its counterpart in cuBLAS \cite{CUBLAS}.

The rest of the paper is organized as follows.
Section~\ref{sec:related_work} presents related work.
Section~\ref{sec:contrib} highlights our contributions. Section~\ref{sec:gpu}
gives a general overview of GPU architectures.
Section~\ref{sec:kblas_lib} describes the KBLAS framework 
and presents its different features and functionalities.
The implementation details of the high performance MVM kernels 
are given in Section~\ref{sec:impl}. The performance model to support
our empirical data is introduced in Section~\ref{sec:model}. 
Section~\ref{sec:perf} shows KBLAS performance results on various
systems and compare against the state-of-the-art commercial and open-source 
high performance MVM implementations. Section~\ref{sec:tuning} 
identifies critical parameters necessary to tune the KBLAS kernels
on different GPU architectures. Section~\ref{sec:app} illustrates
the performance impact after integrating KBLAS into DLA libraries 
and we conclude in Section~\ref{sec:summary}.

%% file: rw.tex
%!TEX root = kblas.tex
DLA solves problems that are usually regular and well structured. Such 
problems are suitable for acceleration using GPUs. In fact, 
the GPU accelerated DLA literature is rich in research efforts that shows orders of magnitude performance 
gain against multi-core architectures. 

For example, standard BLAS operations are provided 
by vendors \cite{CUBLAS}, while researchers keep providing even more optimized BLAS and 
incidentally LAPACK routines. The early work presented by \cite{volkov_sc08} envisions GPU architectures as multi-threaded 
multicore vector units, capable of achieving high performance dense linear algebra routines, including 
matrix multiplication and matrix factorizations (QR, LU and Cholesky). 
The developers of MAGMA \cite{MAGMA} presented a set of optimization techniques to accelerate 
BLAS operations on GPUs \cite{nath_crc,nath_vecpar10}.  

Level 3 BLAS operations, especially matrix multiplication are targets of many 
research efforts that aim to run these operations as 
close as possible to the theoretical peak performance of the GPU \cite{volkov_sc08,nath_ijhpca,dgemm_sc11}. 
The latter work uses optimization techniques like register and shared memory blocking. 
In addition, it proposes an optimal implementation using a pseudo-assembly language (Parallel Thread Execution), 
which runs very close to the peak performance of the GPU. 

Level 2 BLAS are more challenging to optimize due to the lack of data reuse to compensate for the
data transfer overhead. There are fewer research initiatives in this area.
Part of the MAGMA BLAS library is an optimized symmetric MVM kernel \cite{symv_sc11}. 
This kernel takes advantage of the symmetry, unlike what was provided by cuBLAS at that time. It adopts  
a \emph{recursive blocking} technique to handle block sizes that do not 
fit as a whole in shared memory and a \emph{pointer redirecting} technique that prevents memory access violations when 
the matrix dimension is not divisible by the block size. The kernel uses an extra workspace in the GPU global 
memory to perform a final reduction. Although rich in synchronization overhead, 
this implementation still runs more than twice as fast as the cuBLAS kernel. 

NVIDIA cuBLAS $5.5$ currently provides an optimized kernel for symmetric MVM, 
which takes advantage of the symmetry. 
There is also another implementation for the same kernel provided 
by the commercial library CULA \cite{humphrey2010cula}. 
The work presented in this paper compares our performance implementations 
against similar kernels from the aforementioned software libraries i.e., NVIDIA's cuBLAS, MAGMA BLAS, and CULA.

%% file: contrib.tex
%!TEX root = kblas.tex
% The contribution in this paper builds on top of our previous research efforts. 
% We proposed an optimized implementation for a GPU accelerated SYMV/HEMV kernel \cite{ahmad_vecpar12}. The kernel was an 
% improved version of the kernel proposed by the MAGMA developers \cite{symv_sc11}. 
% Both implementations were optimized for the Fermi GPU, 
% and used to require an additional memory workspace to perform a final reduction step. 
% 
% Hatem2Ahmad: is this a contribution?
% We also proposed an optimized GEMV kernel for GPU accelerators \cite{ahmad_heteropar12}. The kernel used to provide performance 
% speedups for relatively small matrices, a crucial result for matrix reduction algorithms. Our recent experiments show that cuBLAS enhances 
% its performance for small matrices starting version $5.5$. 
% 
Our contributions in this paper can be summarized as follows:
\begin{itemize}
	% \item We leverage the implementation of the SYMV/HEMV kernel to take advantage of the Kepler architecture, 
	% 	remove the workspace requirement, and emphasize performance portability across different GPUs.
	% 	\item The KBLAS GEMV kernel implementation is improved through the introduction of tuning parameters 
	% 	that provide controllable and smoother performance, unlike the oscillatory behavior the old implementation \cite{ahmad_heteropar12}
	% 	used to have. 
	\item The performance of our previous SYMV/HEMV~[\citeANP{ahmad_vecpar12} 2013a] 
	and GEMV~[\citeANP{ahmad_heteropar12} 2013b] kernel implementations
	have been improved on single GPUs and across different generations,
	thanks to the identification of tunable critical performance parameters.
	\item We introduce new interfaces for GEMV and SYMV/HEMV kernels, when operating on submatrices.
	These new kernel implementations remove the resulting non-coalesced memory accesses due to matrix offset and maintain
	the original performance improvement, as seen on regular matrices.	
		% 
		% GEMV and SYMV/HEMV kernel implementation with new interfaces. These kernels deliver better 
		% performance than the kernels with standard interfaces if the multiplication is done by a submatrix. These kernels 
		% can compensate for the non-coalesced memory accesses encountered if the input matrix is part of a bigger one. 
	\item We propose new GEMV and SYMV/HEMV kernels for shared-memory systems equipped with multi-GPUs. 
	These kernels operate on matrices distributed across multi-GPUs. The new interfaces abstract the hardware 
	complexity from end users and transparently handle the memory management between the host and the device. 
	\item All developed kernels have been released into a single open-source library named KBLAS, 
	available for download under the modified BSD license.
\end{itemize}

%% file: gpu.tex
%!TEX root = kblas.tex
%Over the past few decades, computer games and applications related to visualization, 
%graphics, and animations, drived GPUs to deliver more throughput in terms of processed 
%fragments (pixels) per seconds. Since pixels are often totally independent, GPUs are 
%built to have a continuously increasing parallel compute power. This drew the attention of researchers 
%to start looking into general purpose GPU computing. Since graphics APIs are not intended 
%for general purpose computing, new programming languages and APIs were proposed, beginning with 
%early efforts such as Brook for GPUs \cite{brook}, and ending with the state-of-the-art programming models, compilers, 
%and runtimes, such as CUDA, OpenCL, and OpenACC.  
%This section describes the main outlines of a modern 
%CUDA-enabled GPU architecture, and its programming model.     
This section provides an overview of a modern GPU architecture and one of its programming models. 

\subsection{Hardware Architecture}
\label{arch}
GPUs represent an example of a many-core architecture. They are built to deliver levels 
of processing throughput beyond the capability of traditional multi-core architectures, 
which adopt architectural optimizations to minimize execution latency rather throughput. 
A GPU thread executes slower than a CPU thread, but the GPU compensates for 
this by being able to executes orders of magnitude more concurrent threads than what a CPU 
can execute. Therefore, GPUs usually outperform CPUs in executing kernels that include 
large amounts of parallelism. 

From an architectural point of view, GPUs do not have complex hardware components that 
usually achieve fast execution of a single instruction stream. For example, large caches, 
advanced memory prefetchers, and deep instruction pipelines are not found in today's GPU 
architectures. Moreover, multiple threads (as of today, typically 32) share the same instruction 
stream, in order to amortize the complexity of the fetch and decode unit. This group of thread 
is called a \emph{warp}. The execution model of a GPU is a variant of the Single Instruction Multiple 
Data (SIMD) model, called the Single Instruction Multiple Threads (SIMT). Several threads execute 
the same scalar instruction on multiple operands. This is in contrast to the explicit 
vector instructions executed on long registers, which is the case in modern CPUs and Intel's 
Many Integrated Cores (MIC) architecture. 

\begin{figure}[ht]
\centering
\includegraphics[height=0.45\linewidth]{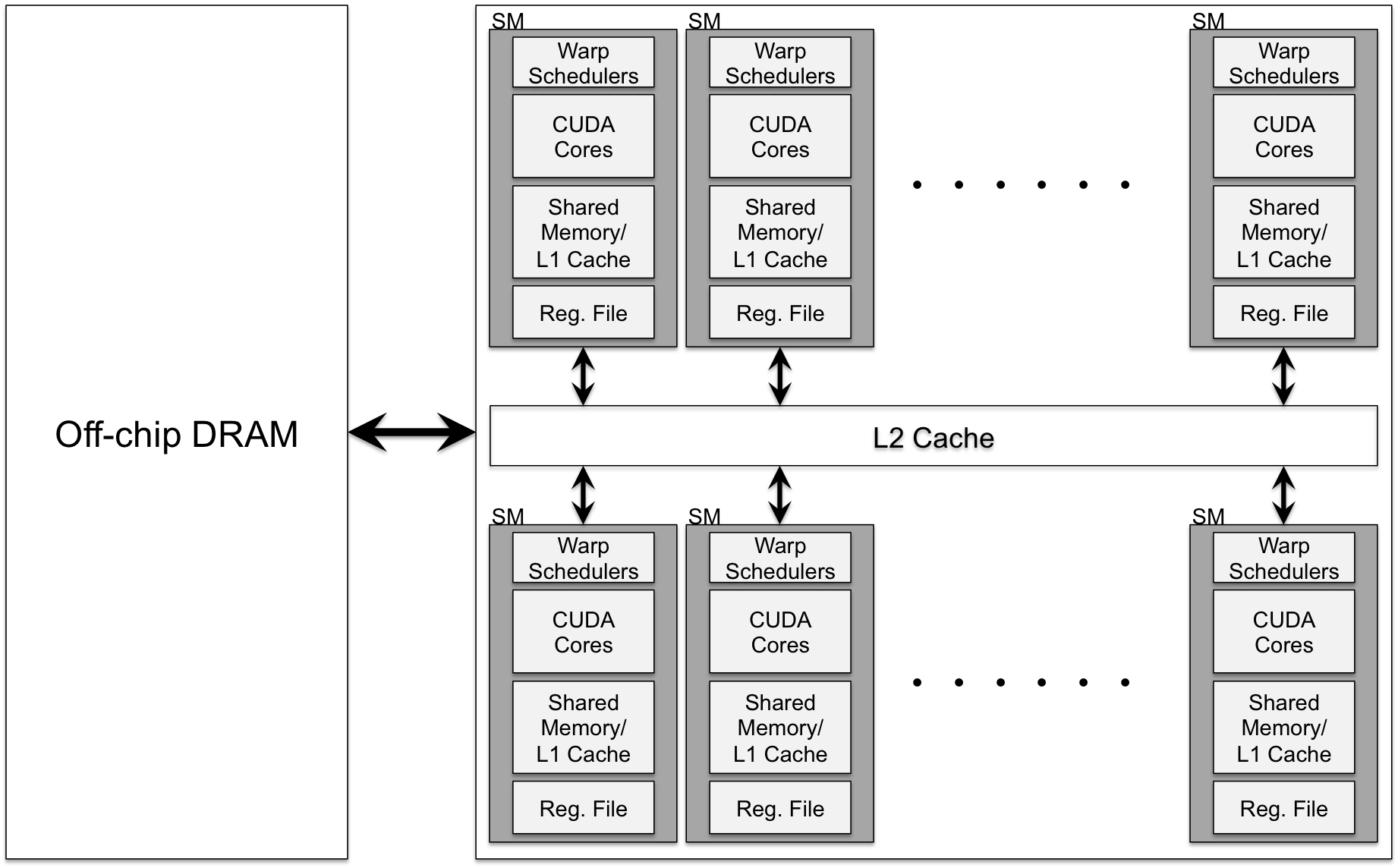}
\caption[]{A modern GPU architecture}
\label{fig:gpu_arch}
\end{figure}

A schematic diagram of a typical GPU architecture is shown in Figure \ref{fig:gpu_arch}. 
As an example, we will mention NVIDIA's Kepler architecture \cite{KEPLERWP}, which is the GPU 
used to show the performance results in this paper.  
The Kepler architecture is similar in many ways 
to its predecessor, the Fermi GPU \cite{FERMIWP}. It mainly consists of Streaming Multiprocessors (SMs). 
Each SM consists of a number of simple cores (CUDA cores) capable of doing both integer and floating 
point operations. Each SM in a Kepler GPU has 192 CUDA cores, equipped with 64 double precision units. 
There is also a number of special function units (SFUs) that can compute nonlinear functions on 32 bit floating 
point numbers. The memory hierarchy on a Kepler GPU is simple. Each SM has its own L1 cache/shared memory module, 
register file, and constant cache. All SMs share an L2 cache and the global DRAM. Texture memory is also globally 
accessible to SMs. 

One of the hardware features used by KBLAS is atomic operations. Different threads can do certain operations 
atomically on shared and global memories. Atomic operations are much faster on Kepler than on Fermi. As we will point 
out later, atomic operations are used in almost every kernel in KBLAS. 
%--------------------------------------------
\subsection{Programming Models}
\label{prog_model}
A GPU programming model tends to leverage the GPU capabilities in general purpose computing. 
One of the early efforts in this area was Brook for GPUs \cite{brook}. GPUs now have many programming models, compilers, 
and runtimes, such as CUDA, OpenCL, and OpenACC. We focus on the CUDA programming model.

NVIDIA GPUs natively support the CUDA programming model (or simply CUDA), which is 
used for the development of KBLAS. CUDA provides extensions to widely used programming languages 
like C, C++, Fortran, and Java. These extensions enable the development of GPU codes within 
a CPU code in one context. The main entry of execution for a GPU code is called a \emph{CUDA kernel}. 
A CUDA kernel is launched by the CPU, or by another CUDA kernel (a feature only in Kepler GPUs, called 
\emph{Dynamic Parallelism}). All kernel launches on the GPU are asynchronous, as the control returns 
immediately to the host thread. It is the programmer's responsibility to ensure the completion of any CUDA 
kernel through the CUDA runtime APIs. 

A CUDA kernel is organized as a Grid of thread blocks (TBs). Each TB consists of an array 
of threads. Both the grid and the thread array can have up to three dimensions. 
The number of TBs is independent from the number of SMs. A TB is executed exactly by one SM, 
while one SM can execute multiple TBs, if it has the sufficient resources. TBs in the same kernel cannot 
communicate or share data. Threads within the same TB, however, can share data and synchronize with each 
other. 

Concurrent kernel execution is also possible on one GPU through \emph{CUDA Streams}, which act as queues 
where independent kernels are submitted to different streams. Managing dependencies and synchronization 
among concurrent kernels is the responsibility of the programmer. If no stream is specified for the 
launch, the kernel is submitted to a default stream. Detailed information about the CUDA programming model 
can be found in the CUDA programming guide \cite{PROGGUIDE}. 
%-------------------------------------------------------------
\subsection{Performance Considerations}
\label{perf_consider}
In order to achieve high performance on a GPU, several considerations have to be taken into account. We will focus on 
considerations that matter most for memory-bound kernels. More details can be found at \cite{greenbook} and \cite{PROGGUIDE}. 

\subsubsection{Memory Coalesced Access}
\label{subsubsec:coalescing}
A memory-bound kernel should access the global memory in a coalesced manner, in order to avoid unnecessary memory traffic. 
Up to 128 bytes can be transferred between an SM and global memory in one transaction. 
Therefore, Memory reads and writes, executed per warp, should be performed in fully aligned 128 bytes. Otherwise, a warp memory request 
will be translated into multiple transactions, which penalizes performance. Matrices stored in global memory are usually 
padded in order to facilitate memory coalesced accesses, as discussed in Section \ref{subsec:layout}.

\subsubsection{GPU Occupancy}
\label{subsubsec:occupancy}
Another consideration is the GPU occupancy. In order to run at full memory bandwidth, 
the number of threads launched within a kernel should be large enough to saturate the memory bus with 
coalesced memory requests. Kernels of low occupancy often fail to operate at the peak bandwidth, even if 
memory coalescing is preserved. However, this does not mean launching the maximum possible number of threads, as 
the kernel might put pressure on other GPU resources, leading to performance drops. For example, kernels that consume 
too many registers do not usually operate at peak performance, since excessive register pressure leads to register spelling 
into global memory. 

Another dimension of increasing occupancy comes at the TB level. If the input matrix is relatively small, the number of TBs 
launched might be too few to fill the GPU resources. It is necessary to ensure that even for small problem sizes, enough 
TBs are launched. For example, through atomic operations, several TBs can collaboratively work on the same output. 
Although there will be an overhead due to the increase in atomic operations, the overall performance is often better than 
launching one TB for the same problem. 

\subsubsection{Data Prefetching}
\label{subsubsec:prefetching}
%A third consideration is data prefetching. 
In order to hide the latency of doing FLOPs, a kernel should appear as if it 
is only performing memory operations. The latency of any computation should be hidden by a memory operation. This can be 
done through prefetching data while a useful computation is taking place. 

%\subsubsection{Instruction Mix}
%\label{subsubsec:mix}
%A last consideration in this section is the instruction mix. This is a common consideration in any kernel, regardless 
%of being memory-bound or compute-bound. Since performance is measured in FLOPs per second, other auxiliary instruction do not count 
%and should minimally contribute to the execution time. For example, integer instructions used in loops and memory address 
%calculations appear very frequently in any computer code. A common optimization is to use compiler technologies like loop unrolling 
%to minimize the number of auxiliary instructions. 
\subsubsection{Performance Oscillation}
\label{subsubsec:oscillation}
In some cases, it is important to 
consider the number of TBs launched with respect to the number of resident SMs on the GPU. Let's assume the 
simple case when all TBs have a balanced workload. We denote by $TB_R$ the number of  remaining TBs after the partial 
kernel execution, where all GPU SMs are fully occupied. Typically $TB_R$=\#TBs$\mod$\#SMs. Performance 
is penalized if it is relatively low ($TB_R$ $<<$ \#SMs), since the GPU will encounter a duration of low utilization 
while executing these remaining TBs. As a result, the performance drops for problem sizes that result in low $TB_R$, leading 
to a near-periodic oscillatory behavior in performance. 

There are two ways to resolve this problem:
\begin{enumerate}
	\item Detect problem sizes with low $TB_R$ values, and reconfigure the kernel by adjusting the number of launched 
	TBs to ensure that the value of $TB_R$ is as close as possible to the number of SMs. This approach requires 
	kernel reconfiguration or even developing a dedicated kernel. 
	\item If the GPU is kept at full utilization for most of the kernel execution time, the impact of the low utilization 
	execution round will be negligible. For example, a kernel launched with 1001 TBs (with balanced workloads) on a 10 SM GPU will 
	result in 100 full execution rounds and 1 round of low utilization. The GPU will be at full utilization for more than 
	99\% of the execution time, and the performance is maintained. This example is valid assuming that an SM executes exactly 
	one TB at a time. In general, a large number of TBs with balanced workloads will ensure that the GPU is fully busy for 
	most of the kernel execution time, and thus performance will be smooth across a wide spectrum of the problem size. 	
\end{enumerate}

%% file: structure.tex
%!TEX root = kblas.tex
KBLAS consists mainly of CUDA source codes and testing codes written in C. The CUDA source 
codes can be categorized into two groups:
\begin{enumerate}
	\item \emph{Kernel Templates}: These are CUDA header files, written in CUDA C/C++ extensions. These files 
	contain CUDA kernels abstracted using C++ templates. The precision as well as the tuning parameters are all 
	template parameters and are never specified in this kind of files. 
	Basic operations like initialization, addition, multiplication, ... etc are all 
	abstracted. 
	\item \emph{Kernel Instances}: These files instantiate, at compile time, a certain precision from the kernel 
	templates. They also specify the tuning parameters with which the instance would be created. These files do not 
	include any CUDA kernel. They provide the APIs that should be used by the user.  
\end{enumerate}

The above categorization has several advantages. It maintains one source file per kernel, with the ability to 
create as many instances as possible. The fact that the tuning parameters are in the instance files makes it 
straightforward to tune a KBLAS kernel, by simply changing the tuning parameters in the instance file. That is, the 
core kernels in the template files are not touched. In addition, the tuning parameters are all known at the compile 
time, which makes it easy for the compiler to optimize the code wherever possible, especially loop unrolling. Finally, 
implementing further optimizations for a certain kernel means only changing its corresponding template file without 
touching the instance files.

%% file: routines.tex
%!TEX root = kblas.tex
KBLAS implements the standard BLAS operation:
\begin{equation}
y = \alpha Ax + \beta y \mbox{ , }
\label{eqn:mv}
\end{equation} 
\noindent where $y$ is the vector to be updated (input/output), $x$ is an input vector, 
$A$ is an input general or Hermitian matrix, and $\alpha$ and $\beta$ are input scalars. 
A general matrix is processed using the KBLAS implementation of the General Matrix Vector 
Multiplication (GEMV) kernel. The Hermitian case is processed using either the Symmetric 
Matrix Vector Multiplication (SYMV), if the matrix is real, or its Hermitian version 
(HEMV), if the matrix is complex. KBLAS provides these kernels in single and multiple GPUs 
that exist on the same node. It also provides sophisticated new interfaces for the single GPU routines 
that can 
perform better for multiplication by a submatrix. 
Table \ref{tbl:kblas_routines} shows all the routines provided by KBLAS. The four supported precisions are 
represented by the letters s, d, c, and z, which represent the single, double, complex, and 
double complex precisions respectively. In any variant of the GEMV routine, $x$ $\in$ \{s, d, c, z\}. 
For the SYMV routines, $x$ $\in$ \{s, d\}, while for the HEMV case $x$ $\in$ \{c, z\}.

\begin{table}[h]
\begin{center}
\begin{tabular}{|m{3cm}|m{1.7cm}|m{3.2cm}|m{0.8cm}|m{0.9cm}|m{1.5cm}|}
\hline
Acronym & Matrix Type & API & single GPU & Multi-GPU & Standard Interface \\
\hline
GEMV & General & \texttt{kblas\_xgemv} & yes & no & yes \\
SYMV & Symmetric & \texttt{kblas\_xsymv} & yes & no & yes \\
HEMV & Hermitian & \texttt{kblas\_xhemv} & yes & no & yes \\
\hline
GEMV-OFFSET & General & \texttt{kblas\_xgemv\_offset} & yes & no & no \\
SYMV-OFFSET & Symmetric & \texttt{kblas\_xsymv\_offset} & yes & no & no \\
HEMV-OFFSET & Hermitian & \texttt{kblas\_xhemv\_offset} & yes & no & no \\
\hline
GEMV-MGPU & General & \texttt{kblas\_xgemv\_mgpu} & no & yes & no \\
SYMV-MGPU & Symmetric & \texttt{kblas\_xsymv\_mgpu} & no & yes & no \\
HEMV-MGPU & Hermitian & \texttt{kblas\_xhemv\_mgpu} & no & yes & no \\
\hline
\end{tabular}
\end{center}
\caption{KBLAS Supported Routines}
\label{tbl:kblas_routines}
\end{table}

%In addition to the routines described in Table \ref{tbl:kblas_routines}, KBLAS implements 
%the standard SCAL routins, which is a standard BLAS operation used for scaling vector by 
%an input constant. Although this routine is exposed to the user, it is used inside KBLAS as 
%an auxiliary routine. 
%We do not claim any contribution regarding this routine. 

There are two groups of KBLAS routines that do not have a standard interface in Table \ref{tbl:kblas_routines}. 
The first one 
provides routines that work on single GPU, and perform better than similar standard routines 
in the case of multiplication by a submatrix. 
%This feature will be discussed in detail in Section \ref{submatrix}. 
The second group of routines works on multi-GPUs that belong to the same node. According 
to our knowledge, there is no standard interface multi-GPU BLAS. KBLAS uses an interface similar to the 
one proposed in \cite{yamazaki_cpe}. 

Every routine in KBLAS has an asynchronous version, which has the same name appended by the "\texttt{\_async}" 
suffix. For example, \texttt{kblas\_sgemv} has a variant called \texttt{kblas\_sgemv\_async}. 
Default routines are synchronous in the sense that they are launched within the default stream on the GPU. 
Kernels that are submitted within this stream do not overlap execution with other kernels in different streams. 
The asynchronous version gives the user the ability to launch KBLAS kernels into a user-defined stream, potentially to 
overlap its execution with other tasks on the GPU. The asynchronous versions have an extra input argument in order 
to specify the launch stream. 

KBLAS also provides few APIs that facilitate using the multi-GPU routines. It provides functions that allocate 
the necessary memory space on each GPU in order to launch the multi-GPU routine. It also provides functions 
that offload the matrix to the GPUs involved in computation, based on the 1D cyclic block column layout. Since the 
size of the block column is required, KBLAS exposes such values through another set of functions. More details about 
KBLAS routines and APIs can be found in the KBLAS User's Guide. 

%% file: layout.tex
%!TEX root = kblas.tex
\subsubsection{Single GPU Routines}
\label{subsubsec:layout_single}
\begin{figure}[ht]
\centering
\subfigure[Unpadded LD]{
\includegraphics[width=0.8\linewidth]{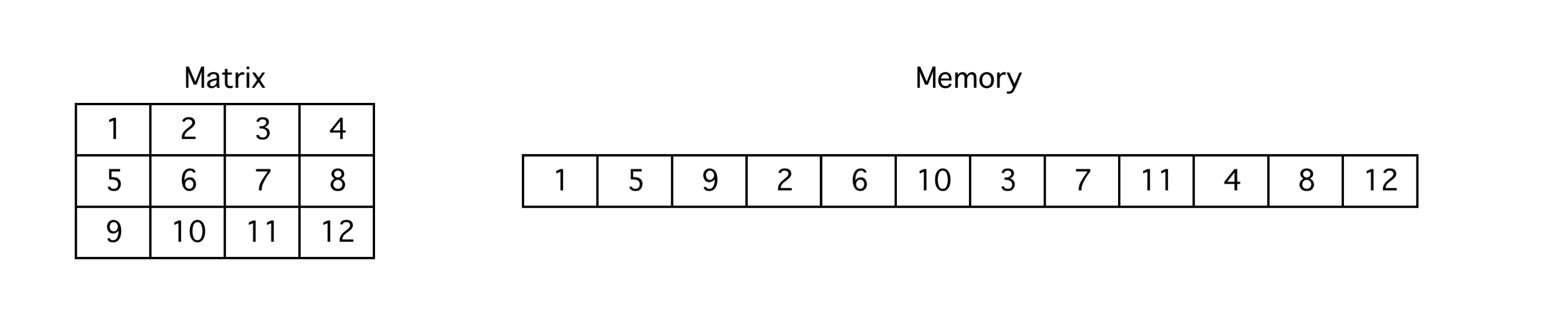}
\label{fig:col_major}
}
\subfigure[Padded LD]{
\includegraphics[width=0.8\linewidth]{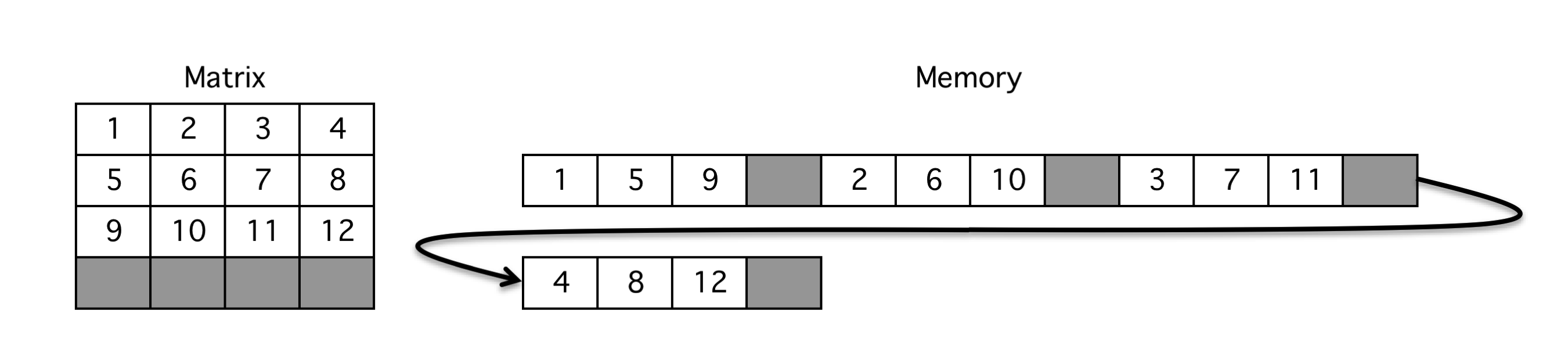}
\label{fig:col_major_lda}
}
\caption[]{Column major layout for single GPU kernels}
\label{fig:layout}
\end{figure}

KBLAS supports the BLAS standard column major format for storing matrices, which 
is the interface used by the standard BLAS and LAPACK libraries.  
%If the matrix is stored in the memory of one GPU, 
%Column elements are continuously stored in memory. Columns are 
%stored continuously with respect to each other, starting by the leftmost one. 
For single GPU routines, a matrix is described using 
its dimensions, a pointer to the first element in the matrix, and the leading dimension (LD), which 
gives the distance (in elements) between two adjacent elements in the same row. 
%The leading dimension is useful when processing a submatrix. 
Unless the matrix is padded, the leading dimension is equal to 
the number of rows of the matrix. Figure \ref{fig:layout} shows the column major storage with both unpadded and padded 
leading dimensions. A padded LD is used to preserve memory coalesced access. 
%Vectors in single GPU routines are stored as a 1D array in memory. 
%In general, a uniform stride exists between every two adjacent elements. 

\subsubsection{Multi-GPU Routines}
\label{subsubsec:layout_mgpu}
\begin{figure}[ht]
\centering
\subfigure[Original matrix]{
\includegraphics[height=0.3\linewidth]{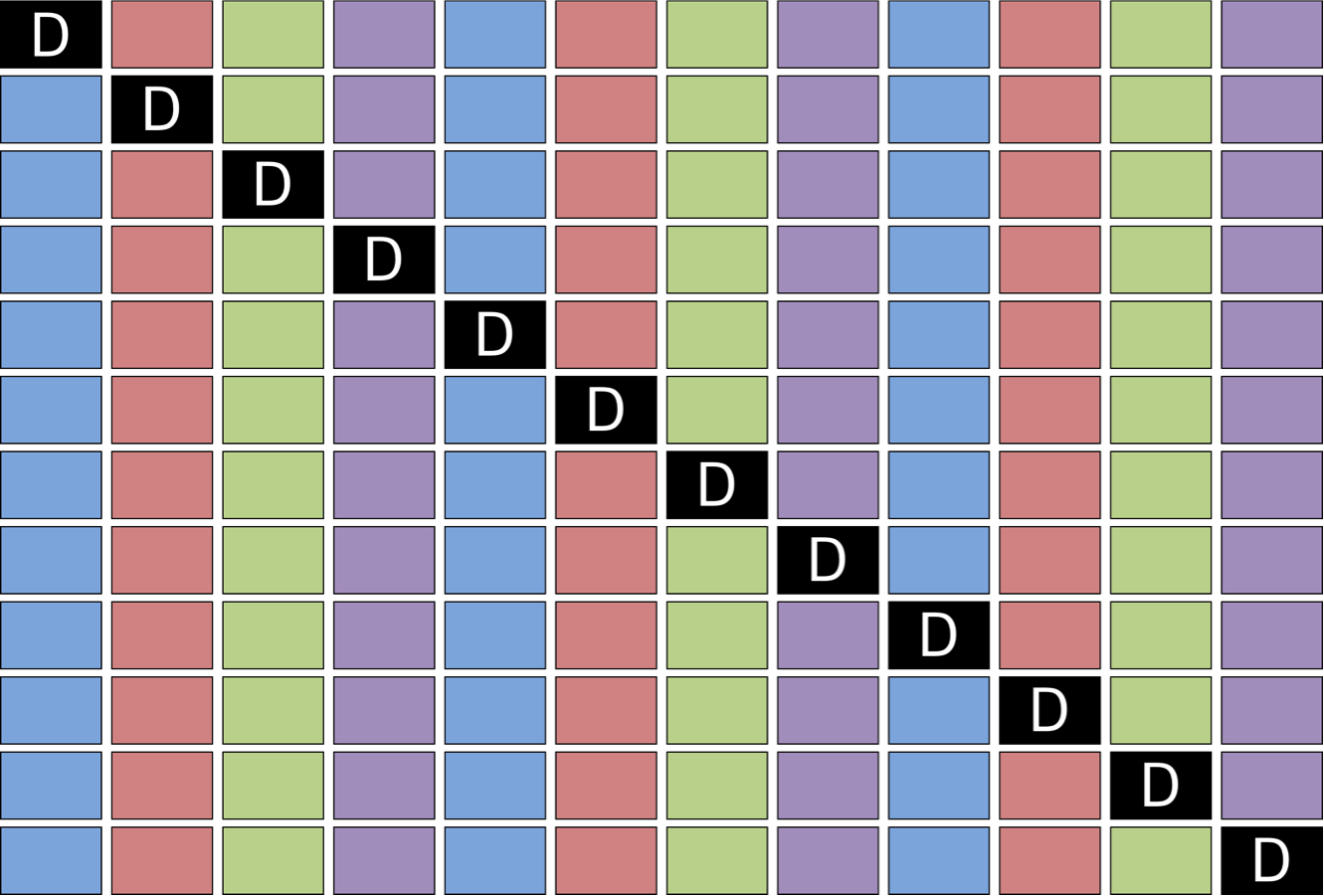}
\label{fig:big_matrix}
}
\subfigure[Distributed matrix]{
\includegraphics[height=0.3\linewidth]{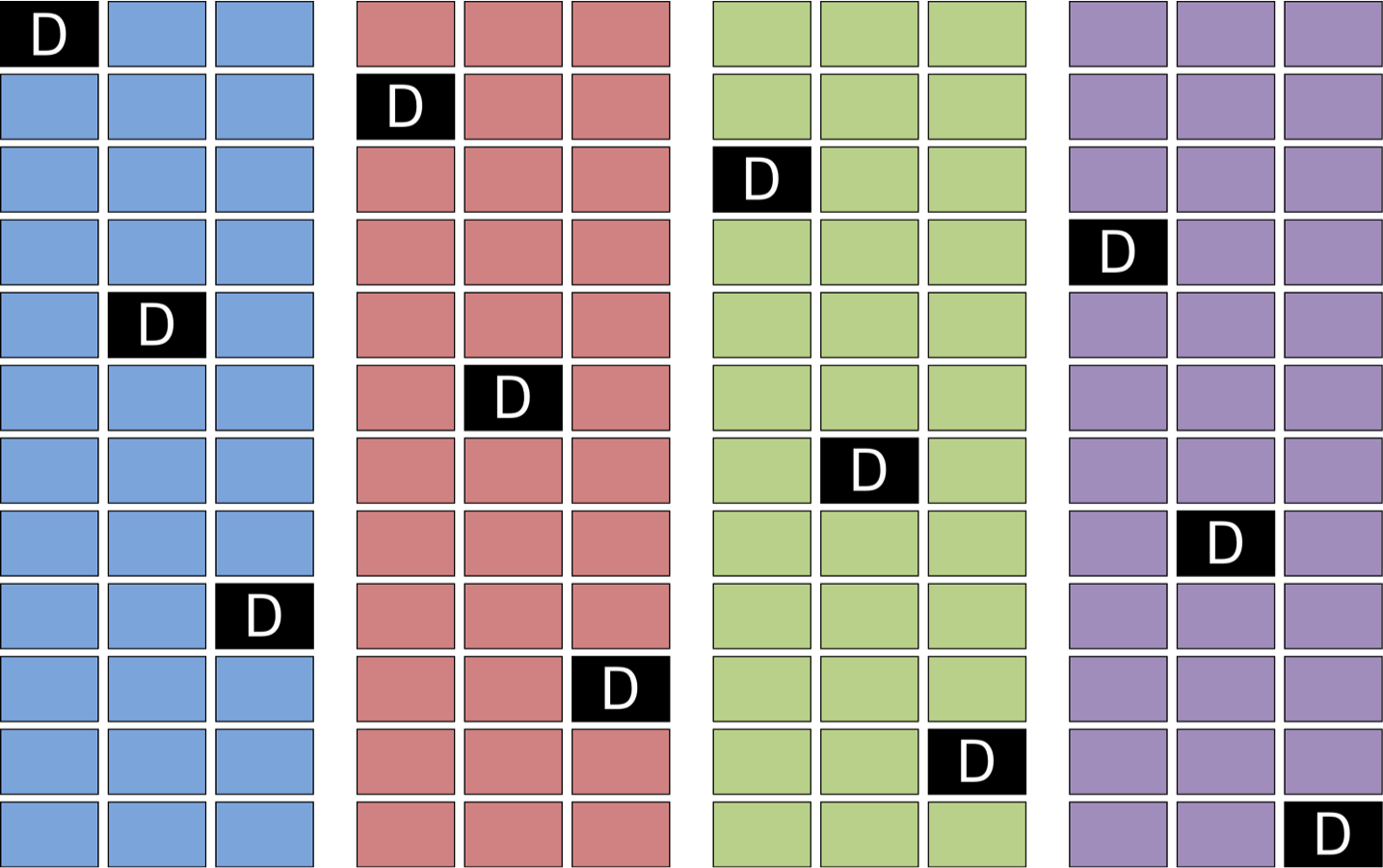}
\label{fig:big_matrix_mgpu}
}
\caption[]{Matrix layout for multi-GPU kernels}
\label{fig:layout_mgpu}
\end{figure}

If the matrix is processed using a multi-GPU routine, it will be distributed among the GPUs involved in 
computation. The data layout has to be suitable for higher-level algorithms utilizing matrix vector multiplications. 
We pay attention to a recent work, which is part of the MAGMA library, 
%These block algorithms are well implemented on a single GPU \cite{magma_brd} \cite{magma_trd}. 
%However, multi-GPU reduction is not yet mature. 
which provides a multi-GPU tridiagonal reduction for dense 
symmetric matrices \cite{yamazaki_cpe}. 
%The matrix layout proposed is the block-column 1D cyclic. 
In the MAGMA implementation, the matrix is distributed among GPUs in a 1D cyclic manner, by block columns. 
KBLAS uses the 
same layout for its multi-GPU routines, since it seeks to accelerate such reduction algorithms in open source libraries. KBLAS provides 
auxiliary routines for the allocation and distribution of matrices over multi-GPU for the aforementioned layout. 
Figure \ref{fig:layout_mgpu} shows an example for a dense symmetric matrix distributed over four GPUs. Each cell (block) represents a 
square submatrix. The black cells marked with `D' are the diagonal blocks. 

%Input and output vectors in multi-GPU routines are different from the single GPU case. 
By default, 
each GPU keeps a local copy of the original vectors $x$ and $y$. However, in some KBLAS kernels, $y$ 
is distributed, in segments, in a 1D cyclic manner. KBLAS provides routines for the 
distribution of the vector among GPUs. Output vectors, upon kernel termination, are sent back to the CPU, 
where a final reduction is performed to obtain the result. 

%% file: impl.tex
%!TEX root = kblas.tex
This section presents a detailed description of the implementation of KBLAS kernels. 
We will begin with single GPU kernels 
and leverage the design idea to the multi-GPU kernels. 
%======================================================================
\subsection{General Outlines}
\label{subsec:outlines}
We focus on describing the processing 
pattern of input matrices, since this is the dominant part when 
compared with reading or writing input and output vectors. 
Input matrices are processed in square blocks. The block size $nb$ is a tuning parameter. 

KBLAS usually launches more than one kernel per a BLAS operation. Sections \ref{grid_design} and \ref{tb_design} discuss 
the design of kernels that dominate the execution time of the BLAS operations. The dominant 
kernels are:
\begin{enumerate}
\item The general non-transposed MV kernel computing the product $\alpha Ax$.
\item The general transposed MV kernel computing the product $\alpha A^Tx$.
\item The symmetric/Hermitian MV kernel computing the product $\alpha \hat{A}x$, where $\hat{A}$ 
refers to the off diagonal blocks of $A$. 
\end{enumerate}
These kernels 
have a common processing strategy, and are designed in almost the same manner. Other kernels will be 
discussed separately in Sections \ref{beta_scaling} and \ref{symv_diag}.
%=========================================================================
\subsection{Grid Design}
\label{grid_design}
In general, the grid of any KBLAS kernel is organized as a 2D array of TBs. 
The size of the grid is ($\bar{X}$, $\bar{Y}$). While $\bar{Y}$ is a tuning parameter that 
the user can pick regardless of the problem size, $\bar{X}$ is decided based on the problem size and $nb$. Given a 
matrix $A_{\scriptscriptstyle m\times n}$, $\bar{X}$ is given by, 
\begin{equation}
\bar{X}=\begin{cases}\lceil \frac{m}{nb}\rceil & \mbox{; A is not transposed} \\ 
\lceil \frac{n}{nb}\rceil & \mbox{; A is transposed }\end{cases}
\label{eqn:grid_x}
\end{equation}
%decided based on two values. The first one is the dimension of the input 
%matrix. If the matrix is general, then one dimension $D$ is picked up depending on the processing 
%mode (transposed/non-transposed). If the matrix dimension is $M$$\times$$N$, then $D$ is equal to 
%$M$ is the matrix is non-transposed, and equal to $N$ otherwise. 
The value of $\bar{Y}$ determines the number of TBs working collaboratively (through atomic operations) on the same 
part of the output vector. From now on, we will consider square matrices, that is, $m=n=d$. 

Each TB is distinguish by its ($\bar{x}$, $\bar{y}$) coordinates. The $\bar{x}$ coordinate indicates the 
block row or the block column on which the TB will work. Eventually, 
TBs having different values of the 
$\bar{x}$ write to different parts of the output vector. TBs with different $\bar{y}$ coordinate 
and share the same $\bar{x}$ coordinate will write to the same segment in the output vector using atomic operations. 
%In such case, 
%writing is managed through atomic operations, mentioned in Section \ref{arch}. 

\begin{figure}[ht]
\centering
\subfigure[GEMV-N (non-transposed)]{
\includegraphics[height=0.25\linewidth]{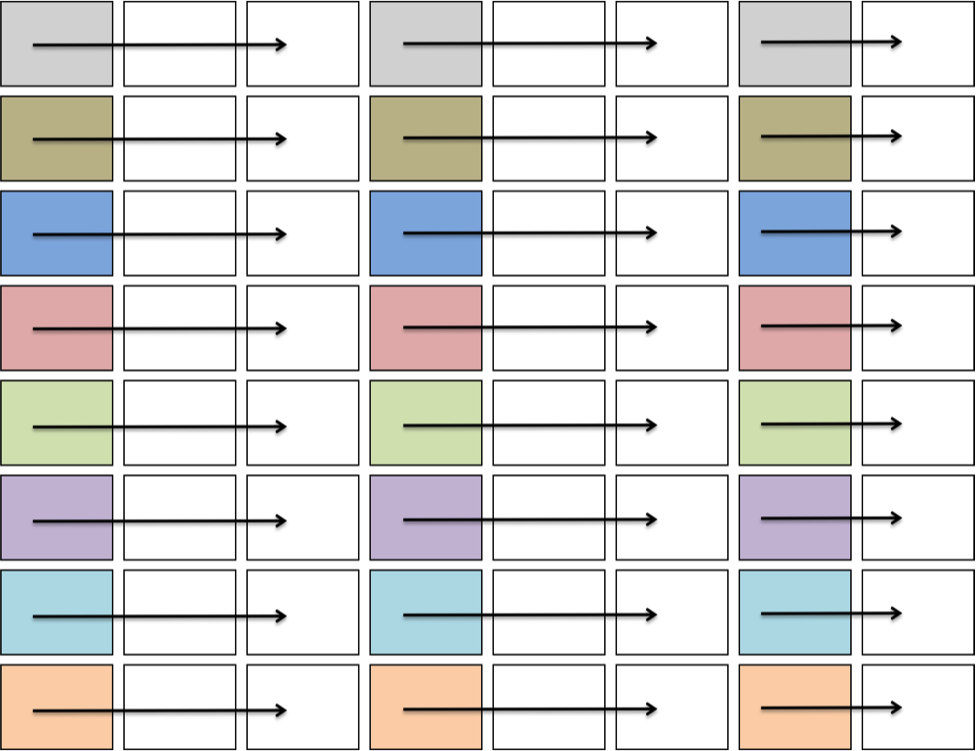}
\label{fig:gemvn_mv}
}
\subfigure[GEMV-T (transposed)]{
\includegraphics[height=0.25\linewidth]{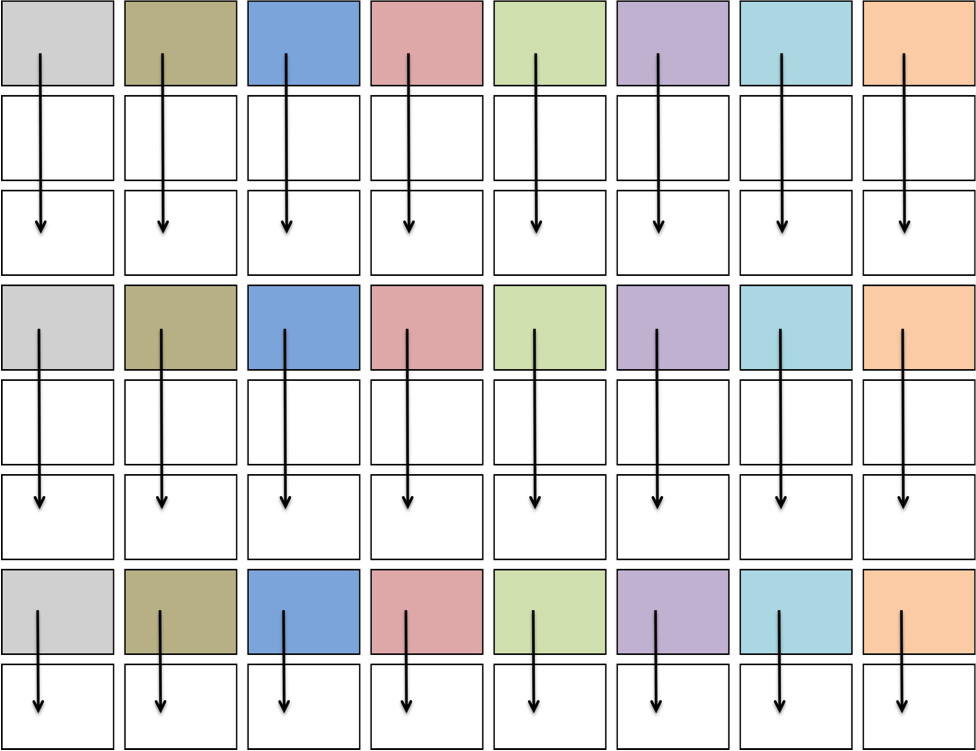}
\label{fig:gemvt_mv}
}
\caption[]{Movement of TBs in a GEMV operation. TBs with the same color share the same value of $\bar{x}$.}
\label{fig:gemv_mv}
\end{figure}

All TBs traverse the input matrix, in blocks, either horizontally or vertically. The movement direction depends 
on the operation to be performed. In the case of a GEMV operation, the movement is decided by the 
input character that specifies whether the matrix is transposed. If it is, then TBs will move vertically. 
Otherwise, the movement is horizontal. Figure \ref{fig:gemv_mv} shows both possibilities. 
In either case, TBs are programmed to traverse the entire matrix. At the beginning of execution, 
each TB must decides its starting point (first block) and its workload (number of blocks to be 
processed by this TB). TBs sharing the same $\bar{x}$ values will collaboratively process an entire block row or and 
entire block column of the matrix. The total workload $W$ for these TBs is given by
\begin{equation}
W = \left \lceil \frac{d}{nb} \right \rceil
\label{eqn:total_workload}
\end{equation}
Then each TB computes its workload share $w$ and its starting point $s$ as follows
\begin{equation}
w = \frac{d}{\bar{Y}} + \left\{\begin{matrix}
0; \; \bar{y} \geq d\bmod{\bar{Y}}\\ 
1; \; \bar{y} < d\bmod{\bar{Y}}
\end{matrix}\right.
\label{eqn:local_workload}
\end{equation}
\begin{equation}
s = \bar{y}\left \lfloor \frac{W}{\bar{Y}} \right \rfloor + min(\bar{y}, W\bmod{\bar{Y}})
\label{eqn:starting_point}
\end{equation}
Equations \ref{eqn:local_workload} ensures a minimum load imbalance among TBs. Equation \ref{eqn:starting_point} allows 
each TB to process adjacent blocks in the matrix. According to the input matrix size and grid configuration, some TBs 
might have their $w$ value equal to zero. In such a case, these TBs terminate immediately, writing nothing to the output 
vector.

\begin{figure}[ht]
\centering
\subfigure[SYMV-L (lower triangular)]{
\includegraphics[height=0.25\linewidth]{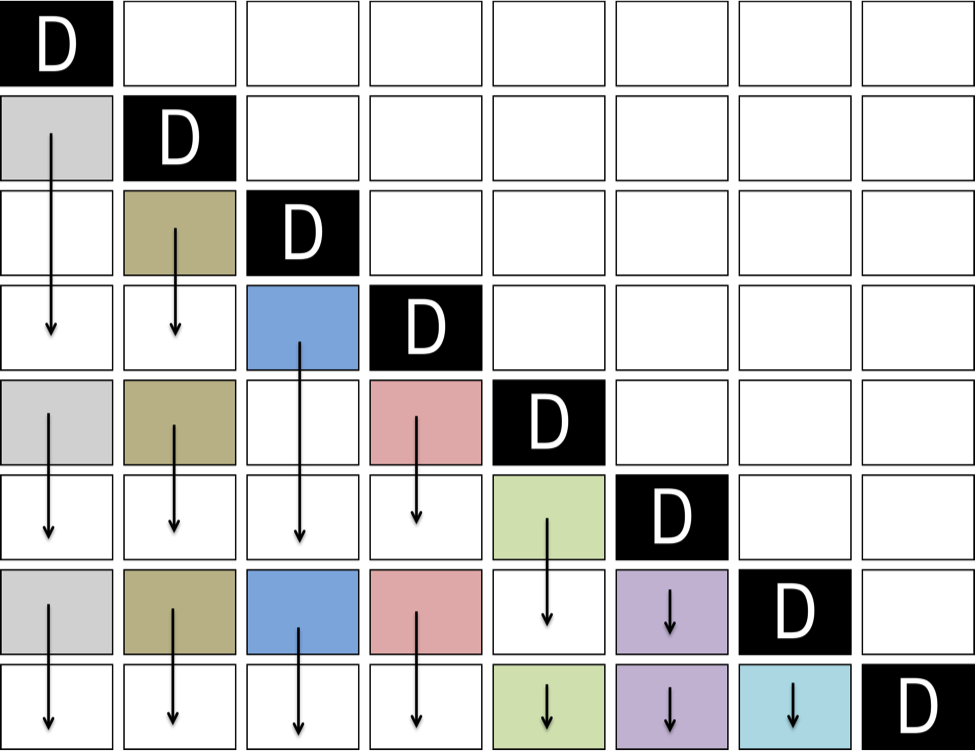}
\label{fig:symvl_mv}
}
\subfigure[SYMV-U (upper triangular)]{
\includegraphics[height=0.25\linewidth]{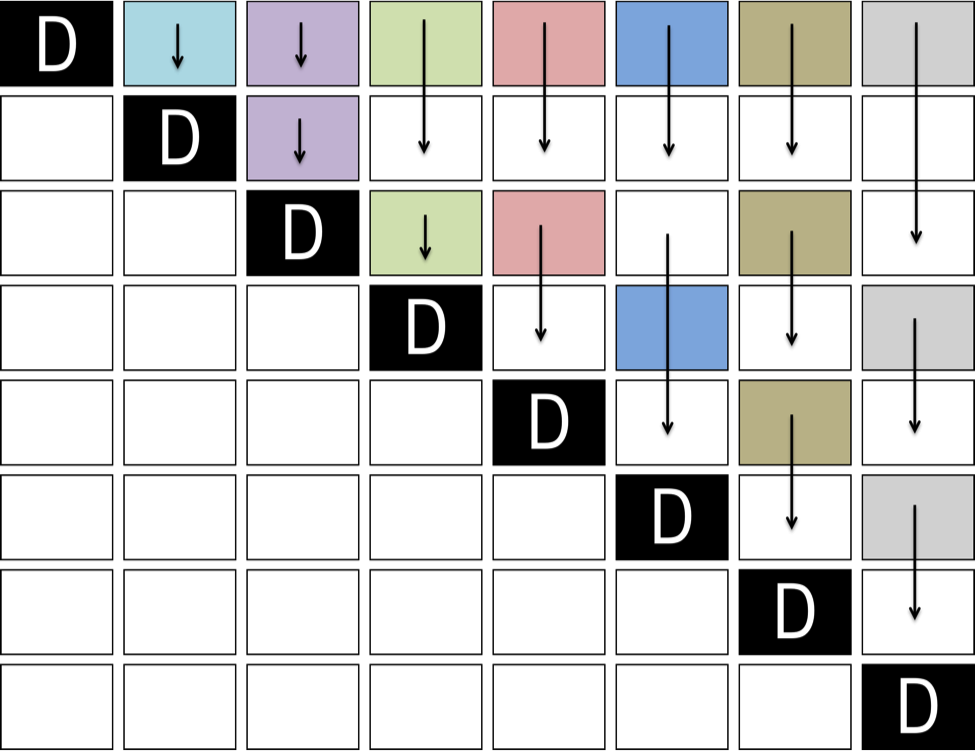}
\label{fig:symvu_mv}
}
\caption[]{Movement of TBs in a SYMV/HEMV operation. TBs with the same color share the same value of $\bar{x}$.}
\label{fig:symv_mv}
\end{figure}

If the matrix is symmetric (SYMV/HEMV kernel), then TBs are programmed to process either the upper or the lower triangular part. As shown in 
Figure \ref{fig:symv_mv}, diagonal blocks, colored in solid black, are processed separately. This is because they have a different 
processing pattern from off-diagonal blocks.
% as discussed in Section \ref{symv_diag}. 
Off-diagonal blocks are processed in a 
similar manner to the GEMV kernel, except for the movement being vertical regardless of which triangular part is being processed. 
The vertical movement ensures contiguous data access for each TB. 
%The difference is that 
The total workload $W$ is not constant across block columns, and is no longer determined by Equation \ref{eqn:total_workload}. In fact, each block row/block column has a unique 
value of $W$ that is given by 
\begin{equation}
W=\left\{\begin{matrix}
\left \lceil \frac{d}{nb} \right \rceil-i-1 \mbox{\hspace{2mm}; lower triangular}
\\ 
\mbox{\hspace{10mm}} i \mbox{\hspace{10mm}; upper triangular}
\end{matrix}\right. \mbox{\hspace{1mm},}
\label{eqn:total_workload_symv}
\end{equation}
where $i$ is the index of a block column, typically 0, 1, 2, $\cdots$, from left to right. 

The same strategy for TB movement applies for the multi-GPU kernels. The movement is horizontal only for the non-transposed GEMV, 
and vertical otherwise. The different is that the movement will be applied to the local submatrix stored on the GPU, as shown in 
Figure \ref{fig:big_matrix_mgpu}, instead of the whole matrix.  
%=========================================================================
\subsection{Thread Block Design}
\label{tb_design}
\begin{figure}[ht]
\centering
\includegraphics[height=0.6\linewidth]{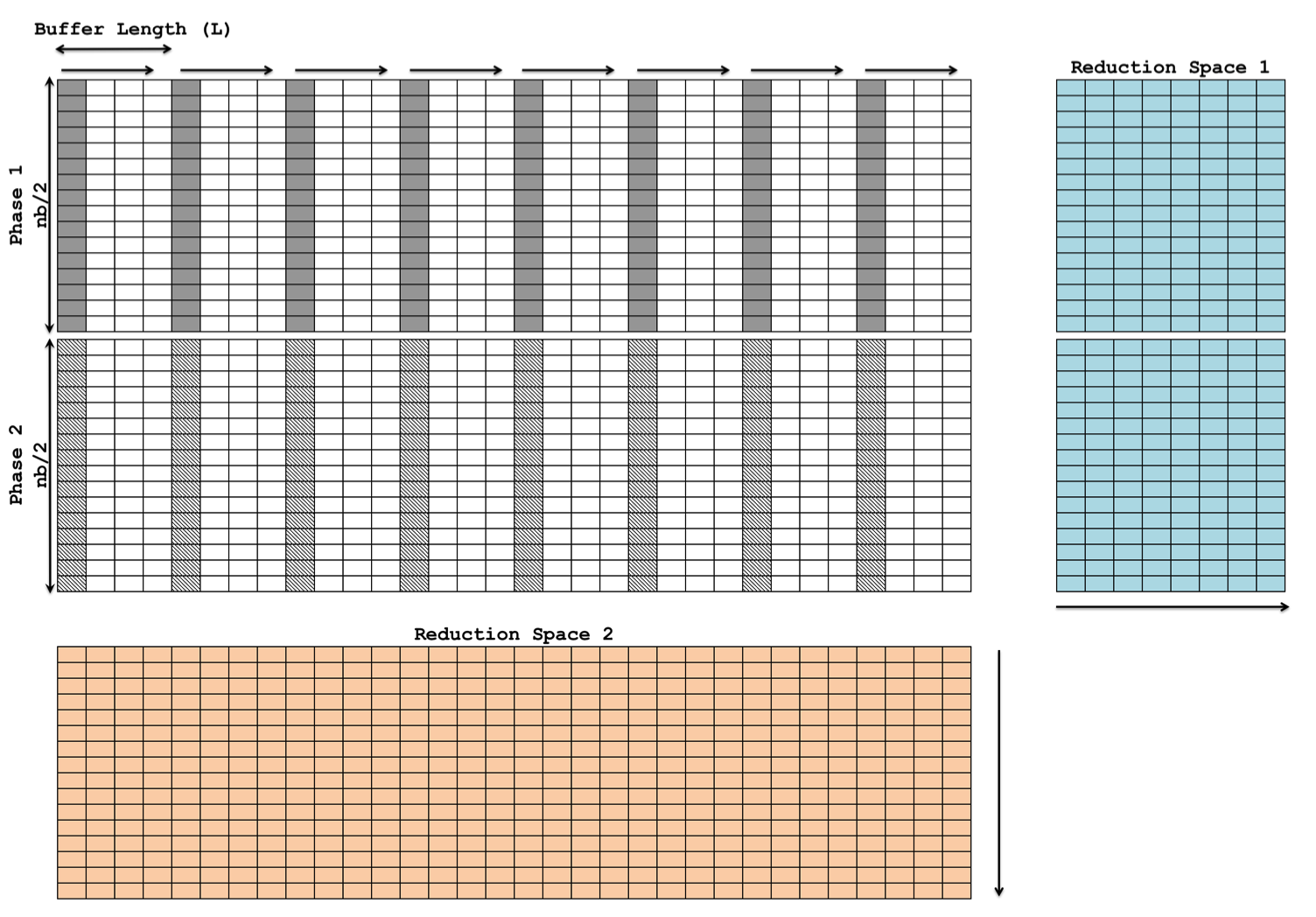}
\caption[]{Thread block design}
\label{fig:tb_design}
\end{figure}

Each TB is designed as a 2D array of threads. The size of the TB is ($\bar{P}$, $\bar{Q}$), where $\bar{P}$=$nb$ and 
$\bar{Q}$ is a third tuning parameter. Each thread is distinguished by its ($\bar{p}$, $\bar{q}$) coordinates. 
For a certain value of $nb$, $\bar{Q}$ controls the total number of threads, the amount of shared memory required 
for local reductions, and most importantly, the register pressure within a single TB. For simplicity, we will discuss an example 
of $\bar{P}$=$nb$=32 and $\bar{Q}$=4, shown in Figure \ref{fig:tb_design}. 

The design of TBs maximizes memory throughput through double buffers. Each square block of the matrix is processed in two phases, as 
shown in Figure \ref{fig:tb_design}. At each phase a half block of dimension $\frac{nb}{2}\times nb$ is processed in register buffers. 
Meanwhile, a new half block is fetched from memory into another set of register buffers. The overlap between computation and memory prefetching 
spans multiple blocks of the matrix, meaning that processing in phase 2 overlaps phase 1 in the next block. 
%computation with a second half block overlaps prefetching the first half of the next block. 
The dark cells in Figure \ref{fig:tb_design} show the original positions of threads in the first half block. The hashed cells indicate the respective positions for 
these threads in the next half block. 
Threads are reorganized as a $\frac{nb}{2}\times 2\bar{Q}$ thread block (16$\times$8). That is, we end up with 8 thread columns, each 
consisting of 16 threads. A thread column is responsible for processing a $\frac{nb}{2}\times L$ (16$\times$4) rectangle, 
where $L$ is the register buffer length required, per thread, per half block. $L$ is given by
\begin{equation}
L = \frac{nb}{2\bar{Q}}
\label{eqn:buffer_length}
\end{equation}
Each thread keeps two register buffers of length $L$ to fetch its corresponding elements. For the example discussed 
in Figure \ref{fig:tb_design}, this means 
thread (0, 0) will be responsible for elements \{(0, 0), (0, 1), (0, 2), (0, 3)\} as well as elements \{(16, 0), (16, 1), (16, 2), (16, 3)\}.
Upon completion of processing, each thread spills its partial products to a reduction space located in shared memory. A synchronization point 
is enforced to make sure that all threads have written their partial products. 
%Afterwards, one thread column performs the necessary 
A reduction is done 
in shared memory before writing the result (using atomic operations) to the global memory. 

Thread columns always read the block horizontally, in order to maintain contiguous memory access. 
However, the number of partial products per thread as well as the frequency of the 
synchronization and reduction depends on the kernel being executed. All possibilities are summarized using a pseudo code in Figure 
\ref{fig:kblas_pseudocode}. In the pseudo code, abbreviations $LHB$ and $UHB$ refer to Lower Half Block and Upper Half Block, respectively.
Variables $ures$, $lres$, and $vres[]$ refer to register accumulators where each thread keeps its local partial result. 
The variables $u[]$ and $l[]$ are register buffers that are used to read elements from $UHB$ and $LHB$ respectively. 

If the kernel is a non-transposed GEMV (Algorithm \ref{alg:gemvn}), then only reduction space 1 is needed for the final reduction. Each thread produces 2 partial 
products, one for each half block. The reduction is performed per row. The total space required for reduction space 1 is given by,
\begin{equation}
\mbox{reduction space 1} = nb\times (2\times \bar{Q})
\label{eqn:rs1}
\end{equation}
The reduction in this case is needed only once. According to the grid design in Section \ref{grid_design}, TBs traverse horizontal adjacent 
blocks, accumulating the partial results of all blocks together. The overhead of synchronization and reduction is, therefore, negligible. 

If the kernel is a transposed GEMV (Algorithm \ref{alg:gemvt}), then only reduction space 2 is needed for the final reduction. Each thread produces $L$ partial 
products, one for each matrix column. The reduction is performed per column. The total space required for reduction is given by, 
\begin{equation}
\mbox{reduction space 2} = \frac{nb}{2}\times nb
\label{eqn:rs2}
\end{equation}
Like the non-transposed GEMV, the reduction is needed only once, since TBs traverse the matrix vertically, and can accumulate partial results 
from one block to another. 

If the kernel is a SYMV/HEMV (Algorithm \ref{alg:symv}), exclusive of diagonal blocks, then both reduction spaces are needed. Each off-diagonal block is processed 
twice: non-transposed and transposed. Each thread will have a total of 2+$L$ partial products. This case combines the two possibilities of the 
GEMV kernel. However, it differs in the frequency of sync-and-reduce steps. Since TBs always traverse half the matrix vertically, partial 
results of the transposed computation can be accumulated, but those of non-transposed computation cannot, since they belong to different parts 
in the output vector. This means that a sync-and-reduce step is required whenever a TB moves from one matrix block to another, in order 
to write the partial products into global memory. The number of reductions is equal to the number of processed off-diagonal blocks plus an extra reduction 
performed only once for transposed computation. 

\begin{figure}
%---------------------------------
\noindent\begin{minipage}[][][t]{0.48\textwidth}
\begin{algorithm}[H]
\KwData{$A$, $x$, $\alpha$}
\KwResult{$y$}
Compute $w$ and $s$ and navigate accordingly\;
$ures=lres=0$ \;
$u[]$ $\leftarrow$ $UHB$\;
\For{$j$ $\leftarrow$ $1$ \KwTo $w$}
{
$l[]$ $\leftarrow$ $LHB$\;
$ures\leftarrow$ $ures+UHB$$\times$$x$\;
\If{$j\neq w$}
{
Move right to the next block\;
$u[]$ $\leftarrow$ $UHB$\;
}
$lres\leftarrow$ $lres+LHB$$\times$$x$\;
}% end of for loop
Write $ures$ and $lres$ to SHMEM\;
Barrier\;
\If{$\bar{q}$ $=$ $0$}
{
$r\leftarrow$reduction in SHMEM\;
Write $\alpha\times r$ into $y$ using atomics\;
}
\vspace{1.4mm}
\caption{Non-transposed GEMV}
\label{alg:gemvn}
\end{algorithm}
\end{minipage}
%---------------------------------
\noindent\begin{minipage}[][][t]{0.48\textwidth}
\begin{algorithm}[H]
\KwData{$A$, $x$, $\alpha$}
\KwResult{$y$}
Compute $w$ and $s$ and navigate accordingly\;
$vres[]=0$ \;
$u[]$ $\leftarrow$ $UHB$\;
\For{$j$ $\leftarrow$ $1$ \KwTo $w$}
{
$l[]$ $\leftarrow$ $LHB$\;
$vres[]\leftarrow$ $vres[]+UHB^T$$\times$$x$\;
\If{$j\neq w$}
{
Move down to the next block\;
$u[]$ $\leftarrow$ $UHB$\;
}
$vres[]\leftarrow$ $lres+LHB^T$$\times$$x$\;
}% end of for loop
Write $vres[]$ to SHMEM\;
Barrier\;
\If{$\bar{q}$ $=$ $0$}
{
$r\leftarrow$reduction in SHMEM\;
Write $\alpha\times r$ into $y$ using atomics\;
}
\caption{Transposed GEMV}
\label{alg:gemvt}
\end{algorithm}
\end{minipage}
%---------------------------------
\begin{center}
\vspace{-1mm}
\noindent\begin{minipage}{0.48\textwidth}
\begin{algorithm}[H]
\KwData{$A$, $x$, $\alpha$}
\KwResult{$y$}
Compute $w$ and $s$ and navigate accordingly\;
$ures=lres=vres[]=0$ \;
$u[]$ $\leftarrow$ $UHB$\;
\For{$j$ $\leftarrow$ $1$ \KwTo $w$}
{
$l[]$ $\leftarrow$ $LHB$\;
$ures\leftarrow$ $ures+UHB$$\times$$x$\;
$vres[]\leftarrow$ $vres[]+UHB^T$$\times$$x$\;
\If{$j\neq w$}
{
Move down to the next block\;
$u[]$ $\leftarrow$ $UHB$\;
}
$lres\leftarrow$ $lres+LHB$$\times$$x$\;
$vres[]\leftarrow$ $vres[]+LHB^T$$\times$$x$\;
Barrier\;
Write $ures$ and $lres$ to SHMEM\;
Barrier\;
\If{$\bar{q}$ $=$ $0$}
{
$r\leftarrow$reduction in SHMEM\;
Write $\alpha\times r$ into $y$ using atomics\;
}
}% end of for loop
Write $vres[]$ to SHMEM\;
Barrier\;
\If{$\bar{q}$ $=$ $0$}
{
$r\leftarrow$reduction in SHMEM\;
Write $\alpha\times r$ into $y$ using atomics\;
}
\caption{Upper/Lower SYMV/HEMV}
\label{alg:symv}
\end{algorithm}
\end{minipage}
\end{center}
%---------------------------------
\vspace{-2mm}
\caption{Pseudo code of the core kernels in KBLAS}
\label{fig:kblas_pseudocode}
\end{figure}

%We emphasize that 
The design of TBs does not change going from one GPU to multi-GPUs. The scope of the design is 
a square block of the matrix, so no major differences in the design exist. 
%=========================================================================
\subsection{Scaling with $\beta$ in GEMV kernels}
\label{beta_scaling}
Sections \ref{grid_design} and \ref{tb_design} show how KBLAS computes the product $\alpha Ax$. However, a standard BLAS 
operation involves scaling $y$ with $\beta$. Although the scaling is a trivial operation, it cannot be performed inside the 
KBLAS kernel that computes $\alpha Ax$. This is because every segment of length $nb$ in the resulting vector is computed by 
multiple thread blocks using atomic operations. Since the CUDA programming model handles the executing 
of TBs transparently, we cannot determine the order of execution for TBs. This means we cannot assign the scaling operation to 
a particular TB, since it is not necessarily executed first. 

The solution is to perform the scaling operation in a separate kernel. The scale and the multiplication kernel are launched 
in order, so the scale operation must finish before the multiplication is performed. The scale kernel is similar to 
the standard level-1 BLAS  operation SCAL. 
%Exactly on TB is executed per a segment of length $nb$ in $y$. 
A KBLAS GEMV operation 
consists, therefore, of two successive kernels called transparently to the user. As discussed in Section \ref{symv_diag}, the scale 
kernel is not needed if the matrix is symmetric/Hermitian.
%=========================================================================
\subsection{Diagonal Blocks Processing in SYMV/HEMV kernels}
\label{symv_diag}
If the matrix is symmetric/Hermitian, then diagonal blocks are different from the off-diagonal blocks, in terms of 
the processing strategy. 
The difference is that only one triangular part of a diagonal block should be read from global memory.
Moreover, only one non-transposed computation per diagonal block is necessary, in contrast with two computations 
per a off-diagonal block. KBLAS processes diagonal blocks in a separate kernel.

One TB is launched per diagonal block. The entire block is fetched into shared memory. Then a mirroring step is performed to copy one 
triangular part to the other. This mirroring step eliminates the triangular part that should not be referenced, meaning that extra 
reads are done from global memory. After mirroring is done, the block residing in shared memory is multiplied by the 
corresponding segment in $x$. The contribution of this kernel to the total execution time of 
a SYMV/HEMV operation is negligible, specially if the matrix is large. 

The reason behind separation is to simplify the programming, since the two kernels obviously need different 
amounts of GPU resources. It is also better for the CUDA runtime to optimize the GPU utilization for each kernel individually.

Furthermore, it is possible to fuse the scaling operation with $\beta$ into this kernel, since exactly one TB is launched per diagonal 
block. The KBLAS SYMV/HEMV operation does not invoke the SCAL kernel mentioned in Section \ref{beta_scaling}. 

%=========================================================================
\subsection{Multiplication by a Submatrix}
\label{submatrix}
\begin{figure}[ht]
\centering
\subfigure[Coalesced access]{
\includegraphics[height=0.43\linewidth]{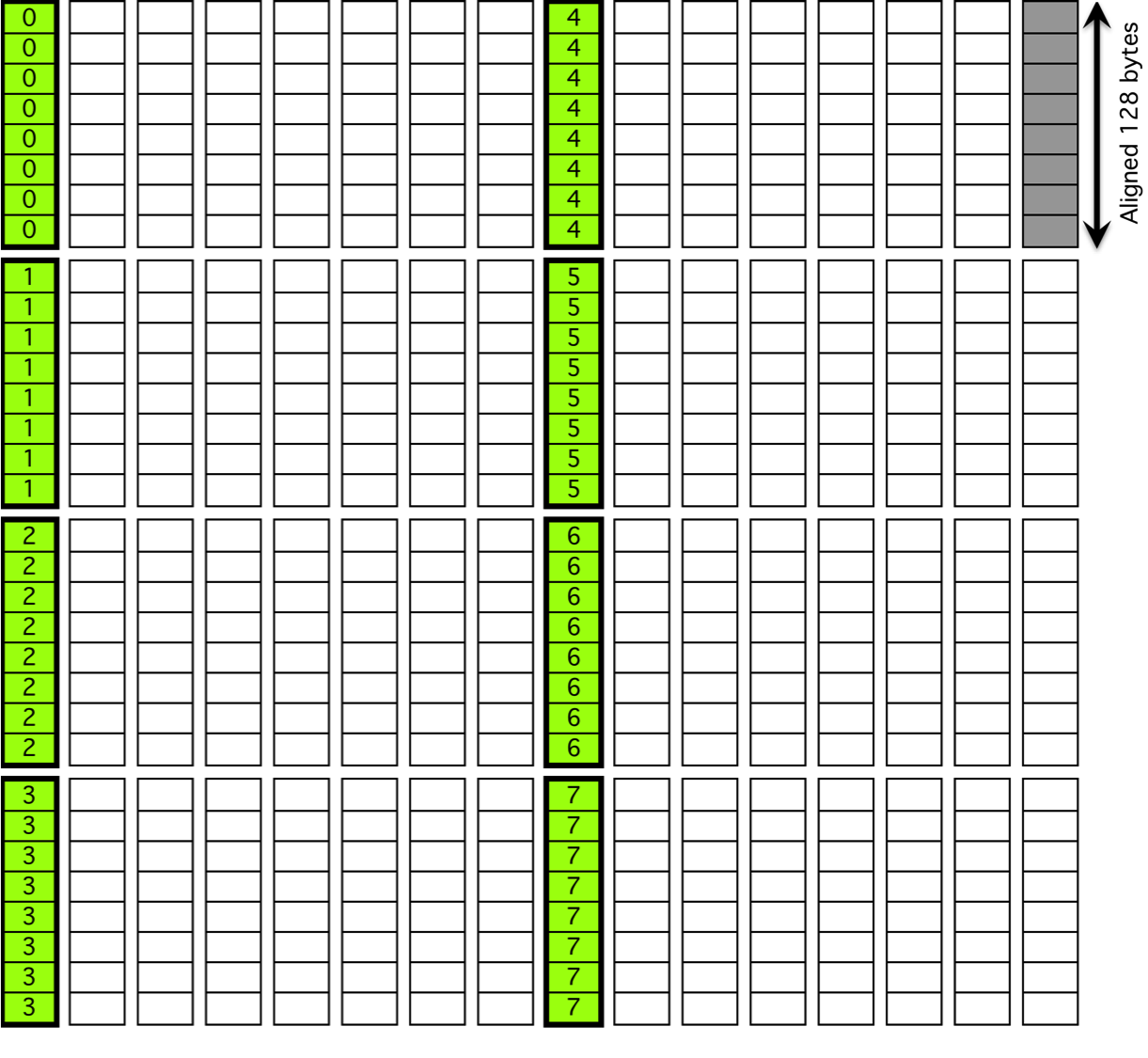}
\label{fig:submatrix_1}
}
\subfigure[Non-coalesced access]{
\includegraphics[height=0.43\linewidth]{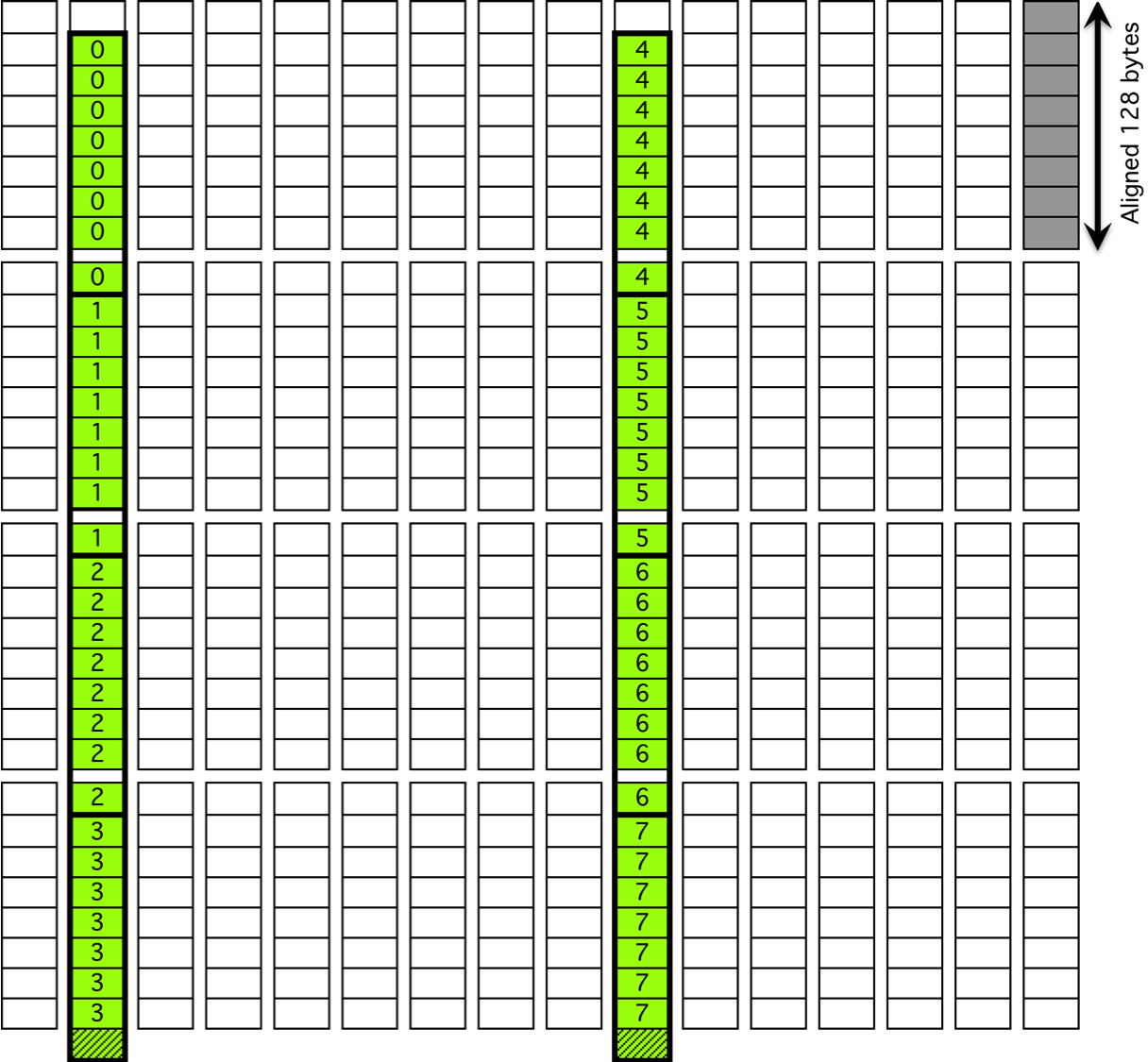}
\label{fig:submatrix_2}
}
\subfigure[Submatrix processing in KBLAS]{
\includegraphics[height=0.43\linewidth]{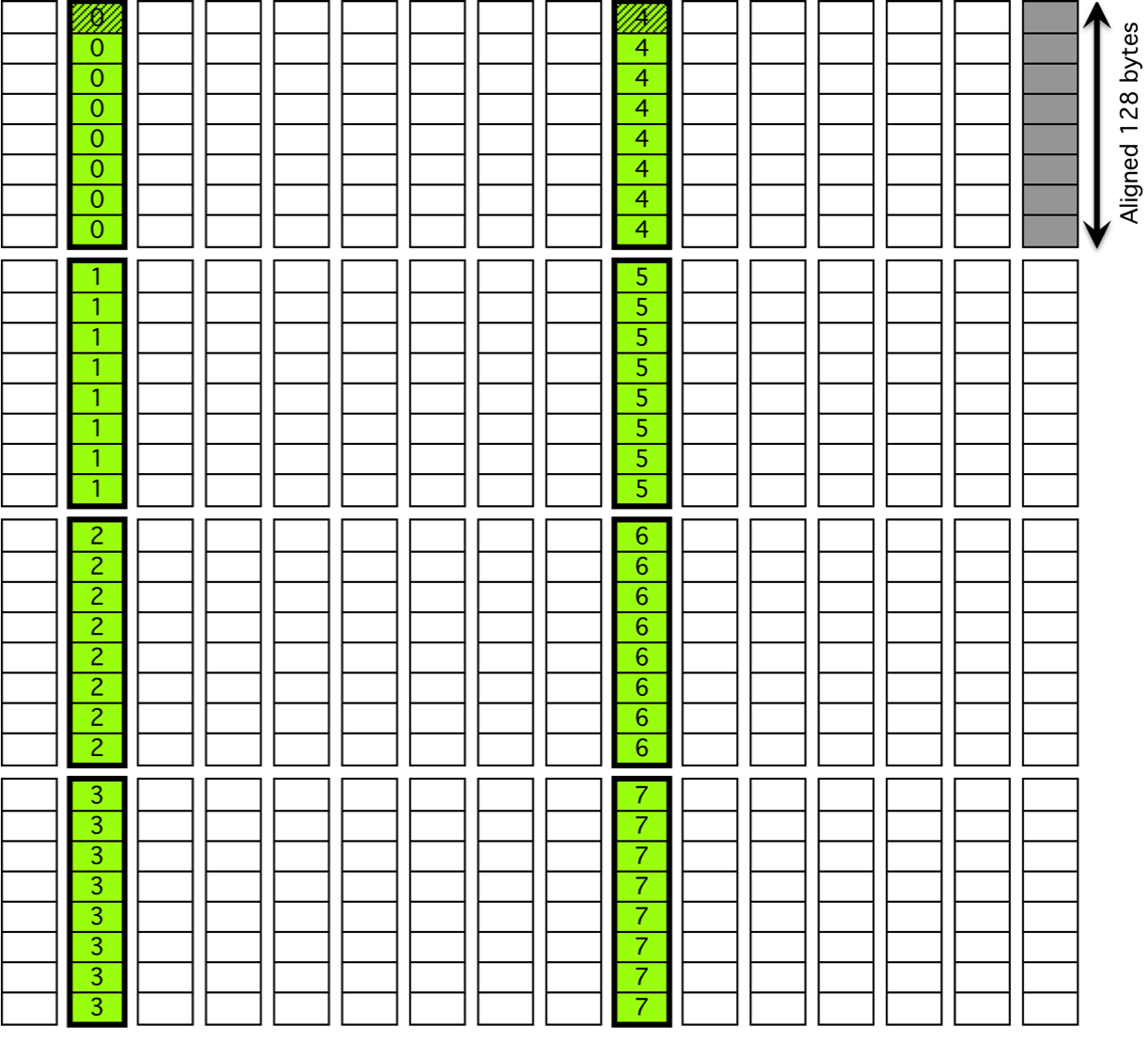}
\label{fig:submatrix_3}
}
\caption[]{Non-coalesced memory access due to processing a submatrix}
\label{fig:submatrix}
\end{figure}

In many LAPACK algorithms, matrix-vector multiplication is performed on submatrices. This is a typical situation in matrix reduction techniques. 
A common performance consideration is to pad the leading dimension so that the matrix can be stored in a fully 128 byte aligned memory space, and 
so, can be accessed in a coalesced manner. However, if the multiplication is done by a submatrix, coalesced access is not guaranteed and depends on the 
shifts in rows and columns. For example, Figure \ref{fig:submatrix} shows a matrix stored in a column major format. Each column consists of segments that 
are stored in aligned 128 bytes. If the entire matrix is to be processed using 8 thread columns, then it is straightforward to write a kernel 
that maps thread columns to access 
the matrix in a coalesced manner as shown in Figure \ref{fig:submatrix_1}. 
Now consider the submatrix obtained by skipping one row and one column of the original matrix, as shown 
in Figure \ref{fig:submatrix_2}. 
%However, processing a submatrix 
%(obtained by skipping the first row and the first column, like in Figure \ref{fig:submatrix_2})
Using the same kernel will result in non-coalesced memory access. 
This is because each memory request by a thread column will translate into 2 memory transactions, 
leading to an overhead in memory traffic. 

KBLAS proposes an additional set of BLAS routines with new interfaces other than the BLAS interfaces. As shown in Figure \ref{fig:submatrix_3}, the routine processes the same original 
matrix, but ignores the top rows and the left most column when computation is performed. This strategy does extra reads, but preserves memory coalesced 
access. The hashed cells refer to matrix elements that are read 
but ignored in computation. This is in contrast with Figure \ref{fig:submatrix_2}, where 
the hashed cells correspond to locations outside the matrix boundary that should be avoided by thread columns. 

In general, given an $m$$\times$$n$ matrix $A$, with a properly padded leading dimension, a multiplication by a 
submatrix $\hat{A}$ with the dimensions $\hat{m}$$\times$$\hat{n}$ can 
be translated into a multiplication by the submatrix $\bar{A}$ with dimensions $\bar{m}$$\times$$\bar{n}$, where 

\begin{equation}
\bar{m} = \mbox{min}(m,\left \lceil \frac{\hat{m}}{nb} \right \rceil \times nb) 
\mbox{\hspace{3mm}and\hspace{3mm}}\bar{n} = \mbox{min}(n,\left \lceil \frac{\hat{n}}{nb} \right \rceil \times nb)
\label{eqn:offset_dim}
\end{equation} 

Equation \ref{eqn:offset_dim} shows that the dimensions $\hat{m}$ and $\hat{n}$ are padded to the nearest values divisible by $nb$. This helps 
minimize the amount of extra global memory reads. 

The standard BLAS interface cannot pass information about the input matrix being a part of a bigger one. This is why we propose a new interface 
to convey these information to the kernel. The user should pass a pointer to the original matrix 
%(or to a submatrix that whose columns are perfectly 
%aligned in memory) 
as well as the offsets in rows and columns. Since KBLAS relies on perfect memory coalesced access while reading the matrix, 
these set of new routines are expected to perform better than standard ones when a submatrix is processed.

%% file: model.tex
%!TEX root = kblas.tex
We conduct a simple roofline model \cite{roofline} to predict the sustained peak performance of MV kernels. 
All the analysis in this section applies to a K20c GPU. 
%The roofline for a given kernel on a specific architecture 
%is identified using three components:
For a memory-bound kernel, the roofline can be identified using two components:
\begin{enumerate}
%\item The sustained peak performance of the architecture ($\bar{P}_{max}$). This can be experimentally obtained by 
%running an optimized compute-bound kernel, like matrix-matrix multiplication. 
\item The sustained memory bandwidth of the architecture ($\bar{B}_{max}$). This can be experimentally 
obtained by running sophisticated micro-benchmarks, like the STREAM benchmark~[\citeANP{stream1995} \citeyearNP{stream1995}; \citeyearNP{stream2007}].
\item The \emph{operational intensity} of the kernel ($I$). This is the ratio of the number of FLOPs executed to the number 
of bytes read or written to the memory. For some kernels, where data reuse is available, estimating this ratio is not 
straightforward, and cache hits and misses must be taken into account. However, in our case, no data reuse exists for 
the input matrix, which is the dominant part regarding both FLOPs and byte counts. The operational intensity of all KBLAS 
kernels are simple to estimate. 
\end{enumerate}
The sustained peak performance for a given kernel ($R$) is, then, given by:
\begin{equation}
%R = \mbox{min(}I\times \bar{B}_{max}, \bar{P}_{max}\mbox{)}
R = I\times \bar{B}_{max}
\label{eqn:roofline}
\end{equation} 
%, for both cases when ECC is on/off. Turning ECC on affects both 
%the bandwidth and the memory space available for the user.
%MV kernels are memory bound. Memory accesses dominate computation. So we can ignore $\bar{P}_{max}$ when estimating 
%the sustained peak performance for KBLAS. 
The limiting factor for performance is 
the sustained peak memory bandwidth $\bar{B}_{max}$. It should be noted that $\bar{B}_{max}$ is different from 
the theoretical peak bandwidth $B_{max}$. The latter is theoretically computed from the memory bus width and clock rate. 
The former is experimentally obtained using a GPU implementation of the STREAM benchmark~[\citeANP{stream1995} \citeyearNP{stream1995}; \citeyearNP{stream2007}]. 
Considering a K20c GPU, the memory bus width is 320 bits, and the memory clock rate is 2.6 GHz. 
The theoretical peak bandwidth is given by:
\begin{equation}
B_{max} = 2.6\mbox{Ghz}\times 40\mbox{ bytes}\times 2\mbox{ (DDR)} = 208\mbox{ GB/s}
\label{eqn:bw_theoretical}
\end{equation}
However, in order to get the sustainable memory bandwidth, a micro benchmark needs to run on the GPU and saturate its 
bus through simple memory operations. A typical STREAM benchmark runs four types of operations: \emph{copy}, \emph{scale}, 
\emph{add}, and \emph{triad}. Through CUDA kernel implementation for each of these operations, we are able to obtain the 
sustainable memory bandwidth on a K20c GPU. Results are summarized in Table \ref{tbl:stream_perf}. The peak bandwidth 
on a K20c is 150.64 GB/s (ECC on), and 175.24 GB/s (ECC off). All performance results in this paper are reported with ECC 
turned off. Therefore, $\bar{B}_{max}$ is equal to 175.24 GB/s from now on. 

\begin{table}
\centering
\begin{tabular}{|c|c|c|c|c|}
ECC & Copy & Scale & Add & Triad \\
\hline
On  & 148.99 & 150.64 & 149.99 & 150.09 \\
Off & 172.44 & 172.33 & 175.24 & 175.24 \\
\end{tabular}
\caption{STREAM performance in GB/s on a K20c GPU}
\label{tbl:stream_perf}
\end{table} 

The operational intensity ($I$) is simply the number of FLOPs executed divided by the number of bytes transferred between 
the CUDA cores and the DRAM. We will consider square matrices only for simplicity. An operation 
$y_{\scriptscriptstyle n\times 1}=\alpha A_{\scriptscriptstyle n\times n}x_{\scriptscriptstyle n\times 1} + \beta y_{\scriptscriptstyle n\times 1}$ 
involves, per output element in $y$, a vector product ($n$ multiplications + $n-1$ additions), two scalar multiplications (for $\alpha$ and $\beta$), 
and one addition for the final output. This sums up to a total of ($n+2$) multiplications and ($n$) additions per element, 
and ($n^{2}+2n$) multiplications and $n^{2}$ additions for the entire operation. This analysis applies to real precisions only. 
When using a complex precision, a single complex addition maps to 2 additions, and a single complex multiplication maps 
to 4 multiplications and 2 additions, summing up to 6 FLOPs in total. The total number of real multiplications and real additions in the 
complex case is, therefore, ($4n^{2}+4n$) and ($4n^{2}+8n$) respectively. 

While the FLOP count is the same for the GEMV and the SYMV/HEMV kernels, the byte count is not. Assuming that $x$ will \emph{ideally} be read once, 
and $y$ is \emph{ideally} read and written once, the total byte count for the GEMV kernel is $(n^2 + 3n)\times b$, where $b$ is the number 
of bytes used to represent one element in a certain precision. If the matrix is symmetric/Hermitian, either 
the upper or the lower triangular part is read. This gives a byte count of $(\frac{n(n+1)}{2} + 3n) \times b$. 

Finally, we can ignore the first order terms in both flop and byte counts, if $n$ is large enough. Table \ref{tbl:op_intensity} summarizes the approximate operational 
intensities for the GEMV and the SYMV/HEMV kernels in all four precisions. 

\begin{table}
\centering
\begin{tabular}{|M{2cm}|M{4cm}|M{4cm}|N|}
Precision & GEMV & SYMV/HEMV \\
\hline
S & $\frac{2n^{2} + 2n}{4(n^{2}+3n)}\approx 0.50$ 	& $\frac{2n^{2} + 2n}{4(0.5n^{2}+3.5n)}\approx 1.00$ &\\ [10pt]
\hline
D & $\frac{2n^{2} + 2n}{8(n^{2}+3n)}\approx 0.25$ 	& $\frac{2n^{2} + 2n}{8(0.5n^{2}+3.5n)}\approx 0.50$ &\\ [10pt]
\hline
C & $\frac{8n^{2} + 12n}{8(n^{2}+3n)}\approx 1.00$ 	& $\frac{8n^{2} + 12n}{8(0.5n^{2}+3.5n)}\approx 2.00$ &\\ [10pt]
\hline
Z & $\frac{8n^{2} + 12n}{16(n^{2}+3n)}\approx 0.50$ & $\frac{8n^{2} + 12n}{16(0.5n^{2}+3.5n)}\approx 1.00$ &\\ [10pt]
\end{tabular}
\caption{Operational intensities for KBLAS kernels}
\label{tbl:op_intensity}
\end{table}

%% file: perf.tex
%!TEX root = kblas.tex
%Notes:\\
%The figures for GEMV/SYMV on a submatrix shows the perf on dimensions of submatrices that 
%are part of a large matrix whose dim is around 16k\\
%As the original matrix gets smaller, we lose performance due to the extra reads and flops 
%we do. \\
%The oscillations in performance when multiplying by a submatrix corresponds to the offset 
%being friendly or not with respect to coalesced memory access. \\
%This effect is shown in the GEBRD performance, as we start to perform better on larger matrices 
%while the performance on small matrices stays similar to MAGMA \\
%I propose to remove the HRD story as I have no clue to the performance not being affected. My 
%guess is that HRD does GEMV on rectangular matrices. In this case cublas does better than KBLAS. \\
%BTW BRD and HRD routines in MAGMA use cuBLAS GEMV \\
%We still have crazy dips for certain dimensions on MGPU (specially ZHEMV at 20k and 40 k). These 
%points are hidden \\
\subsection{System Setup}
\label{subsec:setup}
The single GPU experiments are conducted on a system with 16-core Intel Xeon CPU E5-2650 (2.00GHz) and a 
Tesla K20c GPU (ECC off). The system runs Ubuntu 14.04.1 LTS, CUDA driver version 340.32, and CUDA Toolkit 5.5. 
The multi-GPU experiments are conducted on a system with a 16 core Intel Xeon CPU E5-2670 (2.60GHz), and 
equipped with 8 K20c GPUs (ECC off). The system runs CentOS release 6.3, CUDA driver version 331.62, and CUDA Toolkit 5.5. 
On both systems, we compare the performance of KBLAS against cuBLAS-5.5, MAGMABLAS-1.4.1, and CULA-R17. 
All results are properly averaged among multiple runs.

\subsection{Single GPU Performance}
\label{subsec:single_gpu_perf}
Considering a K20c GPU with ECC turned off, Table \ref{tbl:sustained_perf} translates the operational intensities 
listed in Table \ref{tbl:op_intensity} into their respective sustained peak performances. We will use these values 
to estimate how close KBLAS kernels are to their estimated performance bounds. 

\begin{table}
\centering
\begin{tabular}{|c|c|c|}
Precision & GEMV & SYMV/HEMV \\
\hline
S & 87.62  & 175.24 \\
D & 43.81  & 87.62  \\
C & 175.24 & 338.90 \\
Z & 87.62  & 175.24 \\
\end{tabular}
\caption{Estimated sustained peak performances (Gflop/s) for KBLAS kernels on a single K20c GPU (ECC off)}
\label{tbl:sustained_perf}
\end{table} 

\subsubsection{Performance of GEMV}
\label{subsubsec:gemv_perf}
\begin{figure}[ht]
\centering
\subfigure[SGEMV]{
\includegraphics[width=0.48\linewidth]{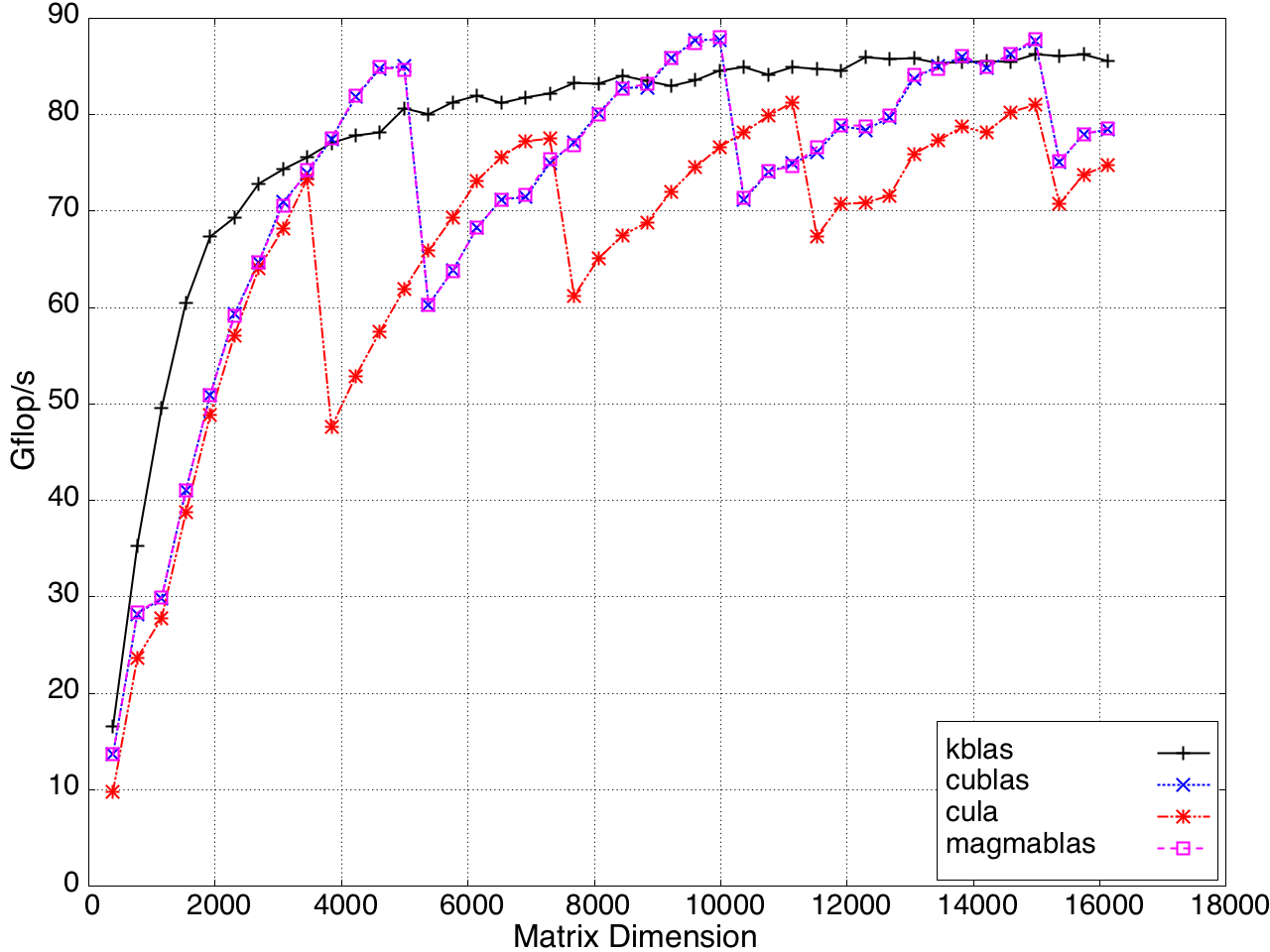}
\label{fig:sgemv}
}
\subfigure[DGEMV]{
\includegraphics[width=0.48\linewidth]{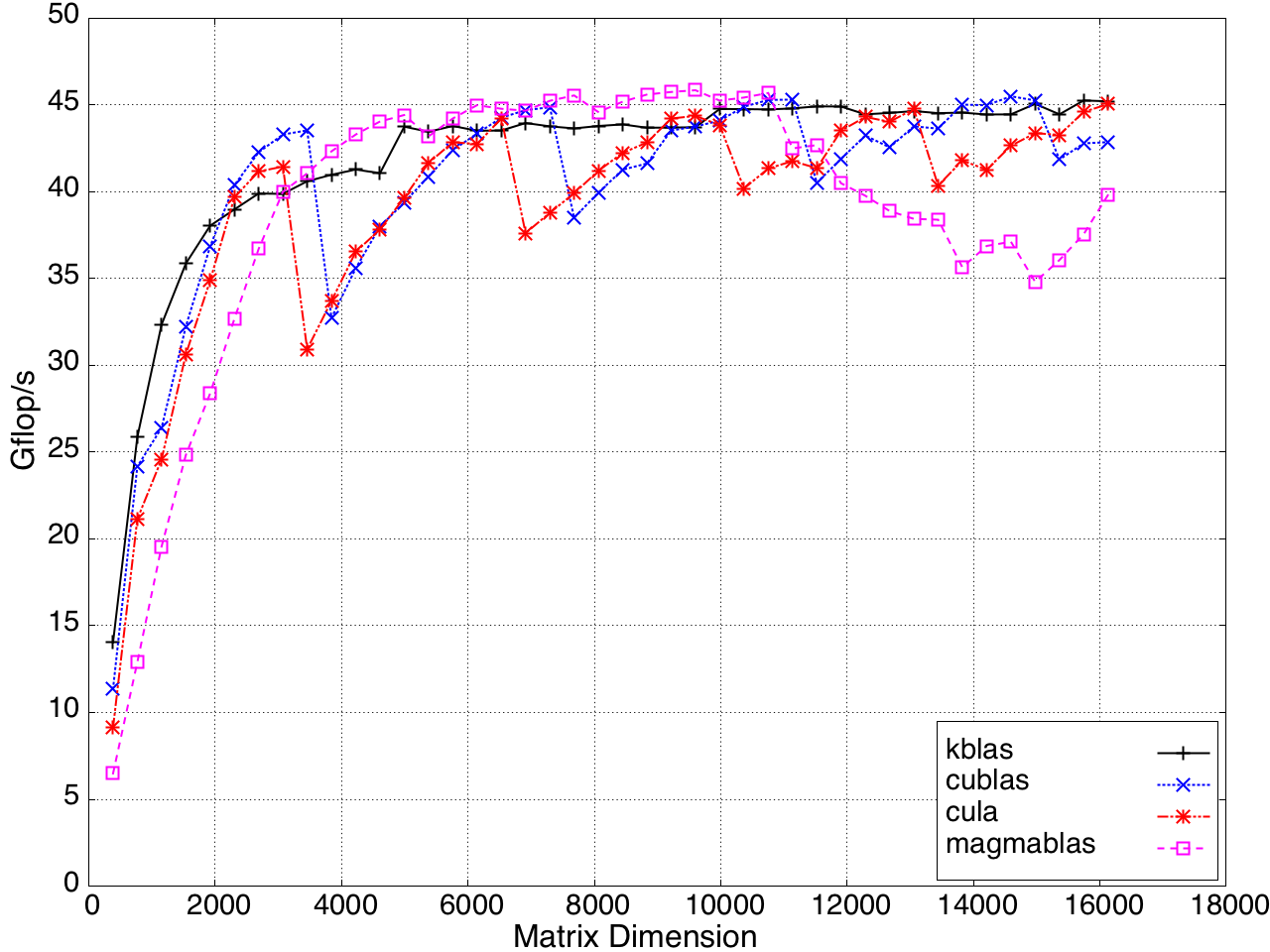}
\label{fig:dgemv}
}
\subfigure[CGEMV]{
\includegraphics[width=0.48\linewidth]{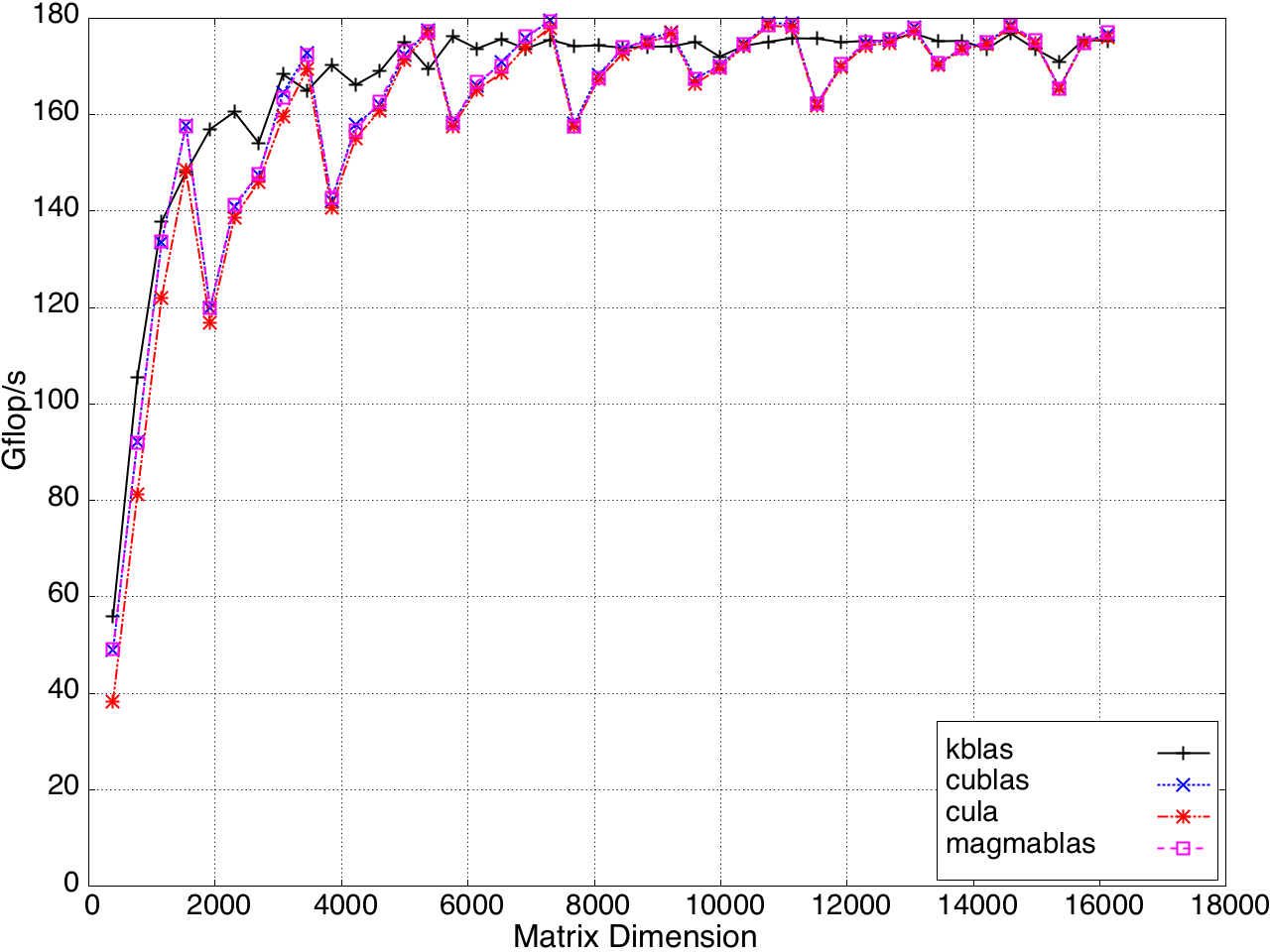}
\label{fig:cgemv}
}
\subfigure[ZGEMV]{
\includegraphics[width=0.48\linewidth]{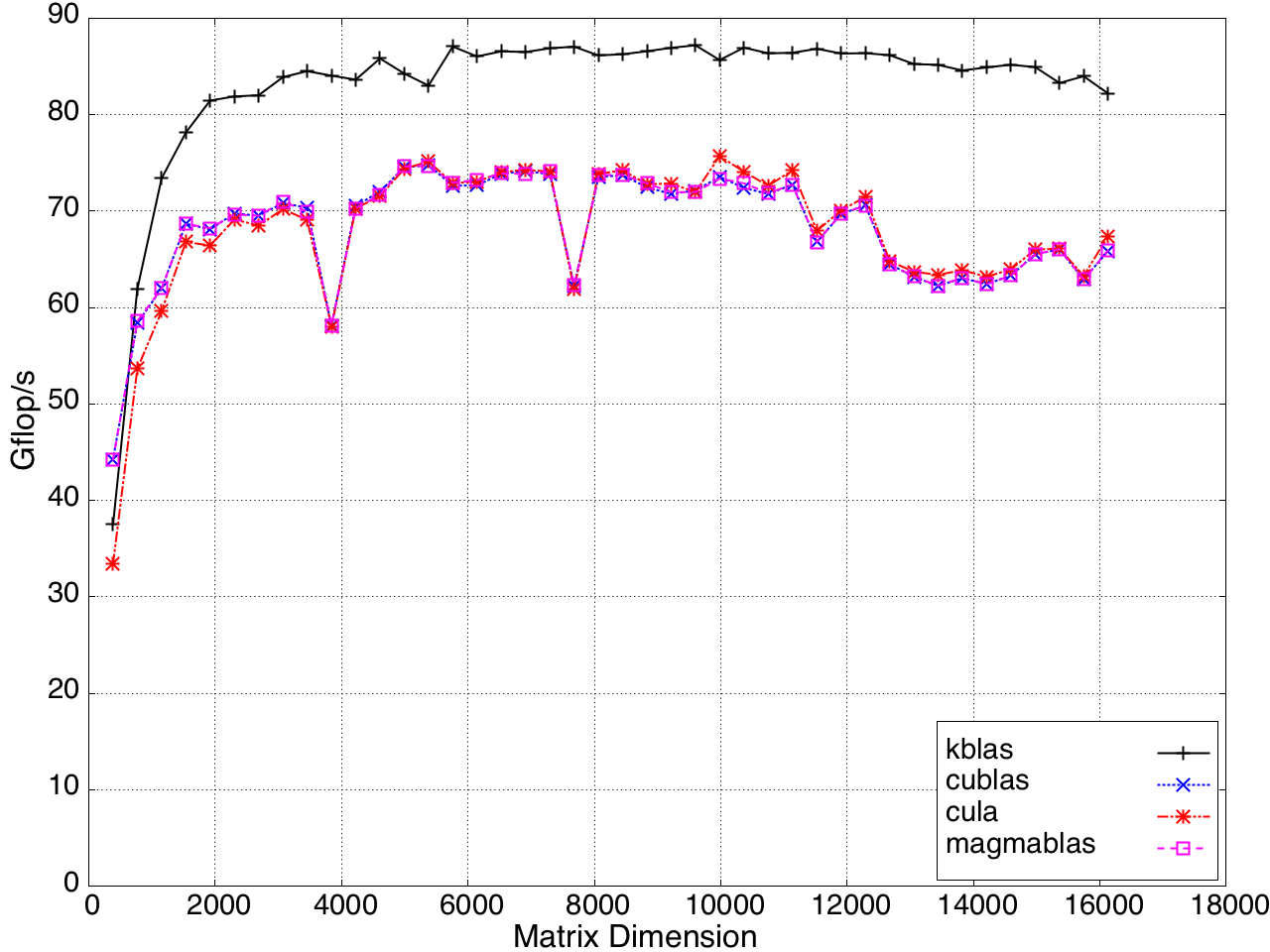}
\label{fig:zgemv}
}
\caption[]{GEMV Performance on a K20c GPU, ECC off}
\label{fig:gemv}
\end{figure}

Figure \ref{fig:gemv} shows, in all precisions, the performance of the GEMV kernel. 
%Apart from the oscillatory behavior, 
KBLAS, along with other libraries, is capable of asymptotically score the sustained peak performance, as determined 
by Table \ref{tbl:sustained_perf}, according to the performance model in Section \ref{sec:model}. 
This is the case in all precisions, except for the double complex 
precision, where KBLAS-ZGEMV has an performance improvement up to 40\% against the best competitor. We also note that 
in some cases, the performance scored by MAGMABLAS or CULA are identical to cuBLAS, which suggests that sometimes the 
vendor's implementation is invoked internally. 
We also observe that KBLAS has a is able to maintain its smooth performance, unlike other implementations 
that suffer from performance oscillations. 
%We observe a periodic oscillatory behavior in almost all implementations. These oscillations are already discussed 
%in Section \ref{subsubsec:oscillation}. However, we will show that, using the appropriate set of tuning parameters, 
%the the KBLAS-GEMV kernels can achiever a smoother performance curve.
%Section \ref{sec:tuning} discusses 
%this behavior from KBLAS, and how it can be recovered using the proper tuning parameters.

\subsubsection{Performance of SYMV/HEMV}
\label{subsubsec:symv_perf}
\begin{figure}[ht]
\centering
\subfigure[SSYMV]{
\includegraphics[width=0.48\linewidth]{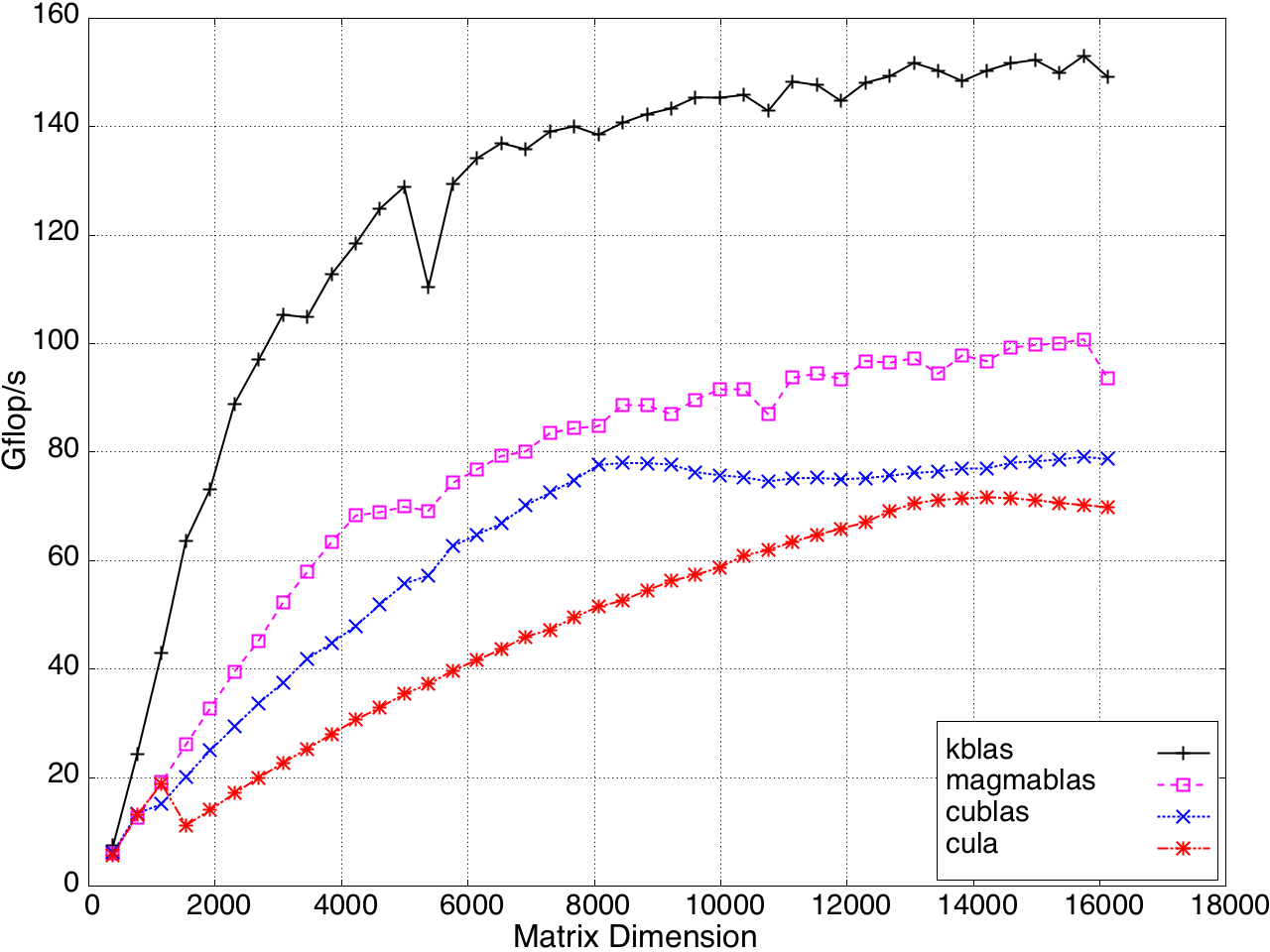}
\label{fig:ssymv}
}
\subfigure[DSYMV]{
\includegraphics[width=0.48\linewidth]{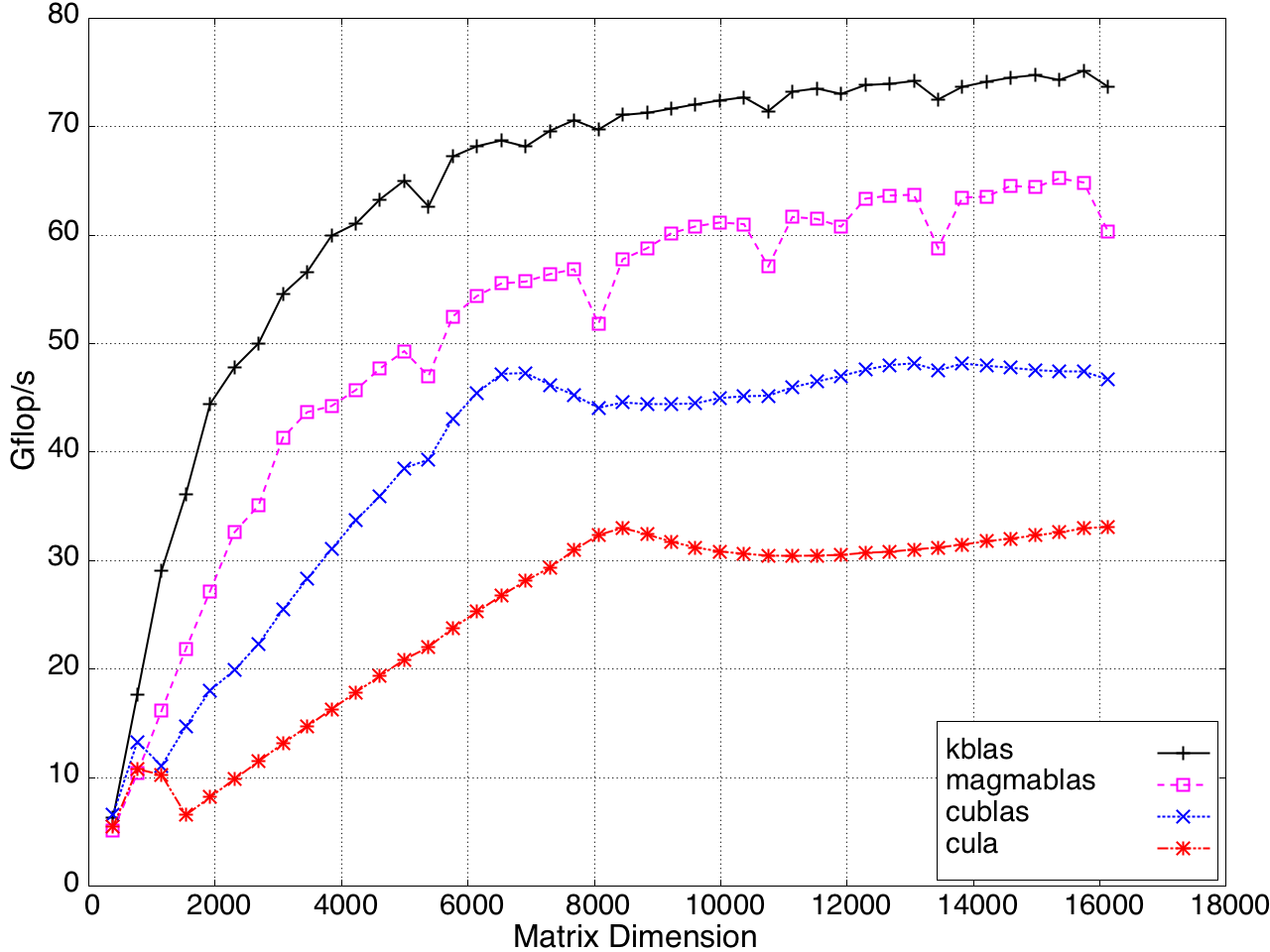}
\label{fig:dsymv}
}
\subfigure[CHEMV]{
\includegraphics[width=0.48\linewidth]{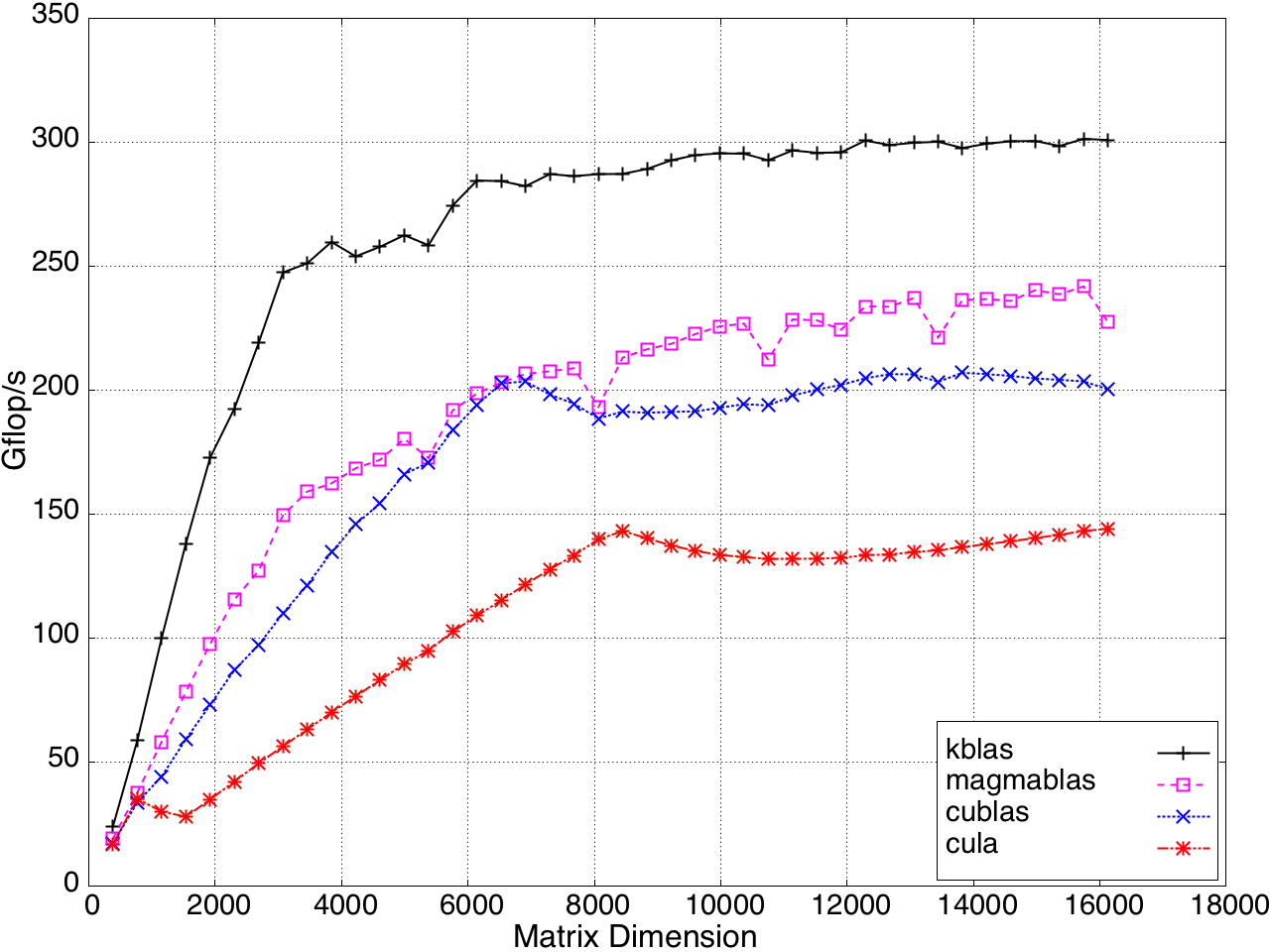}
\label{fig:chemv}
}
\subfigure[ZHEMV]{
\includegraphics[width=0.48\linewidth]{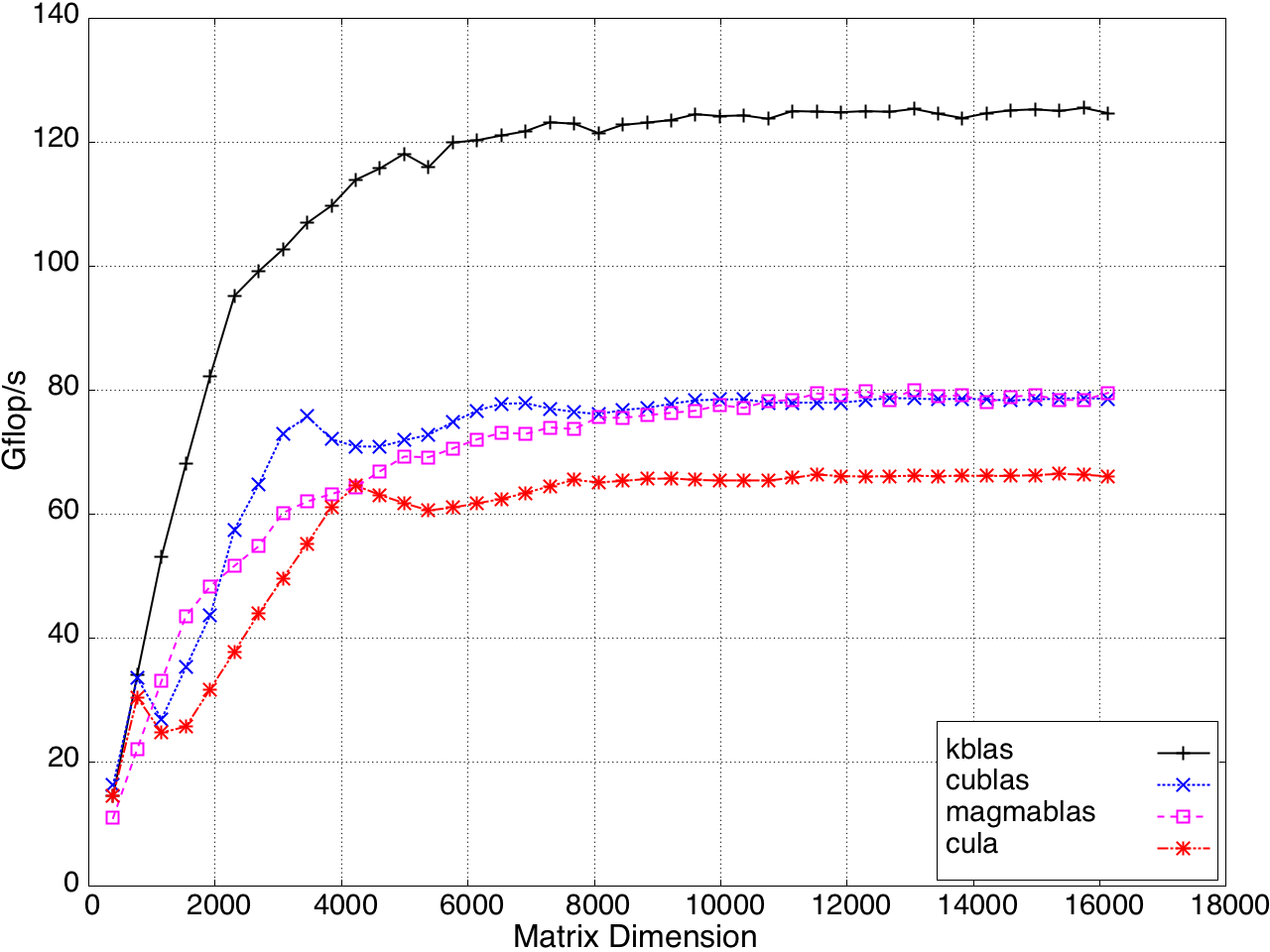}
\label{fig:zhemv}
}
\caption[]{SYMV/HEMV Performance on a K20c GPU, ECC off}
\label{fig:symv}
\end{figure}

When the matrix is Hermitian, KBLAS scores speedups against all other implementations. 
Against the best competitor (MAGMABLAS), KBLAS asymptotic speedup is up to 50\%, 15\%, 24\%, 55\% across 
all precisions, as shown in Figure \ref{fig:symv}. Comparing against the bounds listed in Table \ref{tbl:sustained_perf}, 
the performance is up to 89\%, 87\%, 90\%, and 72\% of the sustained 
peak performance. % estimated in Table \ref{tbl:sustained_perf}. 
Looking at relatively small matrices (4k or less), 
the speedups against MAGMABLAS are up to 2.46x, 1.83x, 1.80x, and 1.89x across all precisions. A variant of the KBLAS 
SYMV/HEMV kernel has been integrated into NVIDIA's cuBLAS library, starting version 6.0.

%------------------------------------------------------------------
\subsubsection{Performance for Submatrix Multiplication}
\label{subsubsec:offset_perf}
\begin{figure}[ht]
\centering
\subfigure[SGEMV]{
\includegraphics[width=0.48\linewidth]{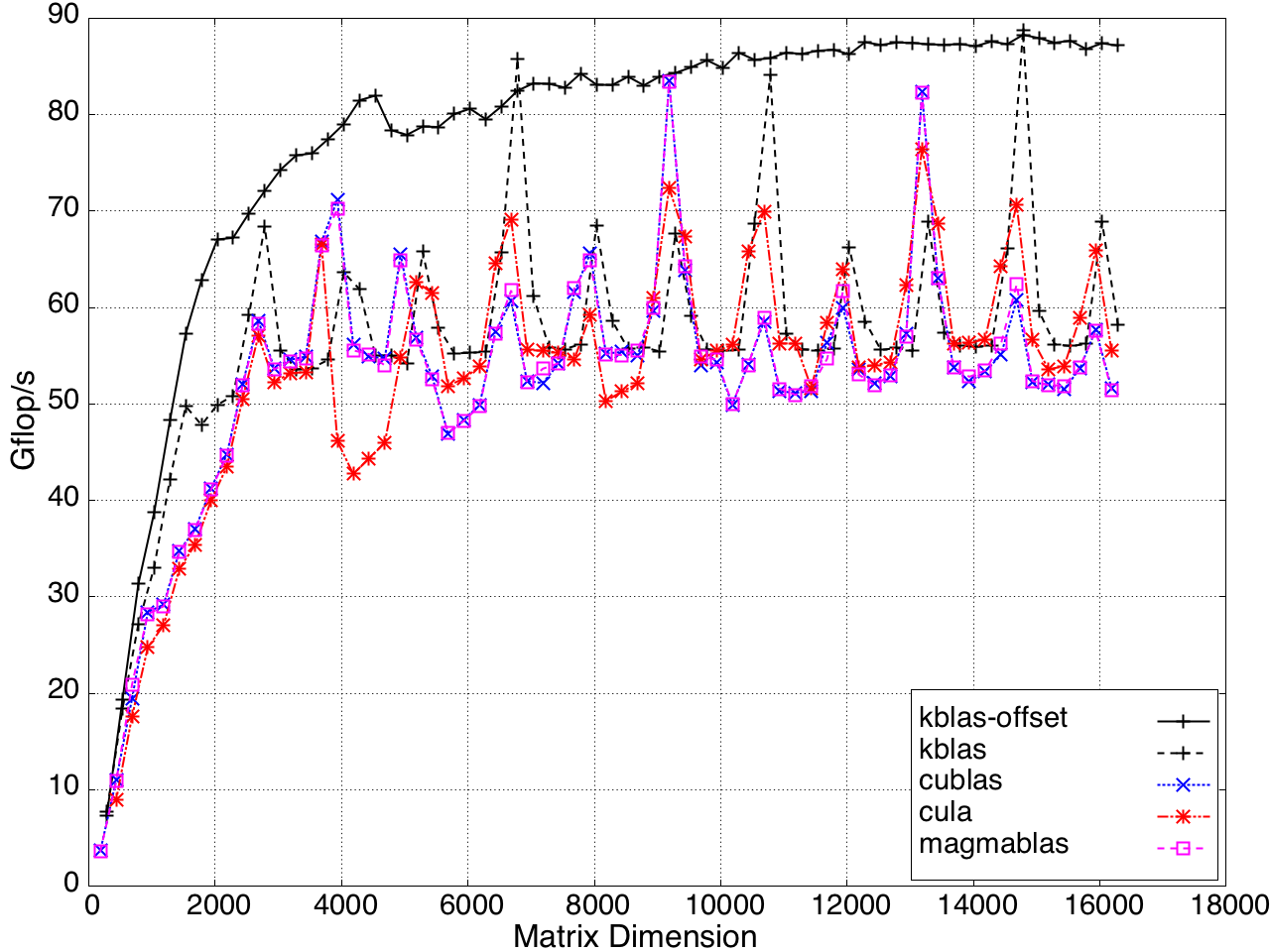}
\label{fig:sgemv_offset}
}
\subfigure[DGEMV]{
\includegraphics[width=0.48\linewidth]{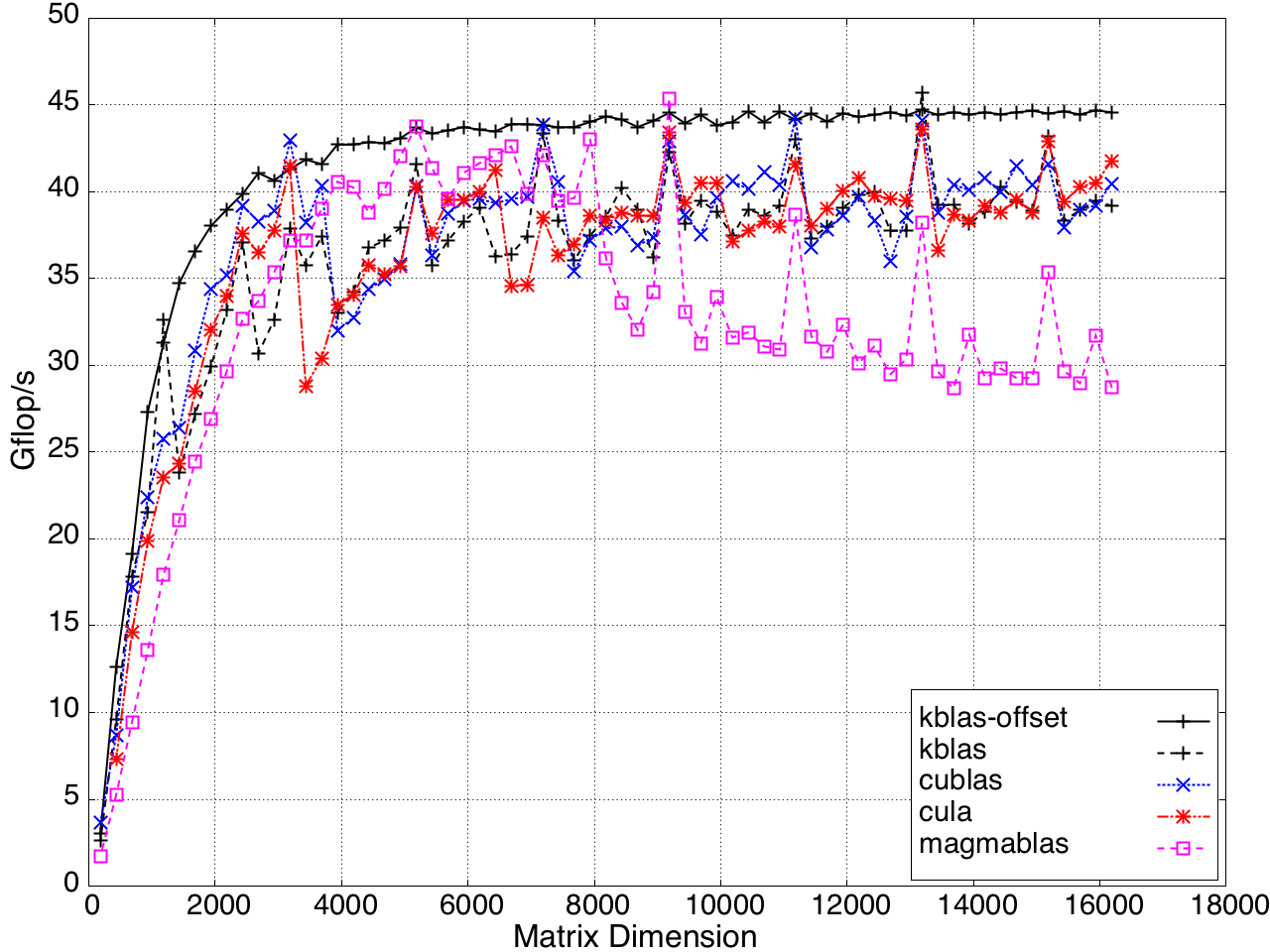}
\label{fig:dgemv_offset}
}
\subfigure[CGEMV]{
\includegraphics[width=0.48\linewidth]{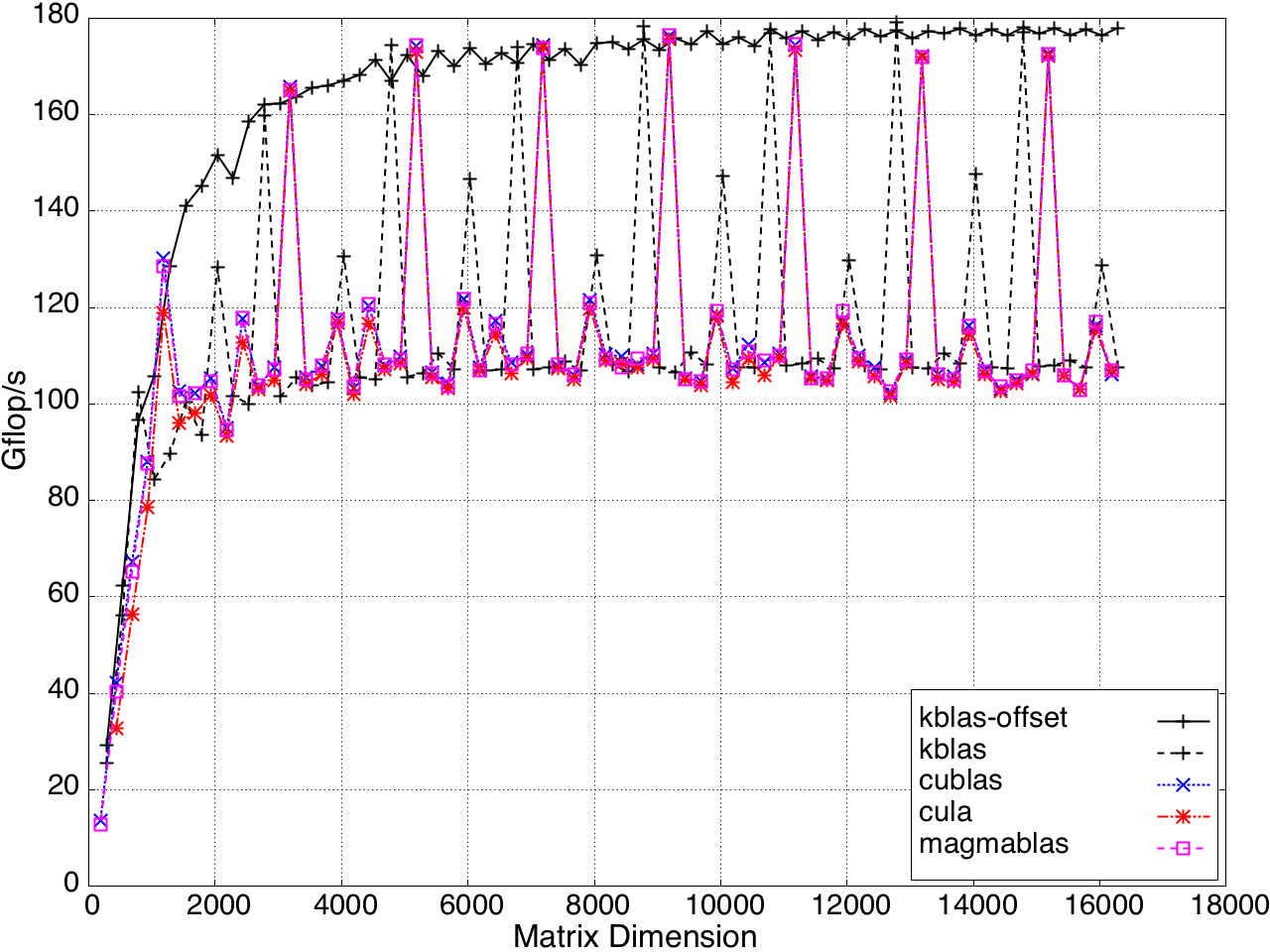}
\label{fig:cgemv_offset}
}
\subfigure[ZGEMV]{
\includegraphics[width=0.48\linewidth]{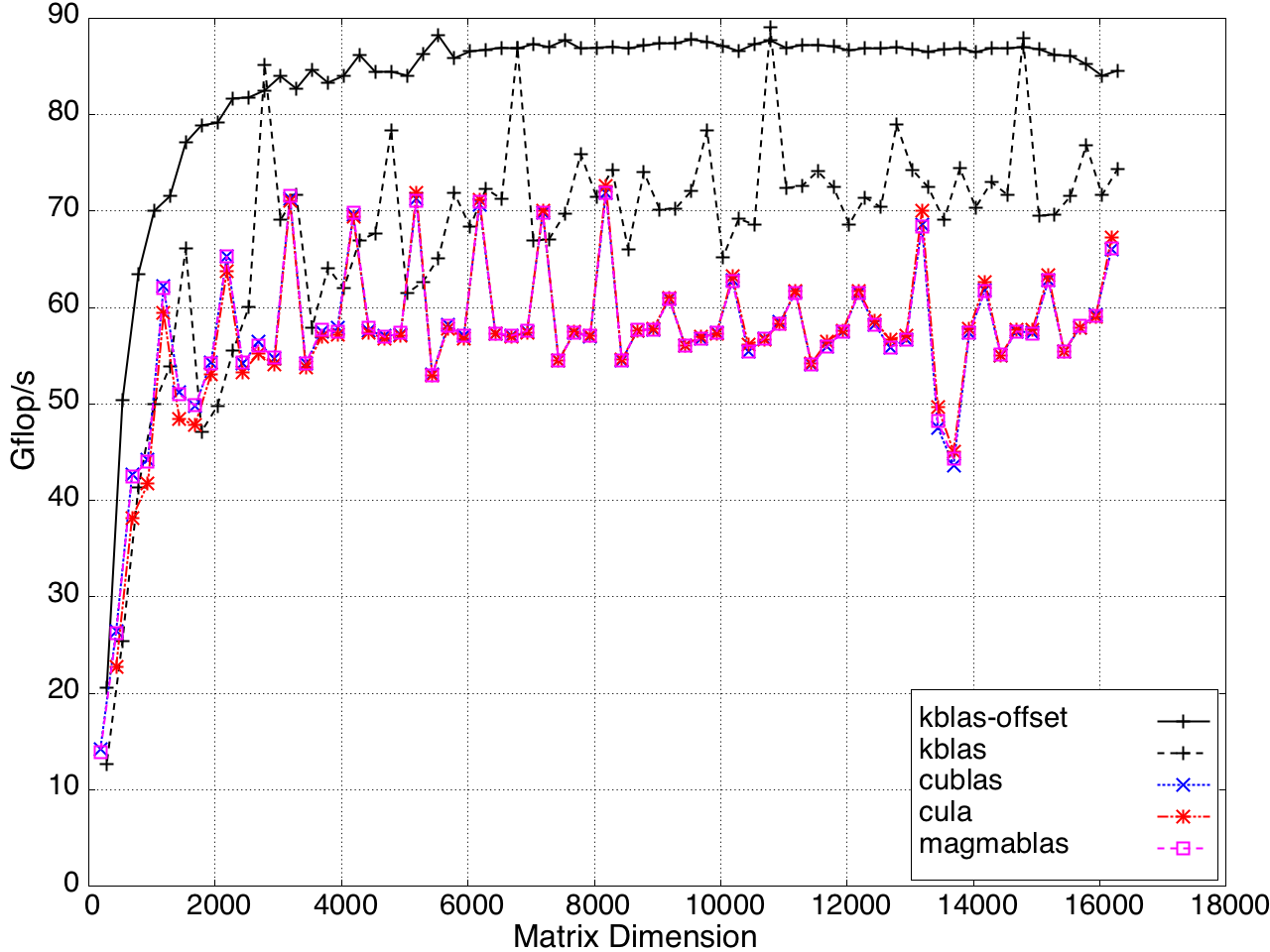}
\label{fig:zgemv_offset}
}
\caption[]{Performance of GEMV by a submatrix on a K20c GPU, ECC off}
\label{fig:gemv_offset}
\end{figure}

\begin{figure}[ht]
\centering
\subfigure[SSYMV]{
\includegraphics[width=0.48\linewidth]{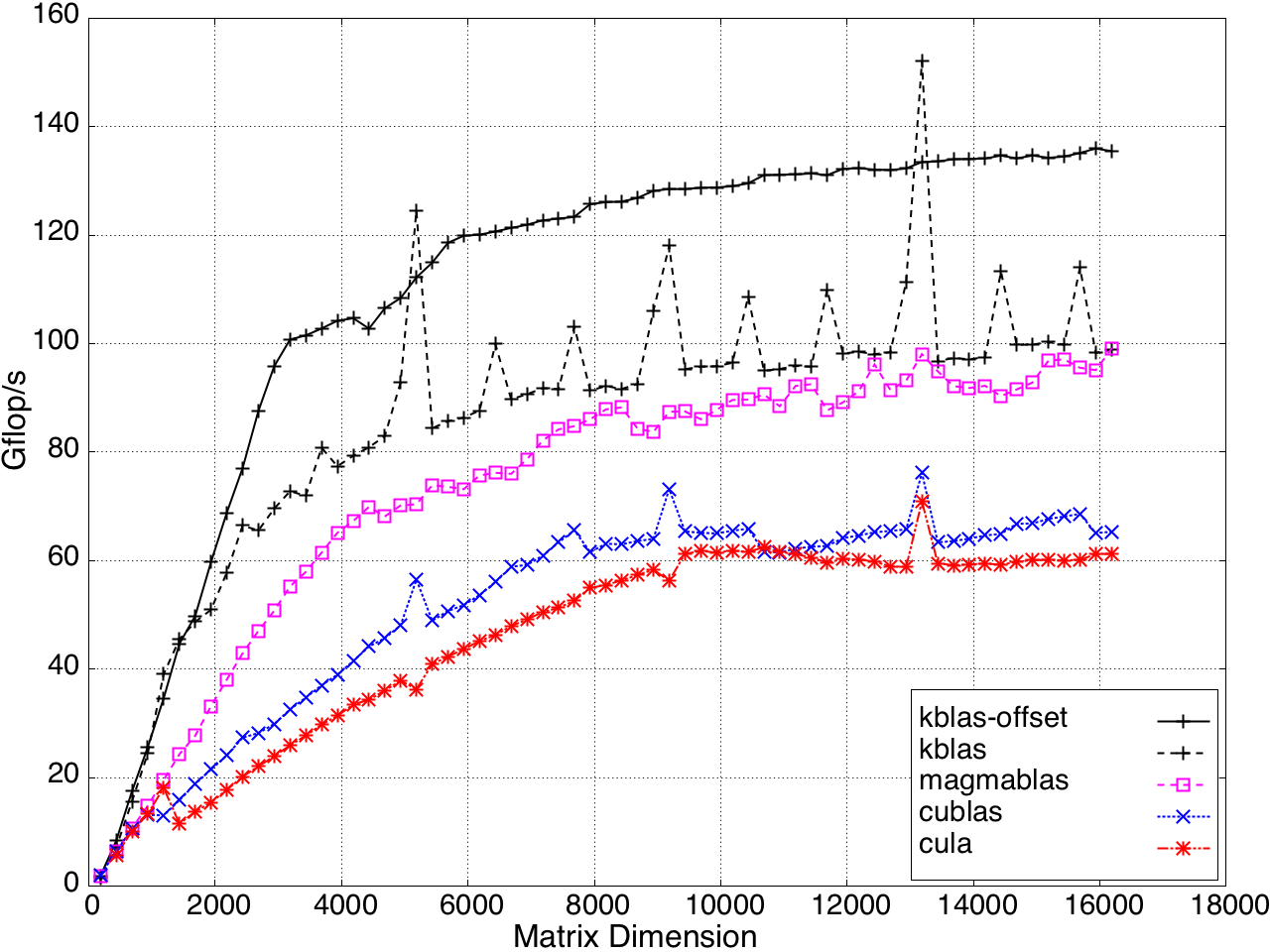}
\label{fig:ssymv_offset}
}
\subfigure[DSYMV]{
\includegraphics[width=0.48\linewidth]{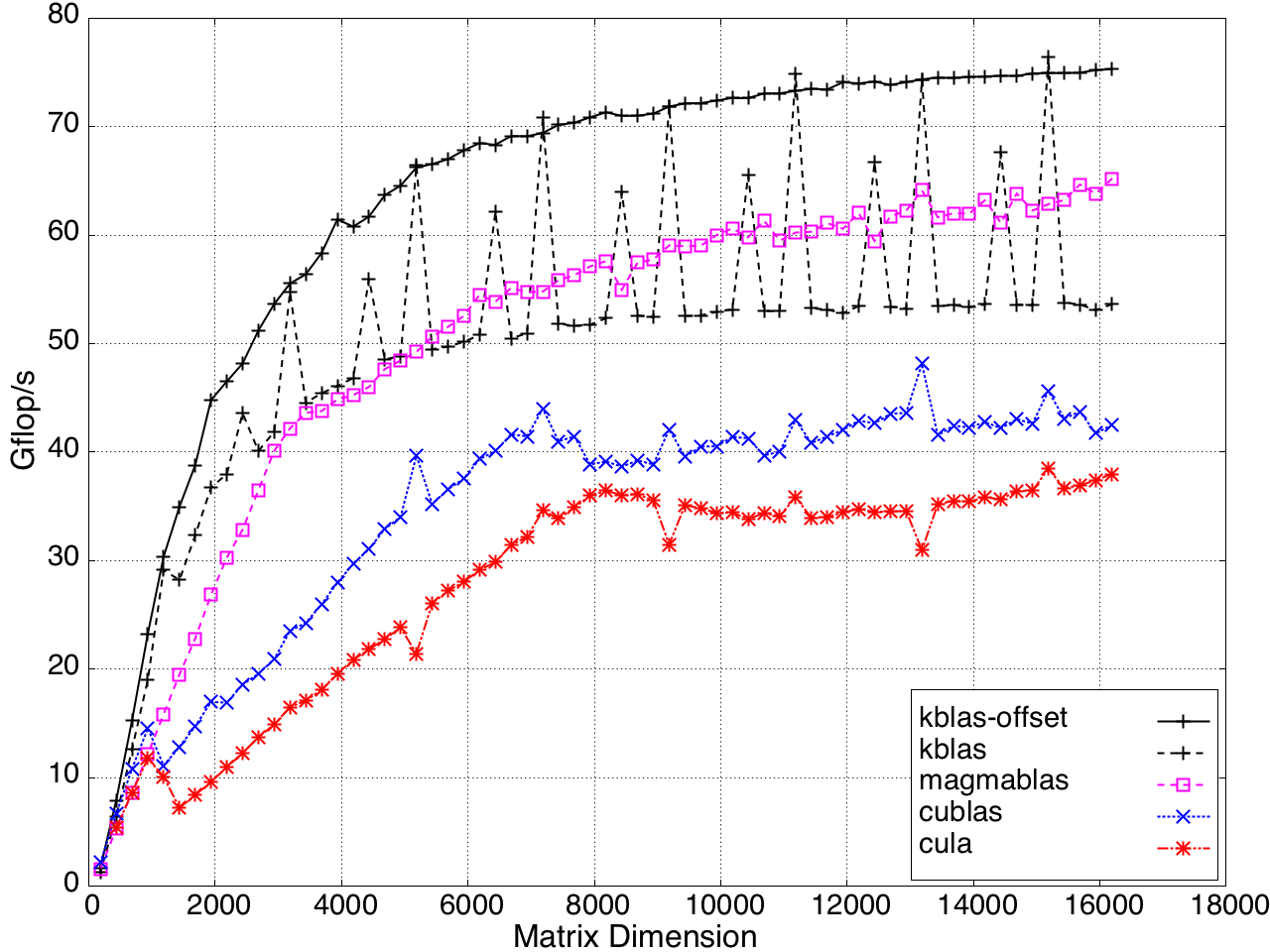}
\label{fig:dsymv_offset}
}
\subfigure[CHEMV]{
\includegraphics[width=0.48\linewidth]{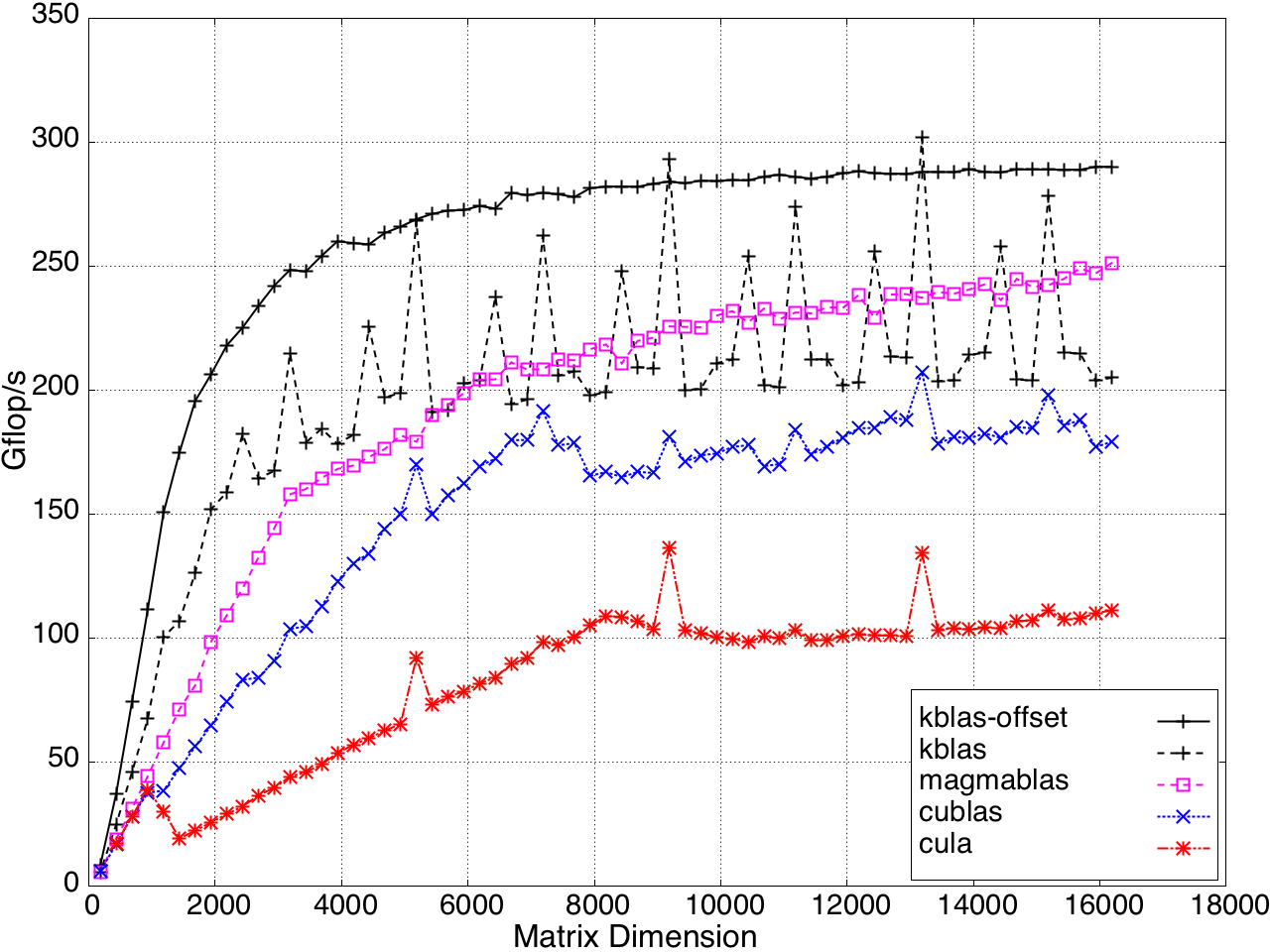}
\label{fig:chemv_offset}
}
\subfigure[ZHEMV]{
\includegraphics[width=0.48\linewidth]{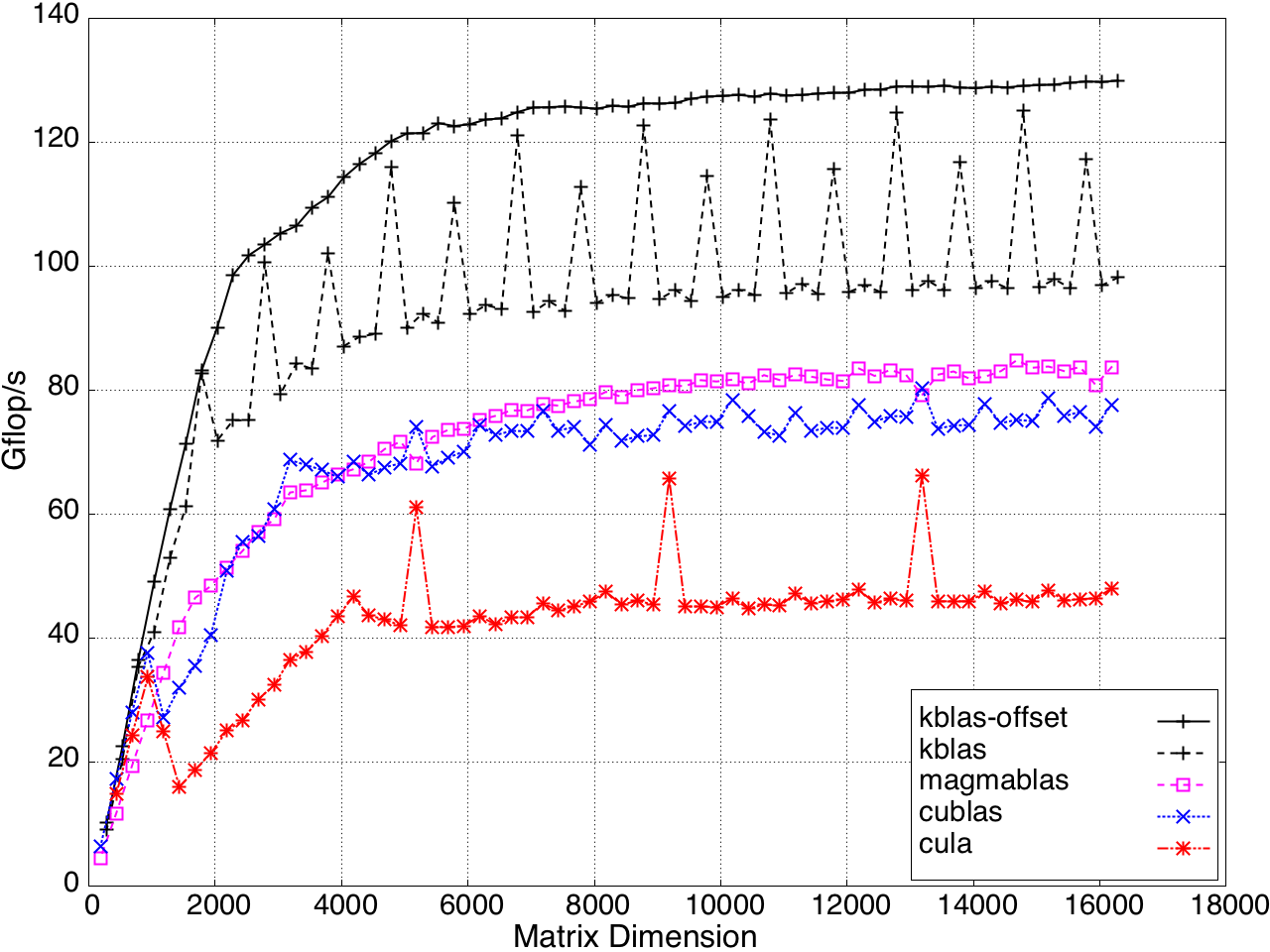}
\label{fig:zhemv_offset}
}
\caption[]{Performance of SYMV/HEMV by a submatrix on a K20c GPU, ECC off}
\label{fig:symv_offset}
\end{figure}

Figures \ref{fig:gemv_offset} and \ref{fig:symv_offset} show the performance 
of the GEMV and the SYMV/HEMV kernels when the multiplication is done on 
a submatrix. The tests represented by these figures are done on submatrices that 
are part of a larger 16384$\times$16384 matrix. These submatrices are obtained 
by skipping certain number of rows and columns from the original matrix. 
We show the performance of the standard 
KBLAS kernels as well as the new-interface kernels that can compensate the memory 
non-coalesced access, as mentioned in Section \ref{submatrix}. Figure \ref{fig:gemv_offset} 
shows that all implementations suffer from the effect of the non-coalesced memory access. 
Only the KBLAS GEMV-OFFSET kernel manages to maintain the same high performance 
shown in Figure \ref{fig:gemv}. A similar behavior is observed for the 
SYMV/HEMV kernel in Figure \ref{fig:symv_offset}.

An interesting observation is the performance spikes for standard implementation that arise 
for certain dimensions. These spikes happens when the offsets in rows and columns 
accidentally result in a submatrix that can read in fully coalesced manner. For example, consider 
Figure \ref{fig:ssymv_offset}, where performance spikes arise at sizes like 5184, 9184, 13184, and others. 
These dimensions represent offsets that are multiples of 32, which represent multiples of 128 bytes in single 
precision. In general, if the offset, expressed  in bytes, is multiple of 128, then the performance 
of the standard KBLAS implementation does not encounter any drops. We also observe minor spikes 
when the offset modulo 32 is either close to 0 or 32. In other precisions, the same 
analysis applies, but 32 should be replaced by 
16 for double and single complex precisions, and with 8 for the double complex precision.

%------------------------------------------------------------------
\subsection{Multi-GPU Performance}
\label{subsec:mgpu_gpu_perf}
\begin{figure}[ht]
\centering
\subfigure[SGEMV-MGPU]{
\includegraphics[width=0.48\linewidth]{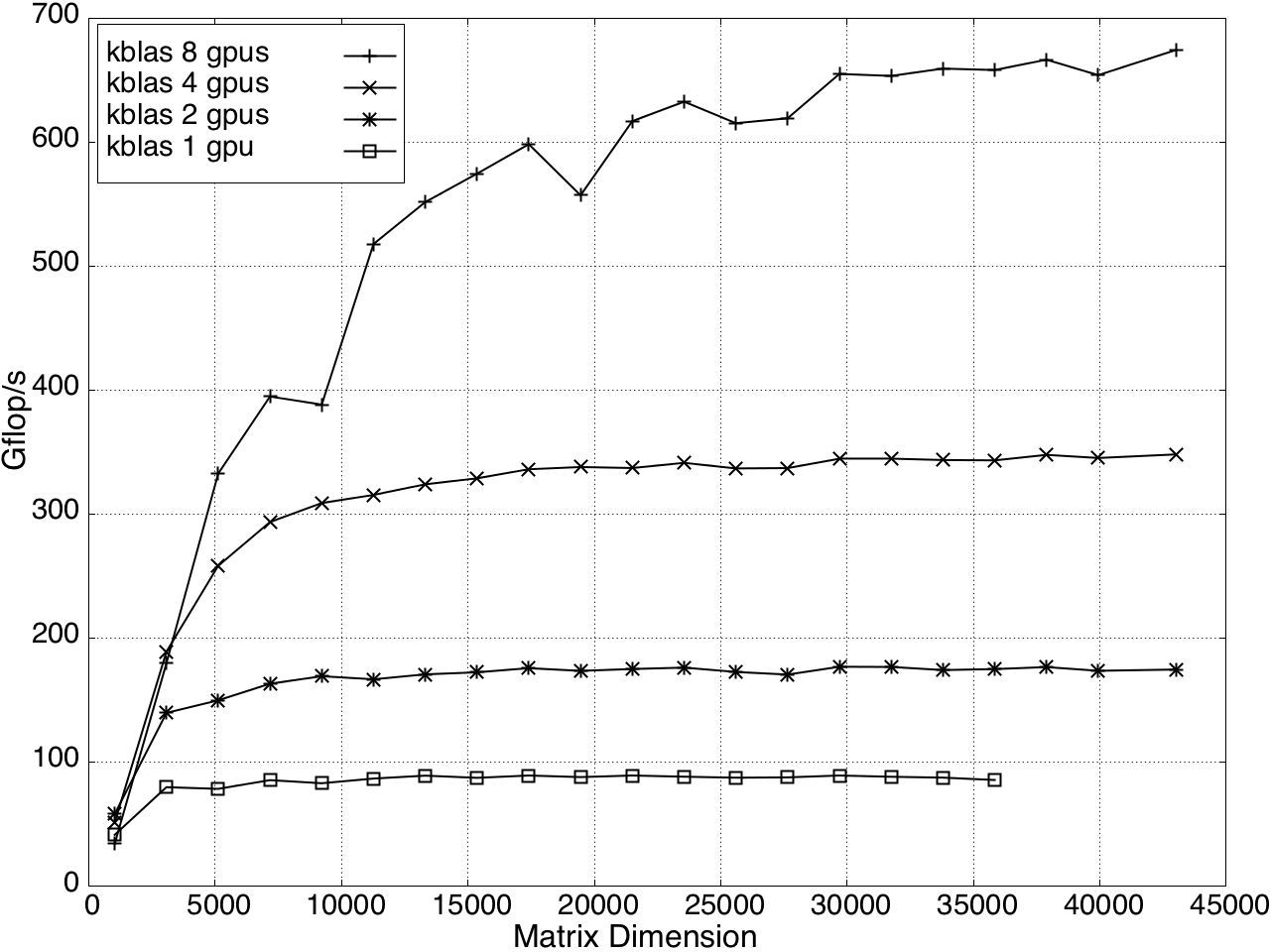}
\label{fig:sgemv_mgpu}
}
\subfigure[DGEMV-MGPU]{
\includegraphics[width=0.48\linewidth]{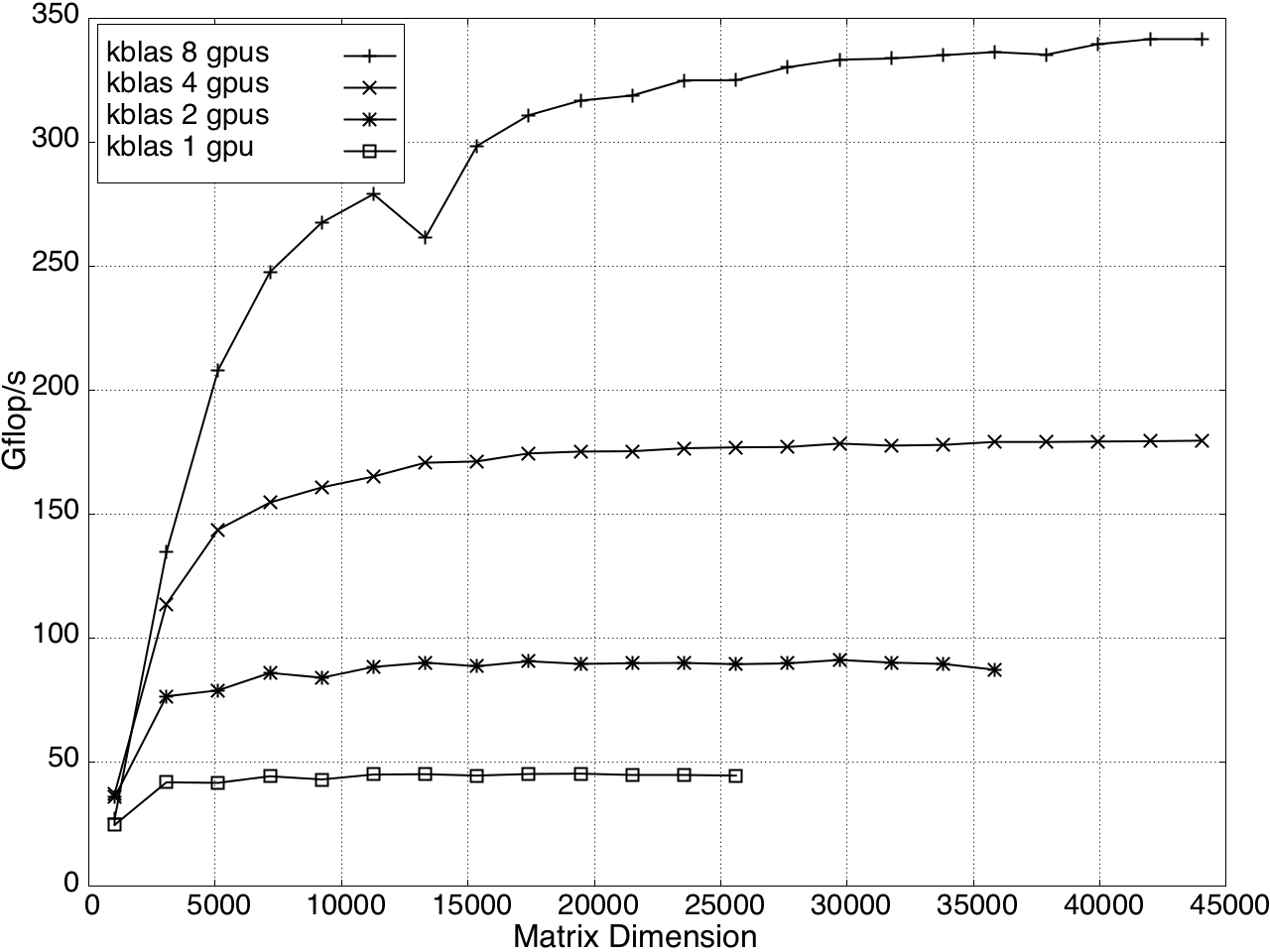}
\label{fig:dgemv_mgpu}
}
\subfigure[CGEMV-MGPU]{
\includegraphics[width=0.48\linewidth]{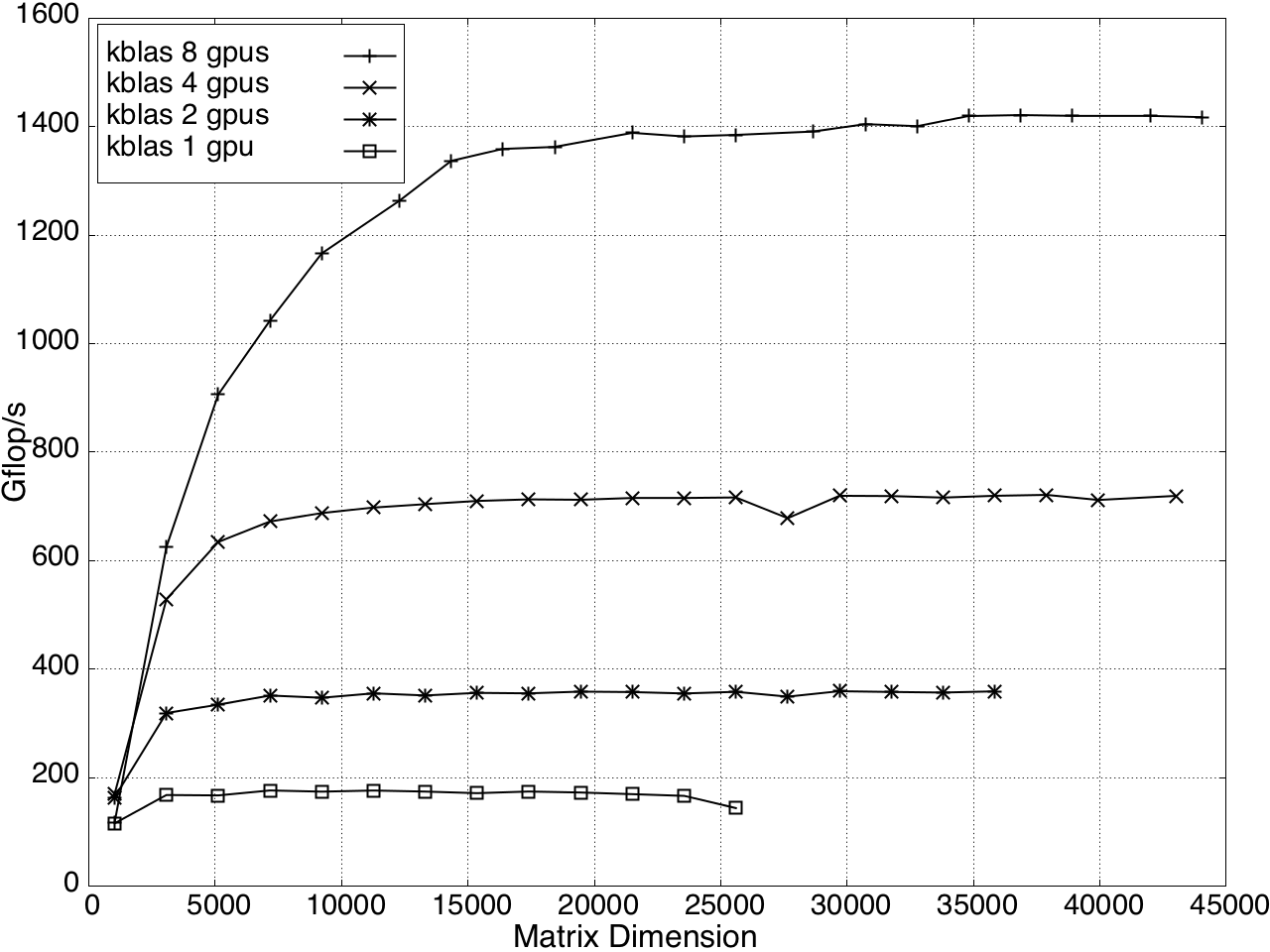}
\label{fig:cgemv_mgpu}
}
\subfigure[ZGEMV-MGPU]{
\includegraphics[width=0.48\linewidth]{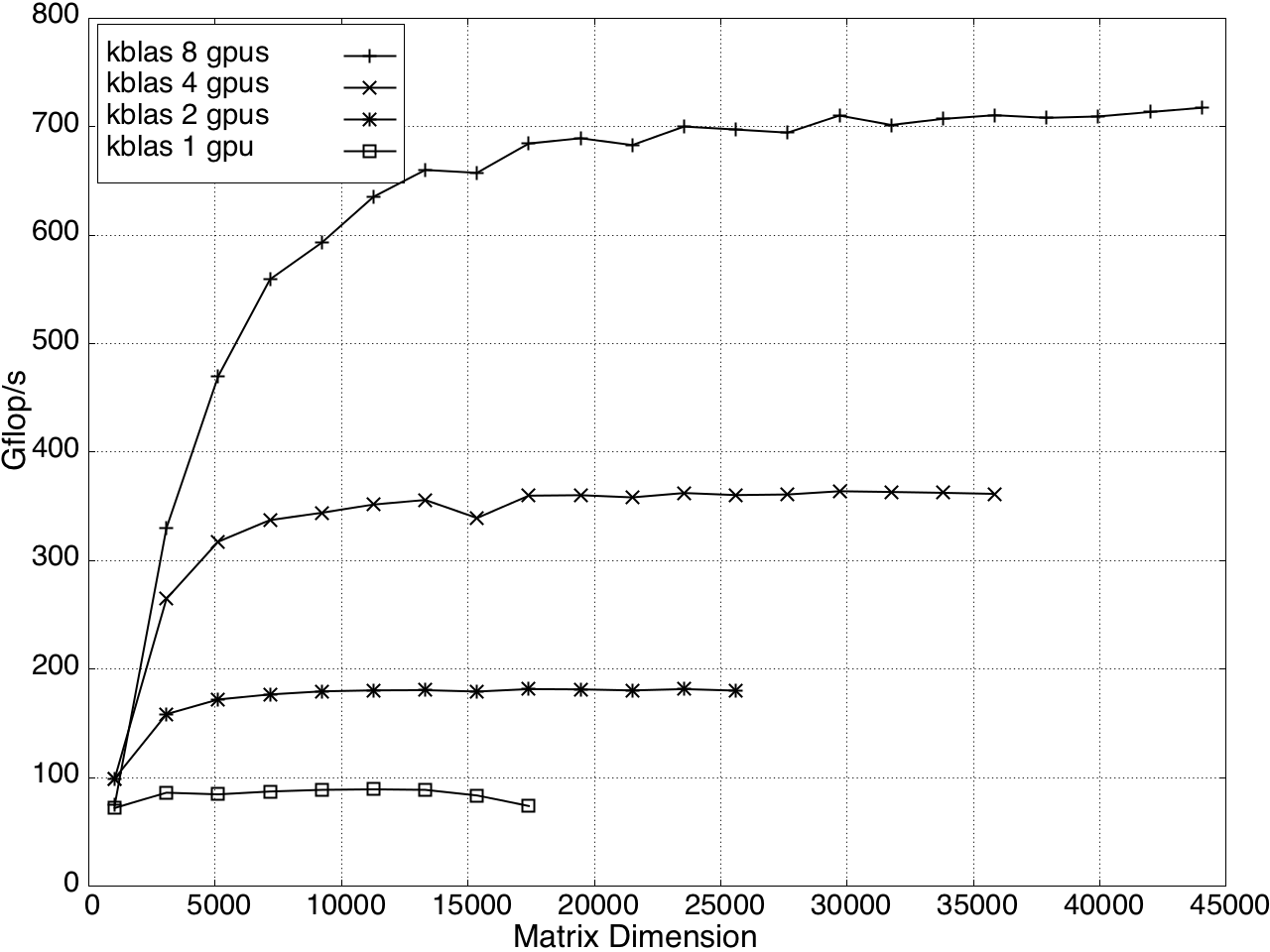}
\label{fig:zgemv_mgpu}
}
\caption[]{GEMV Performance on Multi-GPU, K20c with ECC off}
\label{fig:gemv_mgpu}
\end{figure}

\begin{figure}[ht]
\centering
\subfigure[SSYMV-MGPU]{
\includegraphics[width=0.48\linewidth]{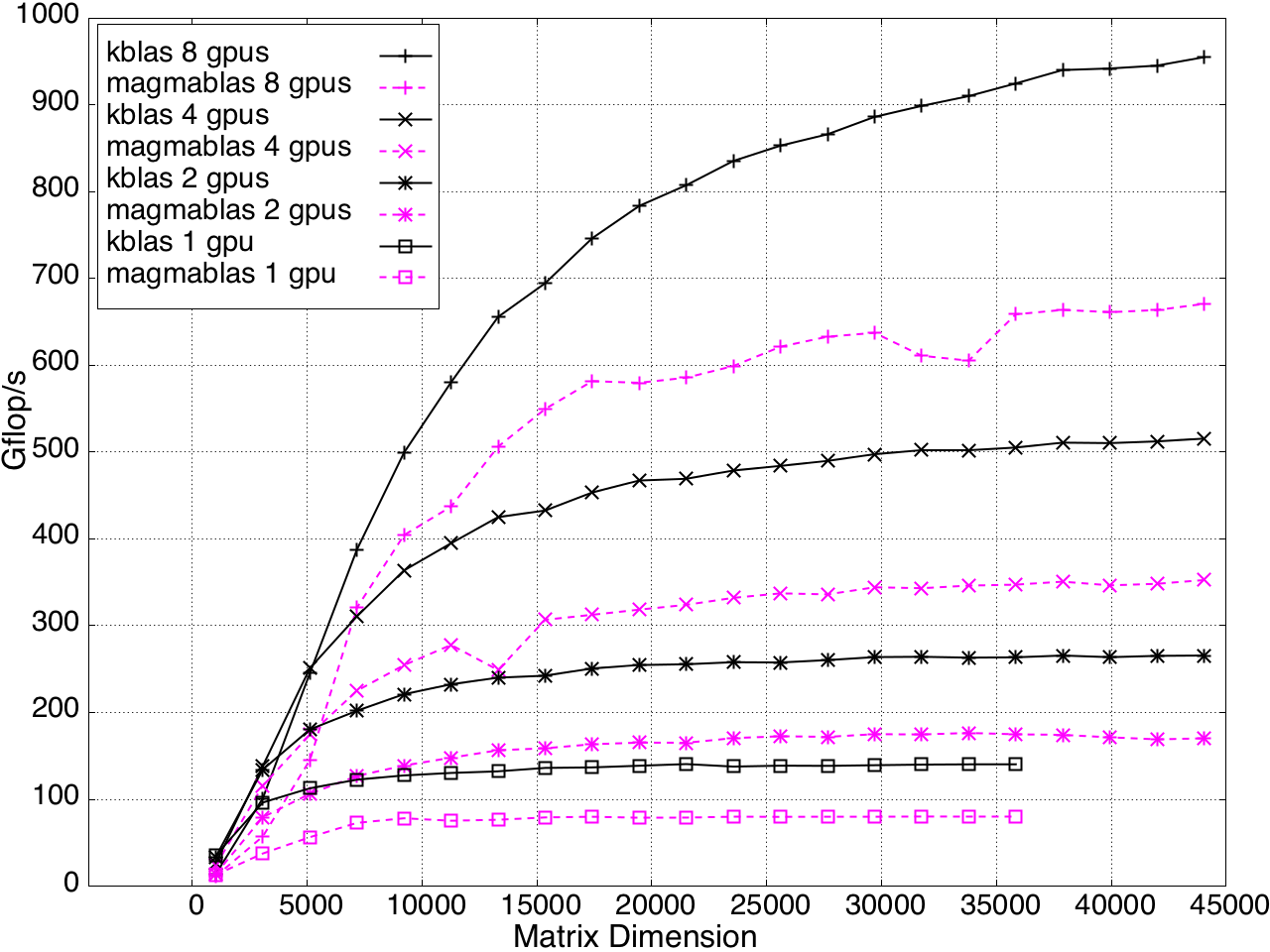}
\label{fig:ssymv_mgpu}
}
\subfigure[DSYMV-MGPU]{
\includegraphics[width=0.48\linewidth]{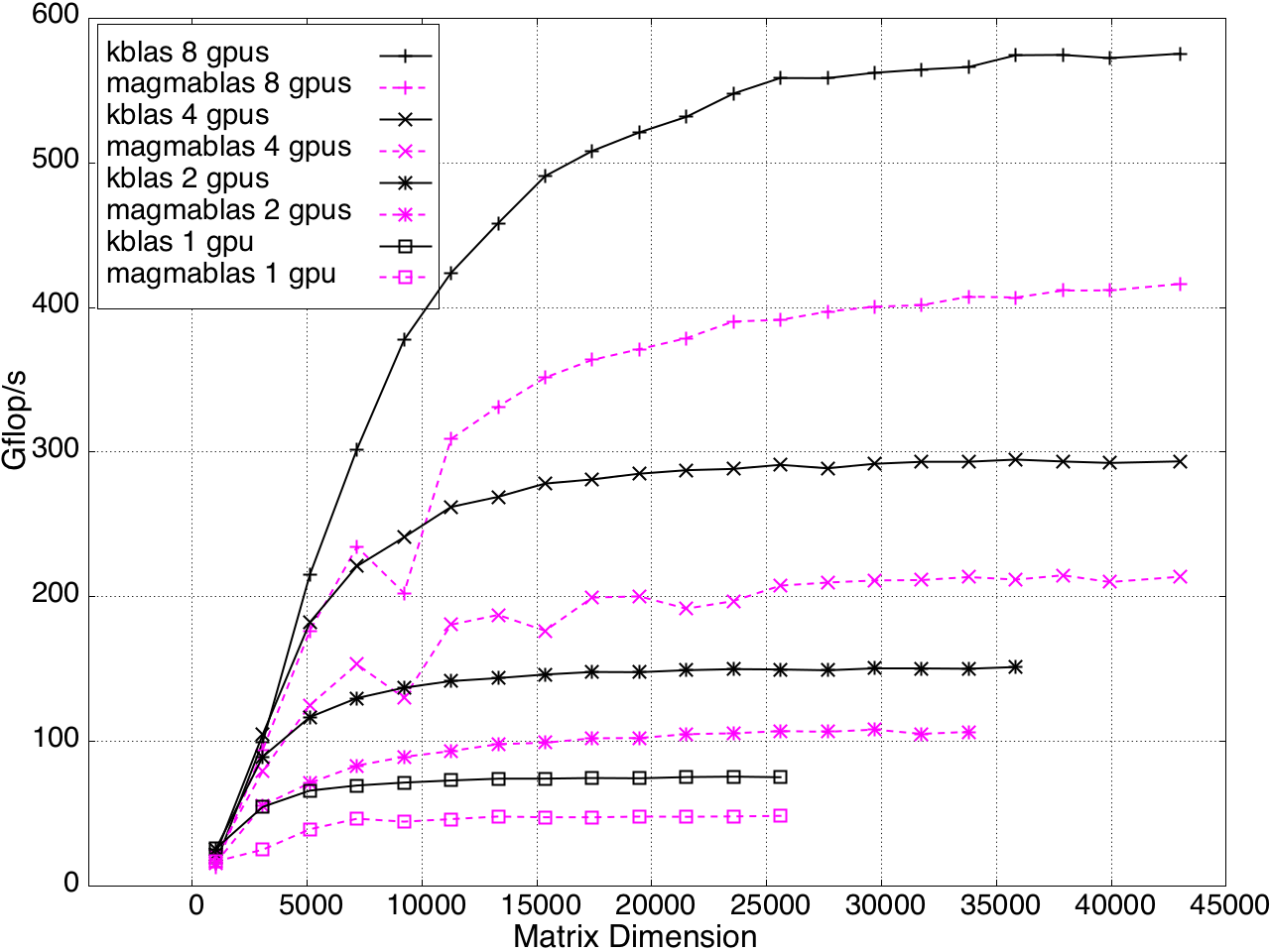}
\label{fig:dsymv_mgpu}
}
\subfigure[CHEMV-MGPU]{
\includegraphics[width=0.48\linewidth]{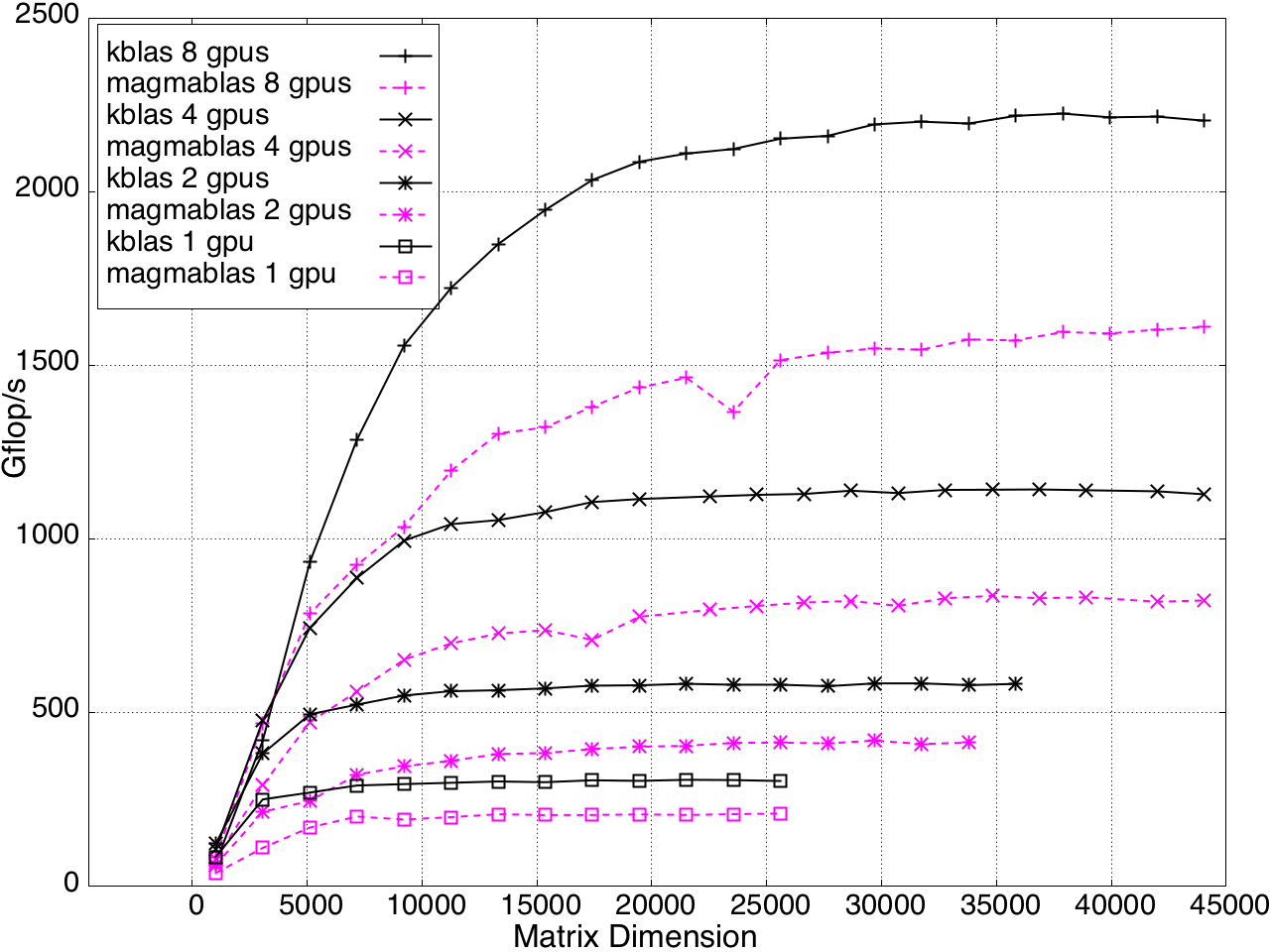}
\label{fig:chemv_mgpu}
}
\subfigure[ZHEMV-MGPU]{
\includegraphics[width=0.48\linewidth]{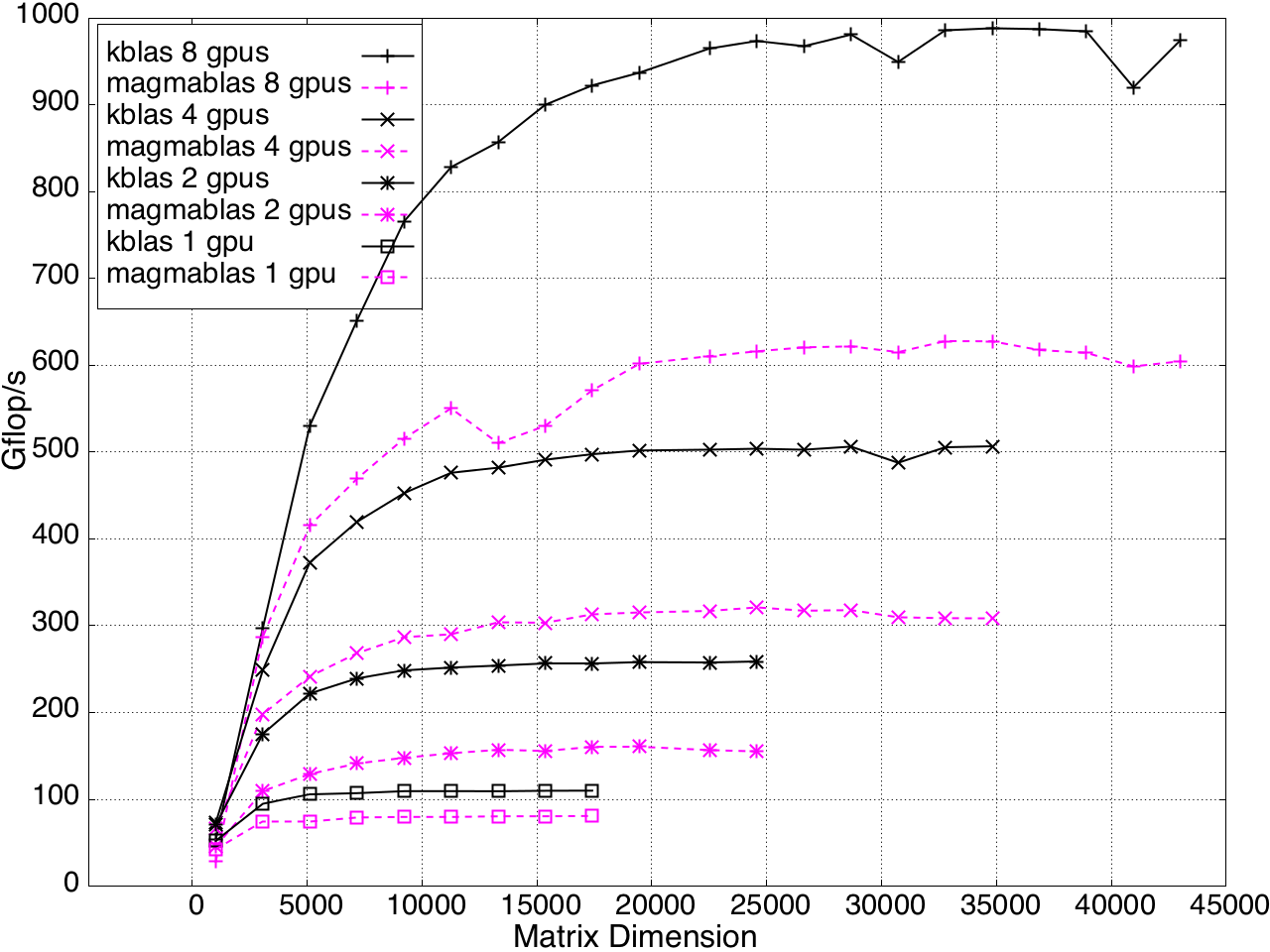}
\label{fig:zhemv_mgpu}
}
\caption[]{SYMV/HEMV Performance on Multi-GPU, K20c with ECC off}
\label{fig:symv_mgpu}
\end{figure}

Figures \ref{fig:gemv_mgpu} and \ref{fig:symv_mgpu} show the performance of the KBLAS GEMV and SYMV/HEMV kernels 
on multi-GPUs. Matrices are stored in the data layout shown in Figure \ref{fig:layout_mgpu}. While a multi-GPU GEMV kernel 
can be factored into calls to the single GPU GEMV kernel, a multi-GPU SYMV/HEMV cannot be factored in the same way. This is 
because the rectangular submatrices lose symmetry, and so a sophisticated kernel is required. 

KBLAS, however, still 
provides a sophisticated GEMV kernel on multi-GPU. This kernel takes into account the multiplication by a submatrix in the same 
manner described in Section \ref{submatrix}. It is expected, therefore, to perform better than calling standard KBLAS-GEMV kernels on 
the local submatrices. The KBLAS GEMV-MGPU kernel is very close to strong scaling on up to 8 GPUs on a single node. This kernel can be used 
in matrix reduction techniques on large non-symmetric matrices that do not fit in single GPU memory. 
Considering the case when the matrix is Hermitian, 
shown in Figure \ref{fig:symv_mgpu}, the KBLAS SYMV/HEMV kernel on multi-GPUs can achieve up to 43\%, 38\%, 38\%, and 
61\% performance improvement against MAGMABLAS on 8 GPUs for all four precisions. According to the authors' knowledge, only 
MAGMABLAS and KBLAS provide a multi-GPU SYMV/HEMV kernel. 

%-------------------------------------------------------------------------

%% file: tuning.tex
%!TEX root = kblas.tex
The performance of KBLAS is controllable through a set of tuning parameters. Such parameters should 
be tuned according to the GPU architecture/model as well as the CUDA runtime version. For best 
performance, KBLAS should be retuned whenever the GPU changes or the CUDA runtime is upgraded. 

\subsection{Tuning Process of a KBLAS Kernel}
\label{subsec:tuning_process}
Any KBLAS kernel has at most three tuning parameters: the block size $nb$, the number of thread columns in a TB $\bar{Q}$ (each of which 
has $nb$ threads), and the 
number of TBs that collaboratively process a block row or a block column of the input matrix ($\bar{Y}$). Both $nb$ and $\bar{Q}$ are used for 
\emph{coarse tuning}, while $\bar{Y}$ is used for \emph{fine tuning}. It is advised, therefore, to tune first $nb$ and $\bar{Q}$. 
We will discuss a case study of tuning the DSYMV kernel on a K20c GPU. All KBLAS kernels are tunable in the same way mentioned 
below.

At the stage of \emph{coarse tuning}, the parameter $\bar{Y}$ should be set to 1. Our experiments show that both values of 
$nb$ and $\bar{Q}$ should be powers of two, for best performance. Additionally, $nb$ is multiples of the warp size in most cases. The value range of 
$\bar{Q}$ is restricted by Equation \ref{eqn:buffer_length}, since $L$ should be an integer greater than zero. This means that 
the parameter space for the ($nb$, $\bar{Q}$) is relatively small, and that hand tuning of KBLAS can be done in a reasonable amount 
of time. 

\begin{figure}[ht]
\centering
\subfigure[Coarse Tuning]{
\includegraphics[width=0.48\linewidth]{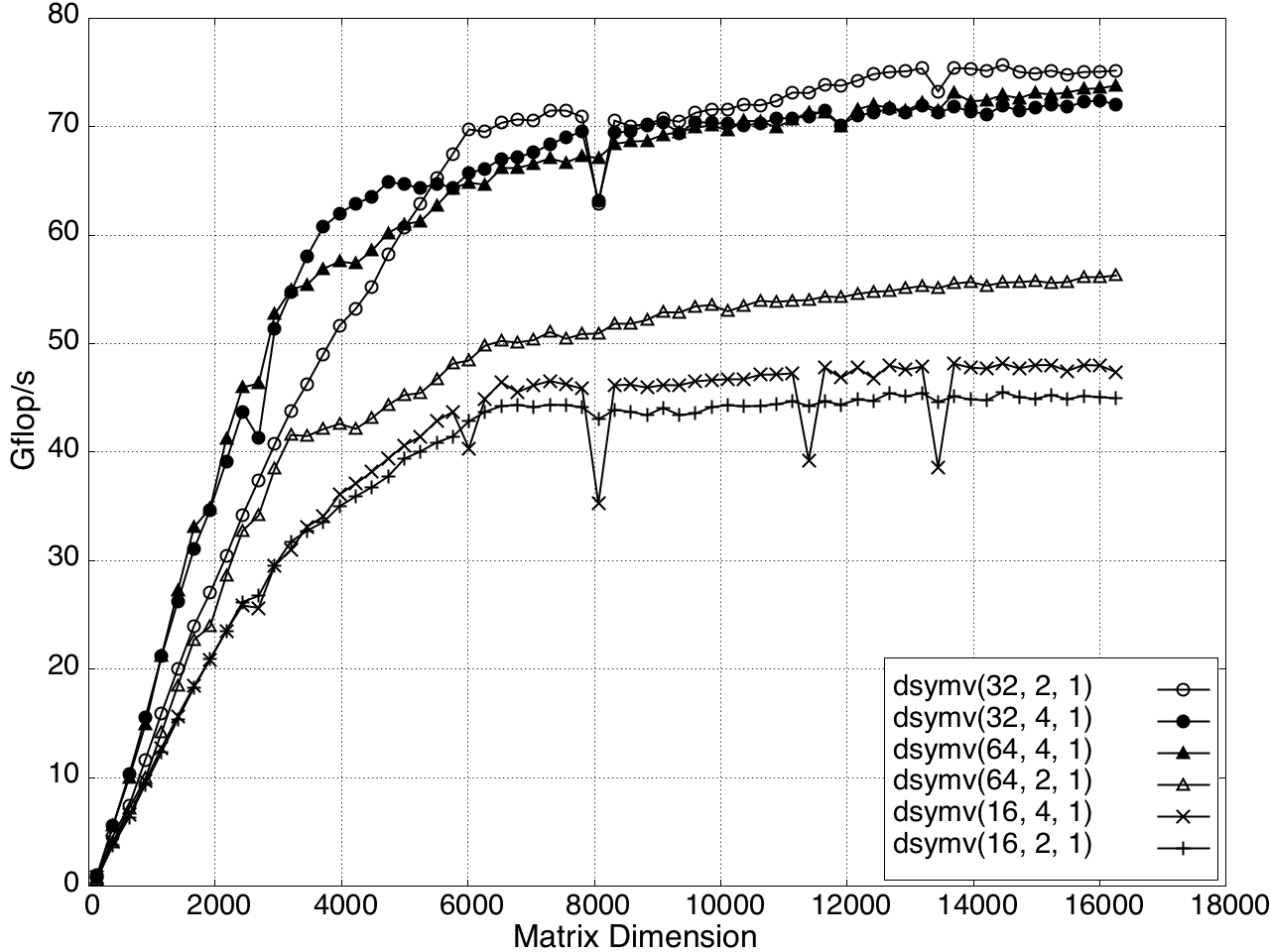}
\label{fig:dsymv_course}
}
\subfigure[Fine Tuning]{
\includegraphics[width=0.48\linewidth]{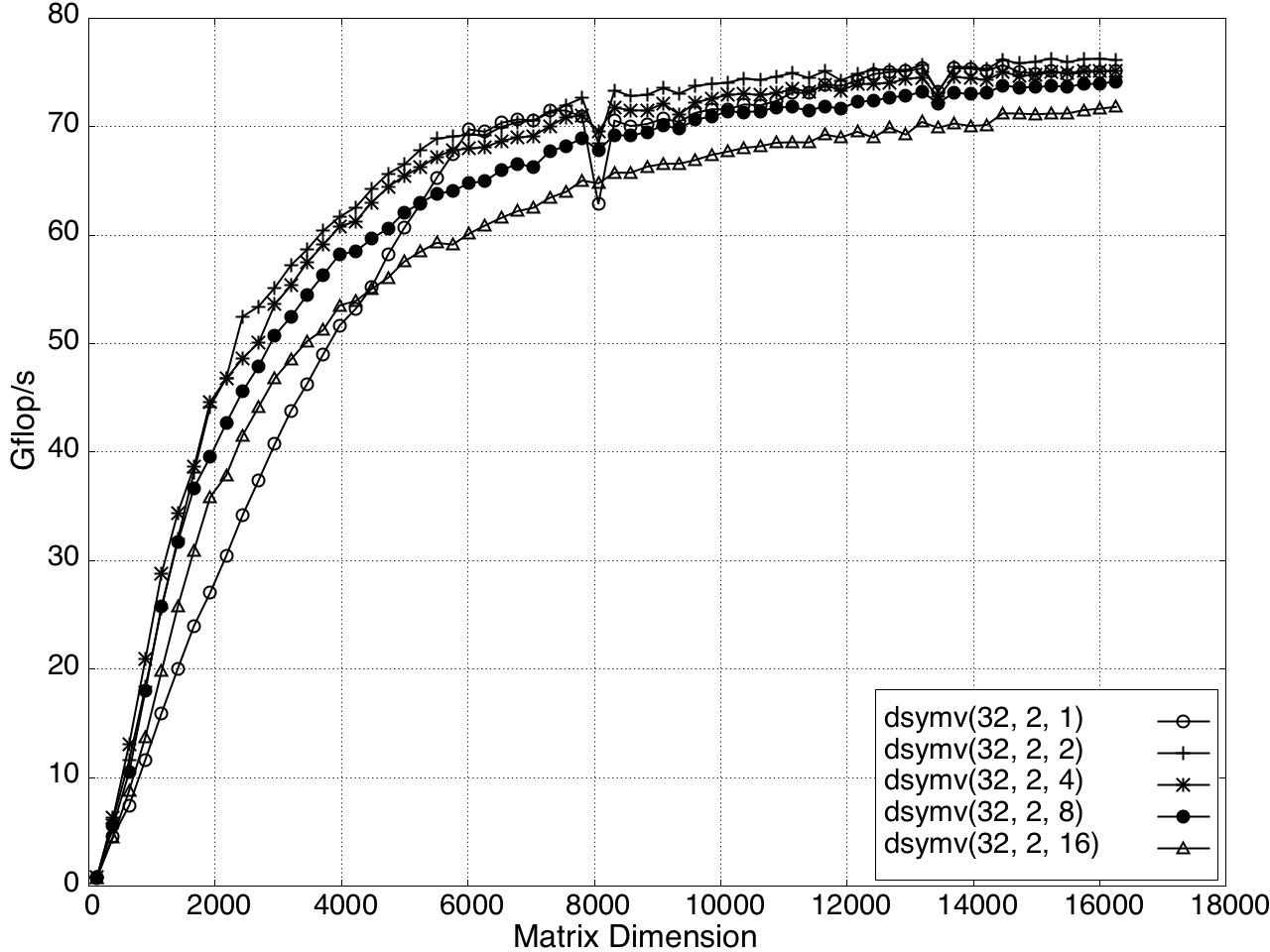}
\label{fig:dsymv_fine}
}
\caption[]{Performance Tuning of KBLAS-DSYMV Kernel on a K20c GPU}
\label{fig:dsymv_tuning}
\end{figure}

Figure \ref{fig:dsymv_course} shows the performance of the DSYMV kernel for different values of the ($nb$, $\bar{Q}$), with $\bar{Y}$ fixed 
at 1. At such stage of tuning, 
the focus should be on the asymptotic performance, since the behavior for relatively small matrices is usually impacted by $\bar{Y}$. Figure 
\ref{fig:dsymv_course} shows that the best asymptotic performance is achieved by the (32, 2, 1) configuration.

The next stage is to try increasing $\bar{Y}$ while fixing ($nb$, $\bar{Q}$). Figure \ref{fig:dsymv_tuning} shows that the best performance is the 
(32, 2, 2) configuration. As expected, increasing the value of $\bar{Y}$ enhances the performance for relatively small matrices. 
But on the other hand, a too large value of $\bar{Y}$ (like the (32, 2, 16) configuration) means more pressure on atomic operations, 
and therefore, the performance becomes negatively influenced. 

\begin{figure}[ht]
\centering
\includegraphics[width=0.6\linewidth]{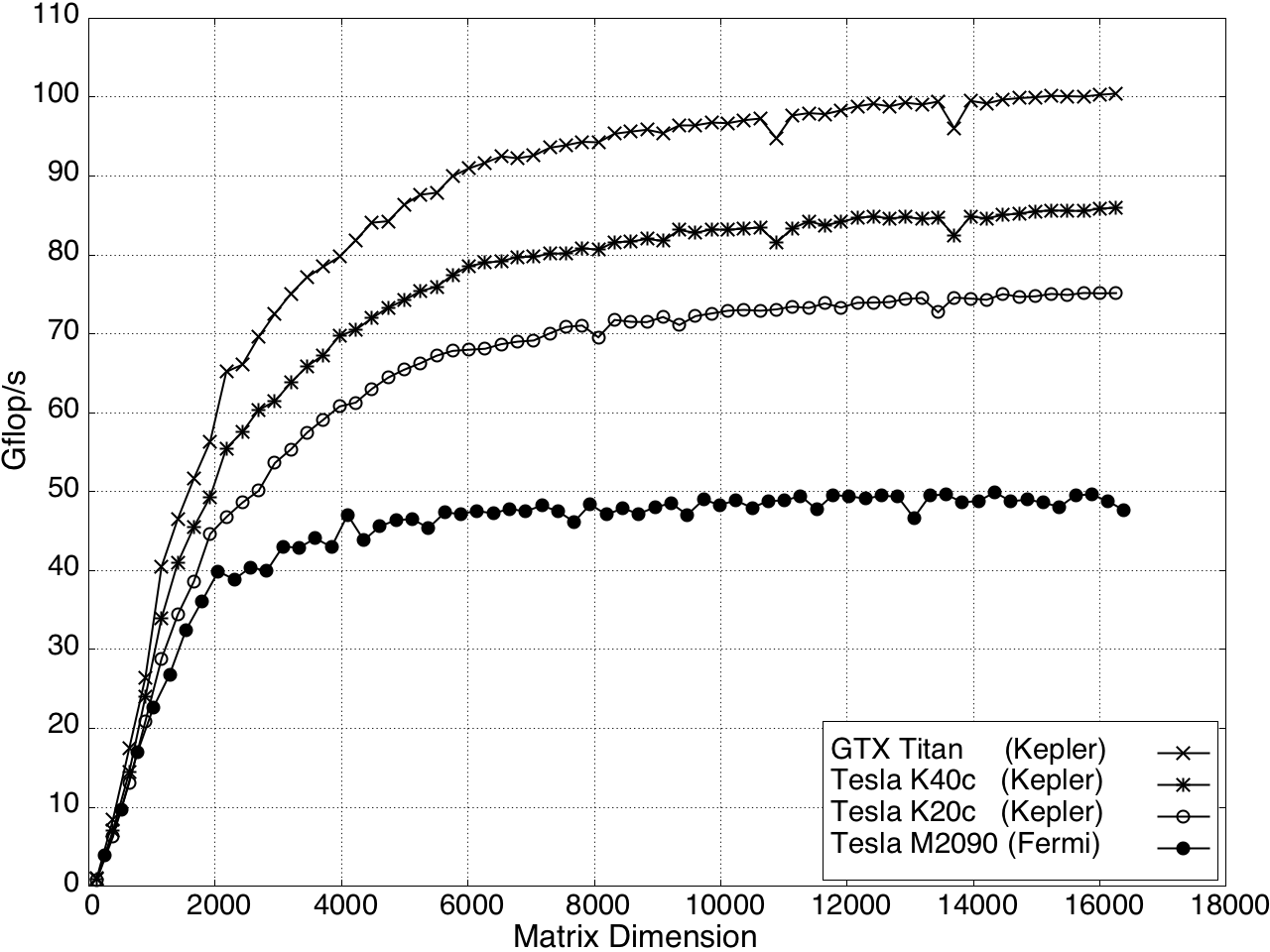}
\caption[]{Performance of a Tuned KBLAS-DSYMV Kernel on Different GPUs}
\label{fig:dsymv_tuned}
\end{figure}

The simple strategy mentioned above works well on several GPU architectures/models. Figure \ref{fig:dsymv_tuned} shows the performance of the 
DSYMV on different GPUs: A Fermi M2090, a Kepler K20c, a Kepler K40c, and a GTX Titan. The respective sustained peak memory bandwidths of these 
GPUs, as scored by the STREAM benchmark, are 130.32 GB/s, 175.24 GB/s, 219.65 GB/s, and 239.87 GB/s, respectively. 
All performance curves in this figure have been 
tuned in the same way. Leveraging the same performance model mentioned in Section \ref{sec:model} on these GPUs, the asymptotic performance 
of the DSYMV kernel is up to 77\%, 87\%, 78\%, and 84\% on the aforementioned GPUs.

\subsection{Smoothing GEMV Performance}
\label{subsec:gemv_smoothing}
\begin{figure}[ht]
\centering
\includegraphics[width=0.6\linewidth]{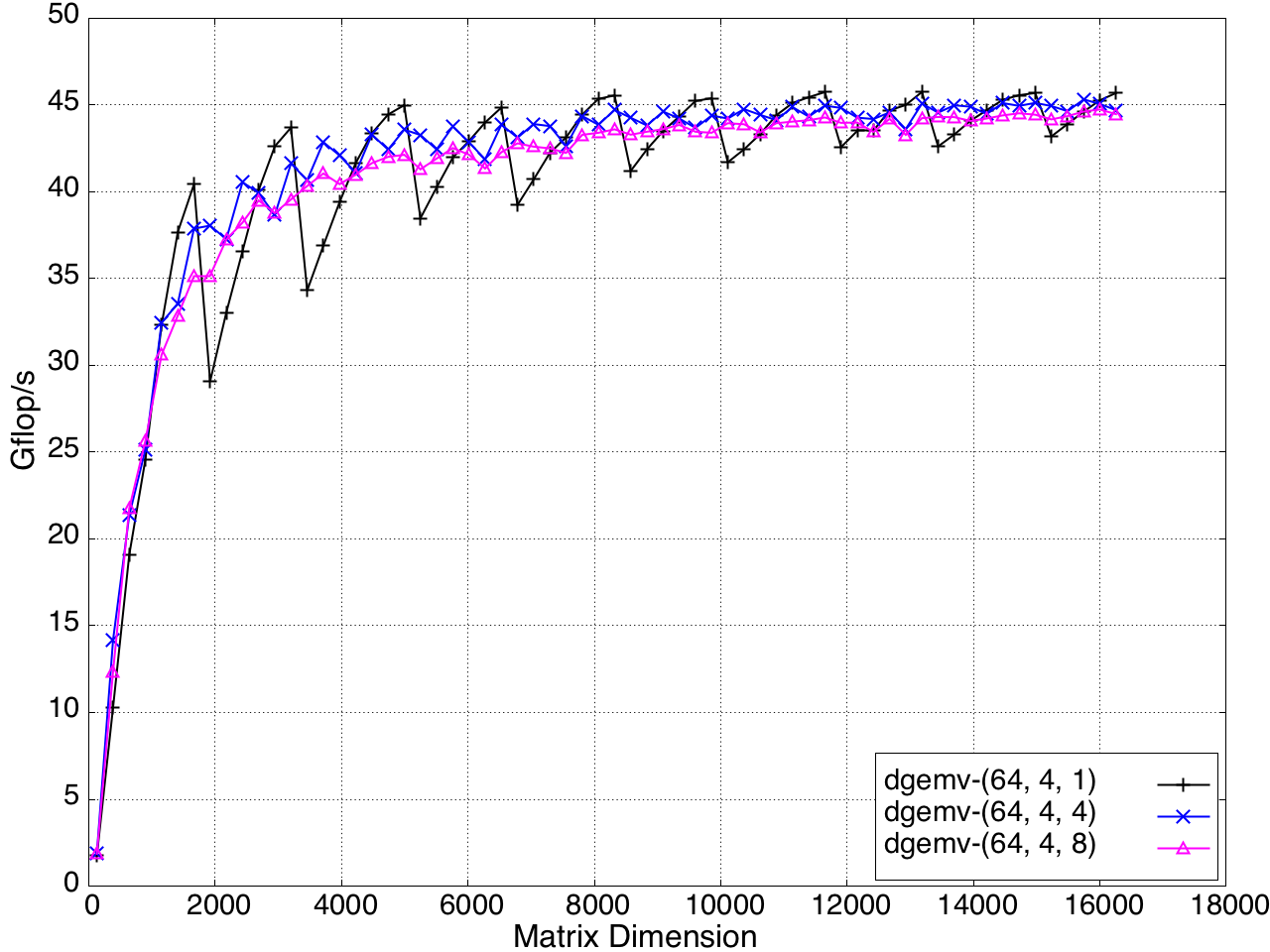}
\caption[]{Impact of $\bar{Y}$ on the Performance of the KBLAS-DGEMV Kernel}
\label{fig:dgemv_tuning}
\end{figure}

Another interesting point is the impact of the $\bar{Y}$ value on the performance of the GEMV kernel. 
%We notice that there is an 
%oscillatory behavior in the performance of the GEMV kernel, as in Figures \ref{fig:sgemv} and \ref{fig:dgemv}. 
%is related to the number of thread blocks launched with respect to the number of resident SMs on the GPU. 
%Recall that the grid dimension of a Kernel is ($\bar{X}$, $\bar{Y}$), where $\bar{X}=\left \lceil \frac{D}{nb} \right \rceil$, and $\bar{Y}$ is a tuning 
%parameter. A TB traverses an entire block row, if $\bar{Y}$=1, or part of it, if $\bar{Y}$$>$1. 
%Consider the case when $\bar{Y}$ is set to 1, which is typically the case in the old designs proposed in \cite{ahmad_vecpar12} and \cite{ahmad_heteropar12}. 
%Given the fact that 
%the load is balanced among the TBs launched within a GEMV kernel, let $TB_R$ marks the number of  remaining TBs after the partial 
%kernel execution where all GPU SMs are fully occupied. Typically $TB_R$=$\bar{X}\mod$\#SMs, and performance 
%is penalized if it is relatively low ($TB_R$ $<<$ \#SMs), since the GPU will encounter a duration of low utilization while executing these remaining 
%TBs. 
Consider the DGEMV performance in Figure \ref{fig:dgemv_tuning} on a K20c 
GPU, which has 13 SMs. We observe an oscillatory behavior in the performance of the 
(64, 8, 1) configuration. Such configuration is typical for the previous GEMV kernel proposed 
by the authors~[\citeANP{ahmad_vecpar12} 2013a] in the sense that it does not expose $\bar{Y}$ as a tuning parameter and 
keeps it fixed at 1. 
The oscillatory behavior 
is already discussed in Section \ref{subsubsec:oscillation}. 
The GEMV kernel launches TBs with balanced workloads, and so it encounters oscillations in performance when 
the input matrix dimension leads to a low $TB_R$. 
The performance drops in the (64, 8, 1) configuration come for dimensions that have low $TB_R$ values. 
For example, dimensions like 
1792, 3456, 5120, 6784, 8448, and 10112 have huge drops in performance. These dimensions all end up with $TB_R$=2. The drops are periodic 
but get smaller, as the number of full utilization rounds gets larger and the GPU becomes closer to a full utilization during most of the kernel 
execution time. 

The ability of KBLAS to incorporate more than a TB per block row can compensate these oscillations. 
Since the execution time of a single TB becomes smaller when $\bar{Y}$ increases,  
the number of rounds of full utilization is significantly increased, while the round of partial utilization 
is always fixed at 1. 
The time spent by the GPU in the partial utilization round with respect to the time spent in full rounds 
becomes smaller and even negligible. 

Figure \ref{fig:dgemv_tuning} shows the impact of increasing $\bar{Y}$. The performance drops gets smaller as $\bar{Y}$ increases. However, the 
asymptotic performance slightly drops due to the overhead of atomic operations. Recall that the GEMV kernel does exactly $\bar{Y}$ atomic 
additions of size $nb$ per a block row or a block column of the input matrix. 

%% file: app.tex
%!TEX root = kblas.tex
In this section, we show the performance improvement that KBLAS can achieve when it is 
integrated into higher level algorithms that extensively use dense matrix vector multiplication. 
This case study focuses on LAPACK algorithms provided by MAGMA. The system setup is the same as described 
in Section \ref{subsec:setup}.

%\subsection{Dense: MAGMA}
%\label{subsec:magma}
%------------------------------------------------------------------
MAGMA is an open source library that provides optimized BLAS and LAPACK routines for multi-core 
architecture accelerated using GPUs \cite{MAGMA} \cite{ADDHKLLLT09}. We pick two algorithms from MAGMA and accelerate 
them using KBLAS: The bidiagonal reduction algorithm for general matrices (GEBRD); and the tridiagonal reduction 
for symmetric/Hermitian matrices (SYTRD/HETRD). These two algorithms are used in Singular Value Decomposition (SVD), 
and Eigenvalue Decomposition (EVD) for dense matrices, respectively. They are implemented using block algorithms, where 
the compute intensive components are offloaded to the GPUs, as proposed in \cite{tomov_jpc}. 
Since KBLAS provides two implementations for each of the GEMV and and the SYMV/HEMV kernels, we will show the impact 
on the performance of the aforementioned algorithms with each implementation. 

\begin{figure}[ht]
\centering
\subfigure[SGEBRD]{
\includegraphics[width=0.48\linewidth]{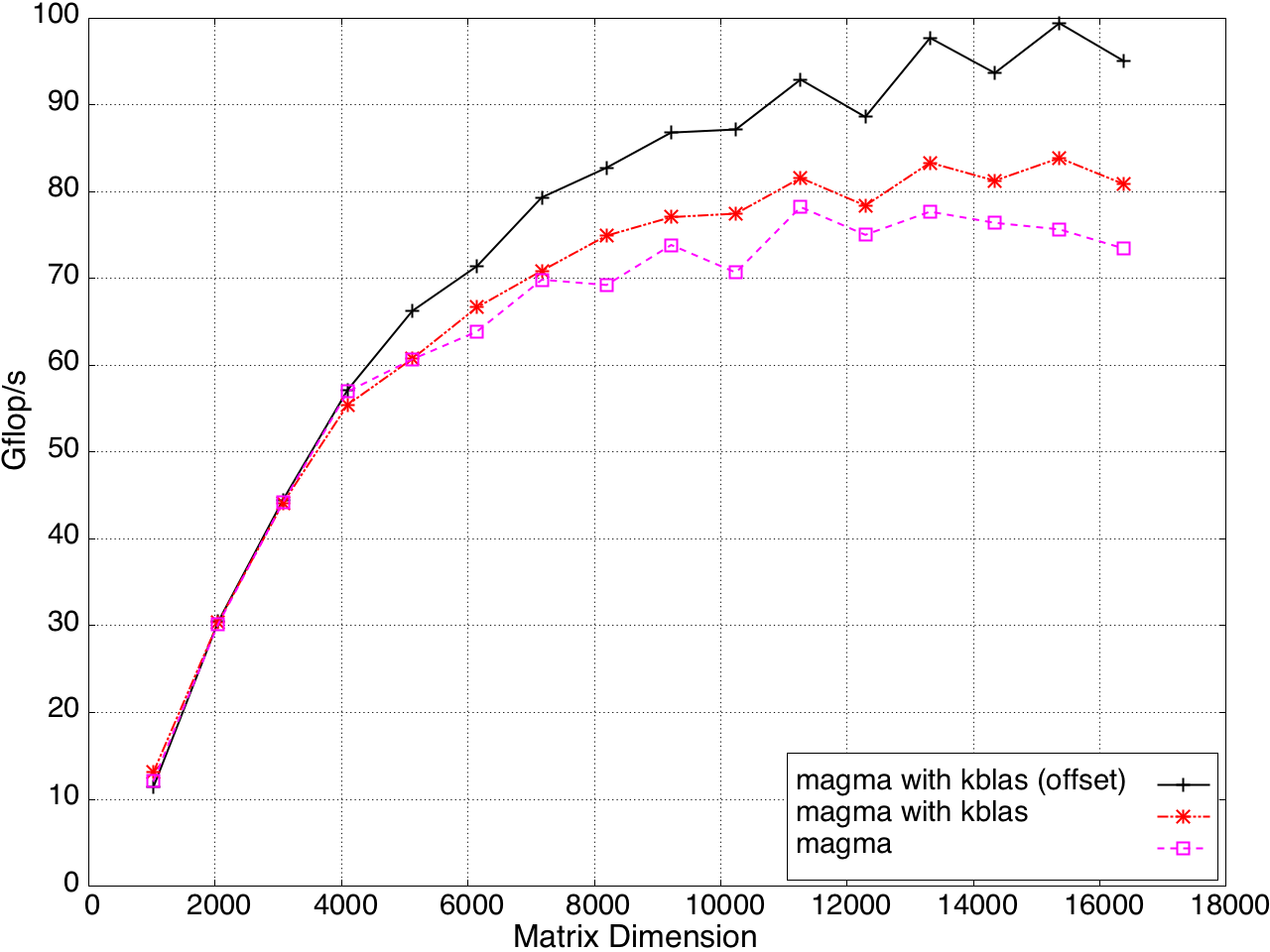}
\label{fig:sgebrd}
}
\subfigure[DGEBRD]{
\includegraphics[width=0.48\linewidth]{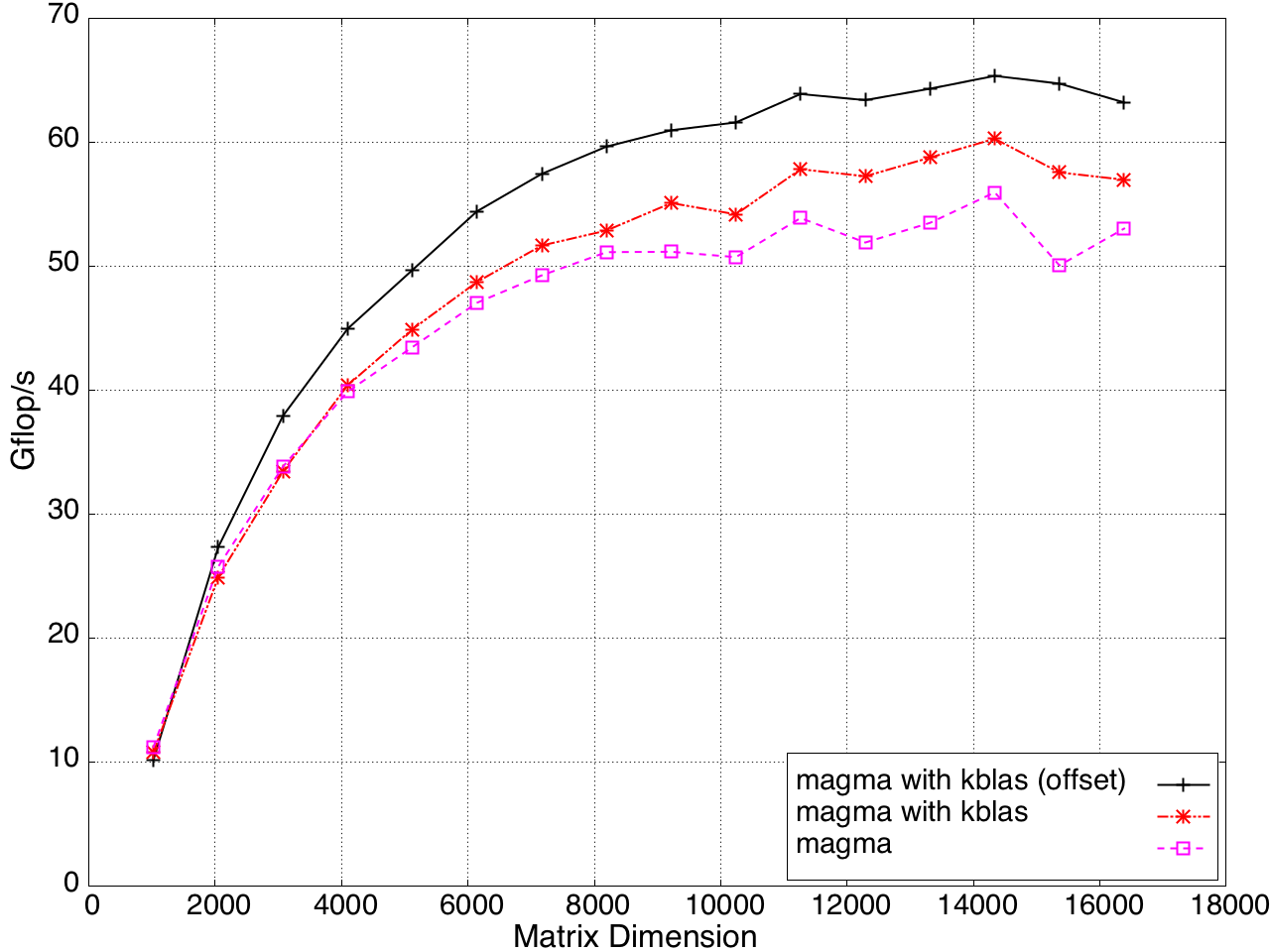}
\label{fig:dgebrd}
}
\subfigure[CGEBRD]{
\includegraphics[width=0.48\linewidth]{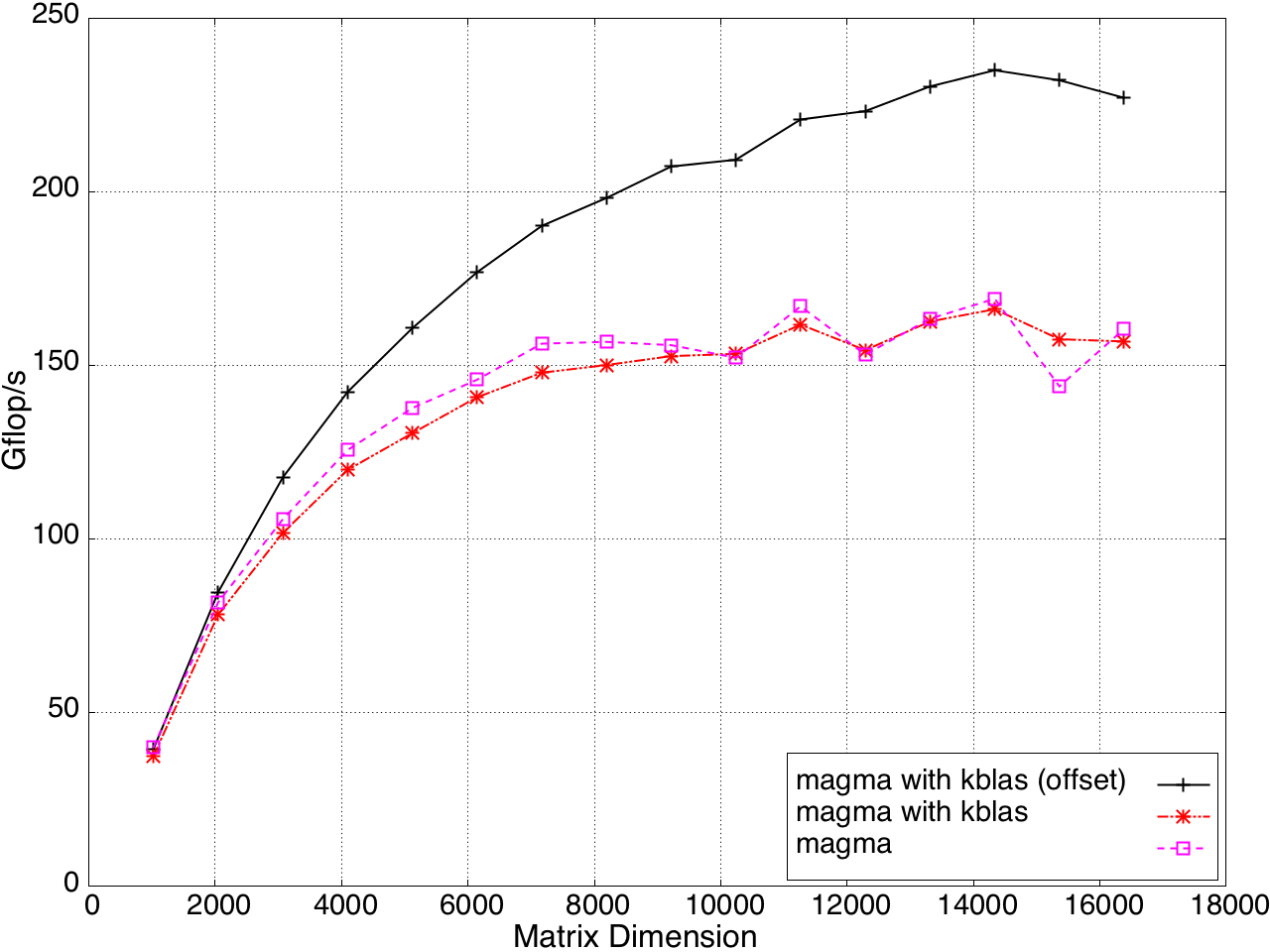}
\label{fig:cgebrd}
}
\subfigure[ZGEBRD]{
\includegraphics[width=0.48\linewidth]{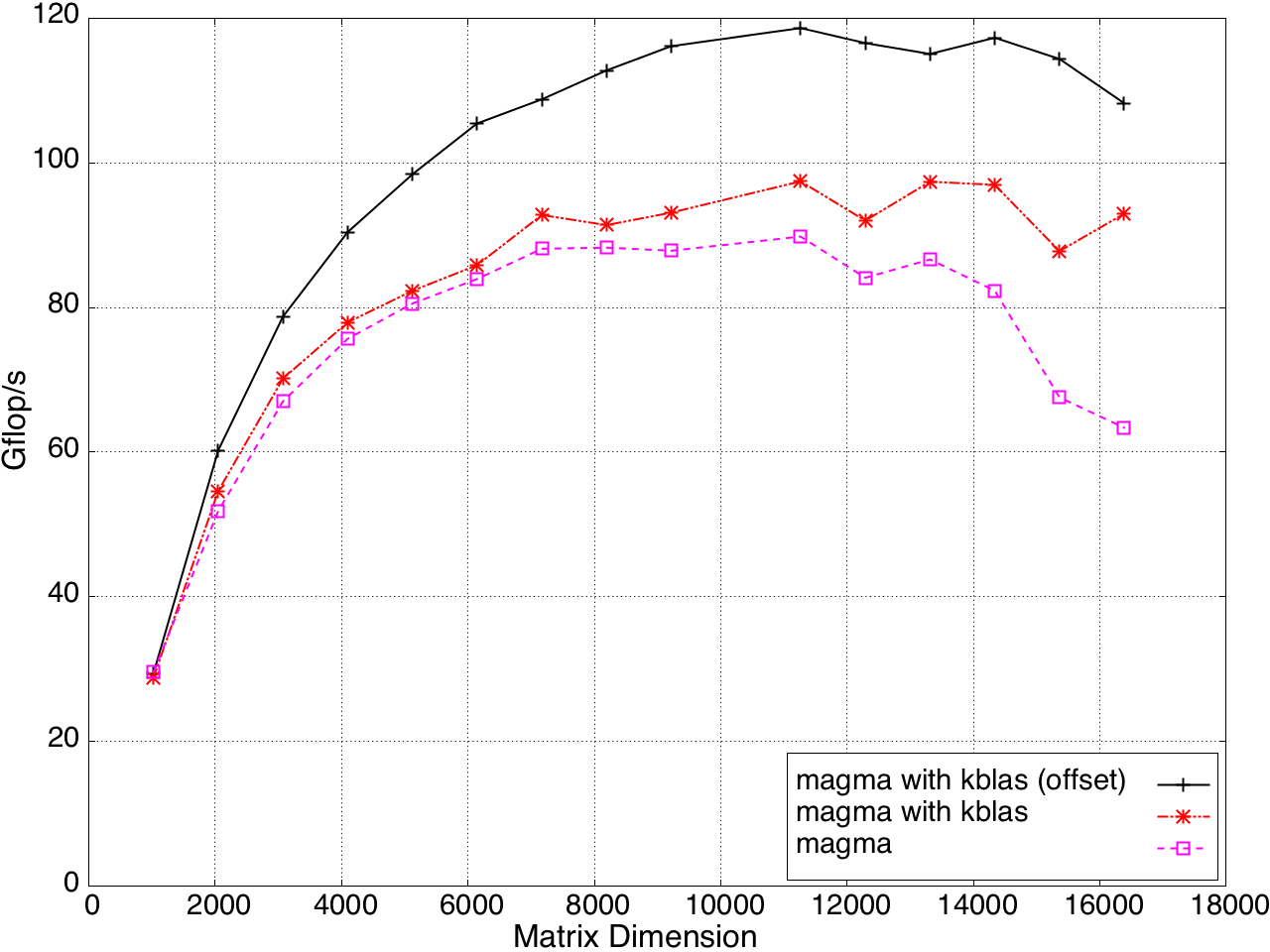}
\label{fig:zgebrd}
}
\caption[]{Bidiagonal Reduction Performance on a K20c GPU, ECC off}
\label{fig:gebrd}
\end{figure}

Figure \ref{fig:gebrd} shows the performance of the MAGMA GEBRD algorithm for all four precisions on a single GPU. 
The original MAGMA implementation uses cuBLAS GEMV kernel to update the unreduced part of the input matrix. 
As pointed in Section 
\ref{subsec:single_gpu_perf}, the performance of KBLAS GEMV is very similar to its cuBLAS counterpart 
as shown in Figure \ref{fig:gemv}. Therefore, the performance of the GEBRD algorithm 
is either very similar or slightly better than the original MAGMA implementation. However, the impact of using 
the KBLAS GEMV-OFFSET kernel is more significant than the KBLAS GEMV. This is due to the fact that the former achieves 
a much better performance if the multiplication is done by a submatrix, as we showed in Figure \ref{fig:gemv_offset}.
%the improvements in the GEBRD performance 
%comes as a result of using the GEMV kernel with the new interface. 
%This kernel shows speedups against cuBLAS, as shown in Figure \ref{fig:gemv_offset}. 
The improvements in the GEBRD performance are up to $31$\%, $29$\%, $61$\%, and $71$\% on all four precisions. Given that the KBLAS GEMV-OFFSET kernel does 
extra reads from global memory, its significance to the GEBRD appears only for relatively large matrices, where the extra reads 
become negligible. 

%\begin{figure}[ht]
%\centering
%\subfigure[SGEHRD]{
%\includegraphics[width=0.48\linewidth]{sgehrd.png}
%\label{fig:sgehrd}
%}
%\subfigure[DGEHRD]{
%\includegraphics[width=0.48\linewidth]{dgehrd.png}
%\label{fig:dgehrd}
%}
%\subfigure[CGEHRD]{
%\includegraphics[width=0.48\linewidth]{cgehrd.png}
%\label{fig:cgehrd}
%}
%\subfigure[ZGEHRD]{
%\includegraphics[width=0.48\linewidth]{zgehrd.png}
%\label{fig:zgehrd}
%}
%\caption[]{Hessenberg Reduction Performance on a K20c GPU, ECC off}
%\label{fig:gehrd}
%\end{figure}

\begin{figure}[ht]
\centering
\subfigure[SSYTRD]{
\includegraphics[width=0.48\linewidth]{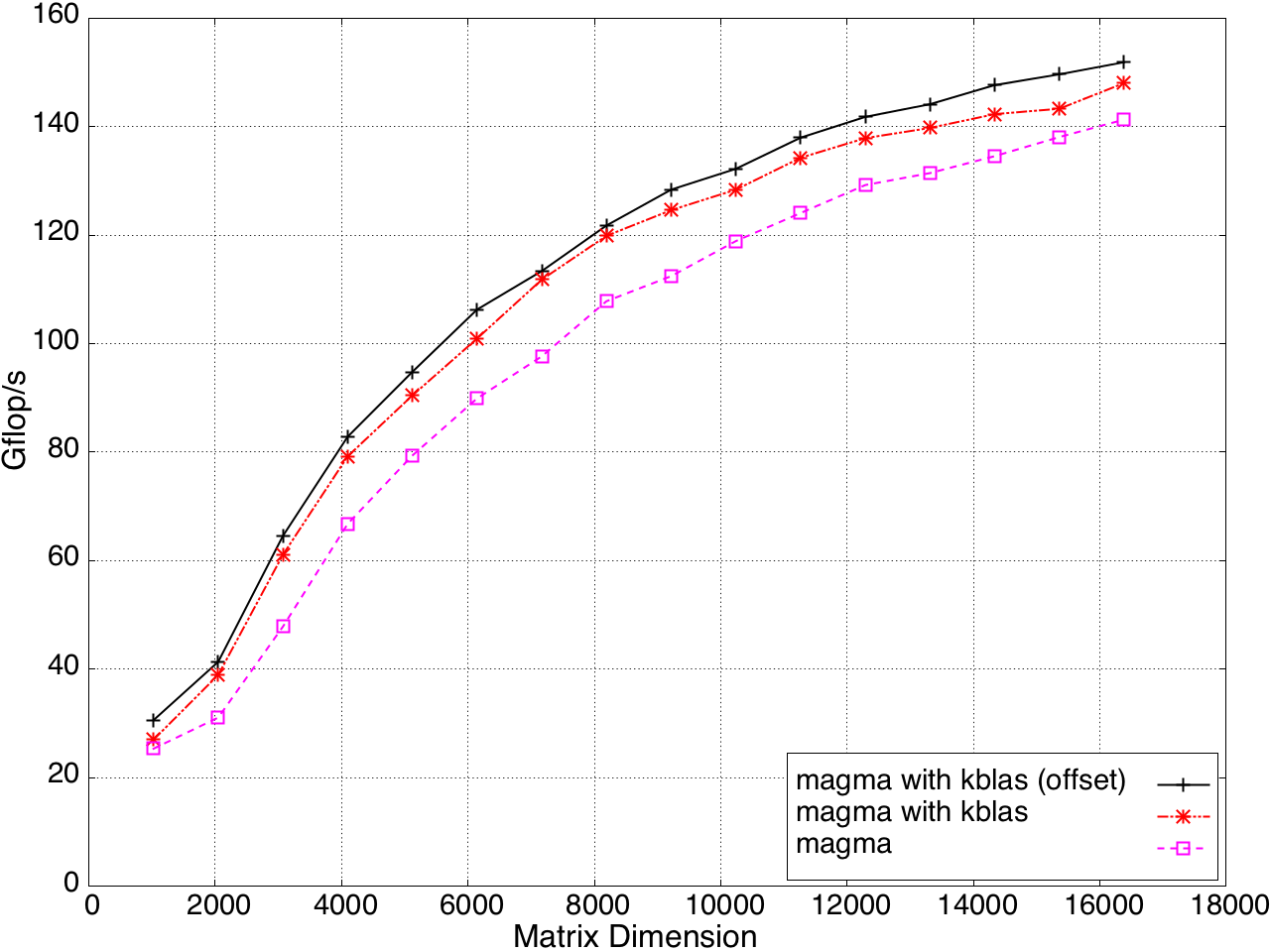}
\label{fig:ssytrd}
}
\subfigure[DSYTRD]{
\includegraphics[width=0.48\linewidth]{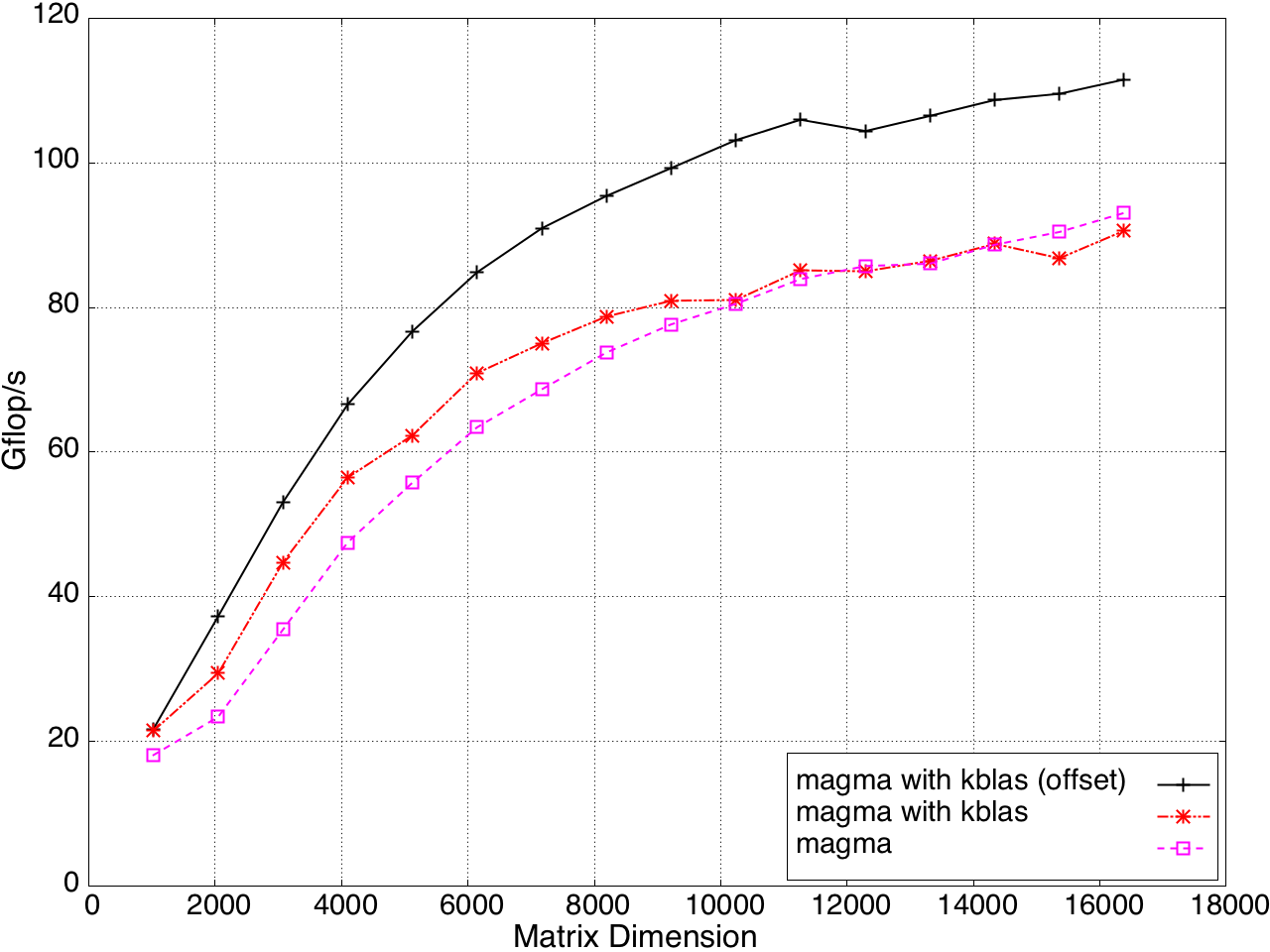}
\label{fig:dsytrd}
}
\subfigure[CHETRD]{
\includegraphics[width=0.48\linewidth]{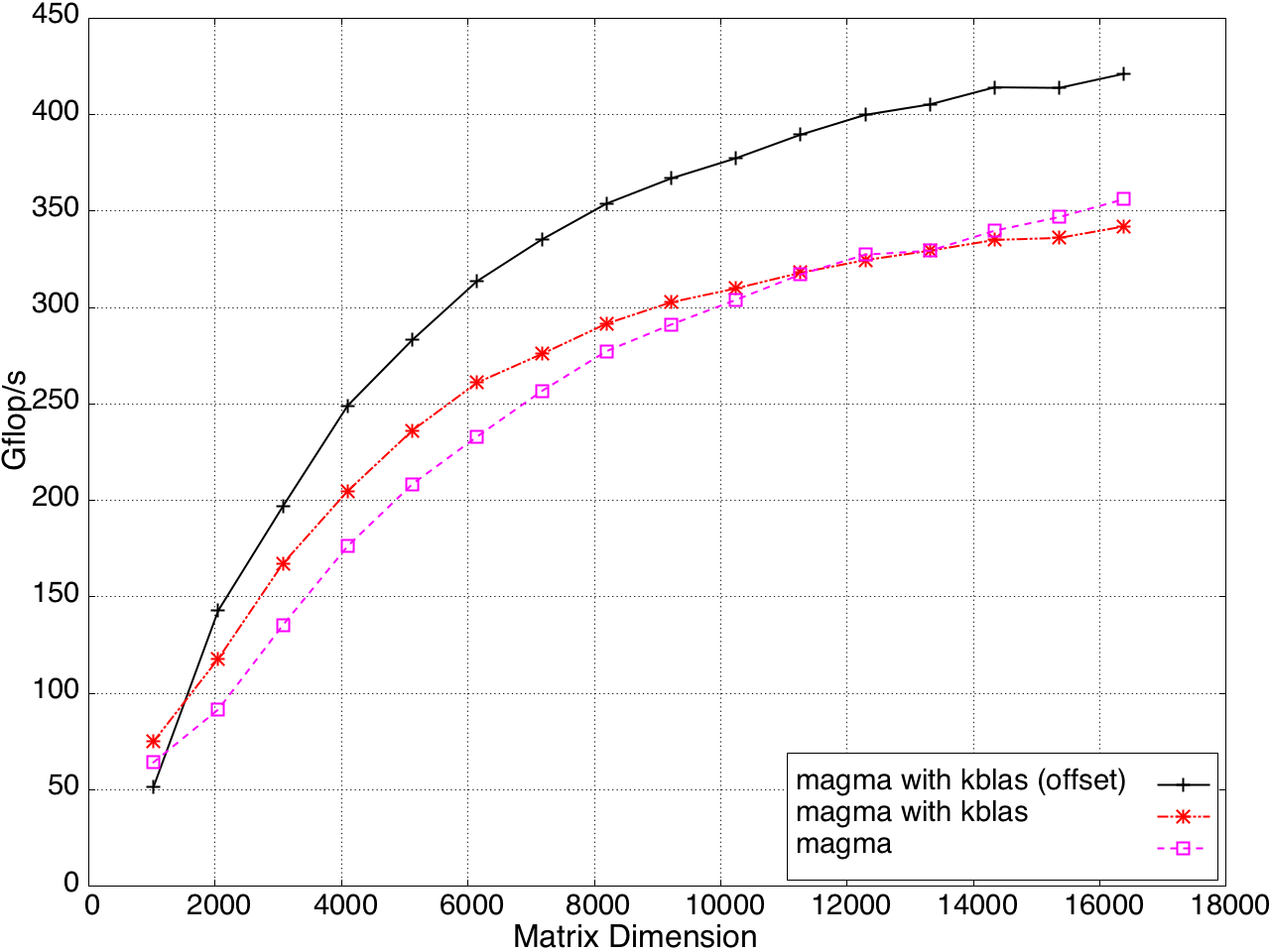}
\label{fig:chetrd}
}
\subfigure[ZHETRD]{
\includegraphics[width=0.48\linewidth]{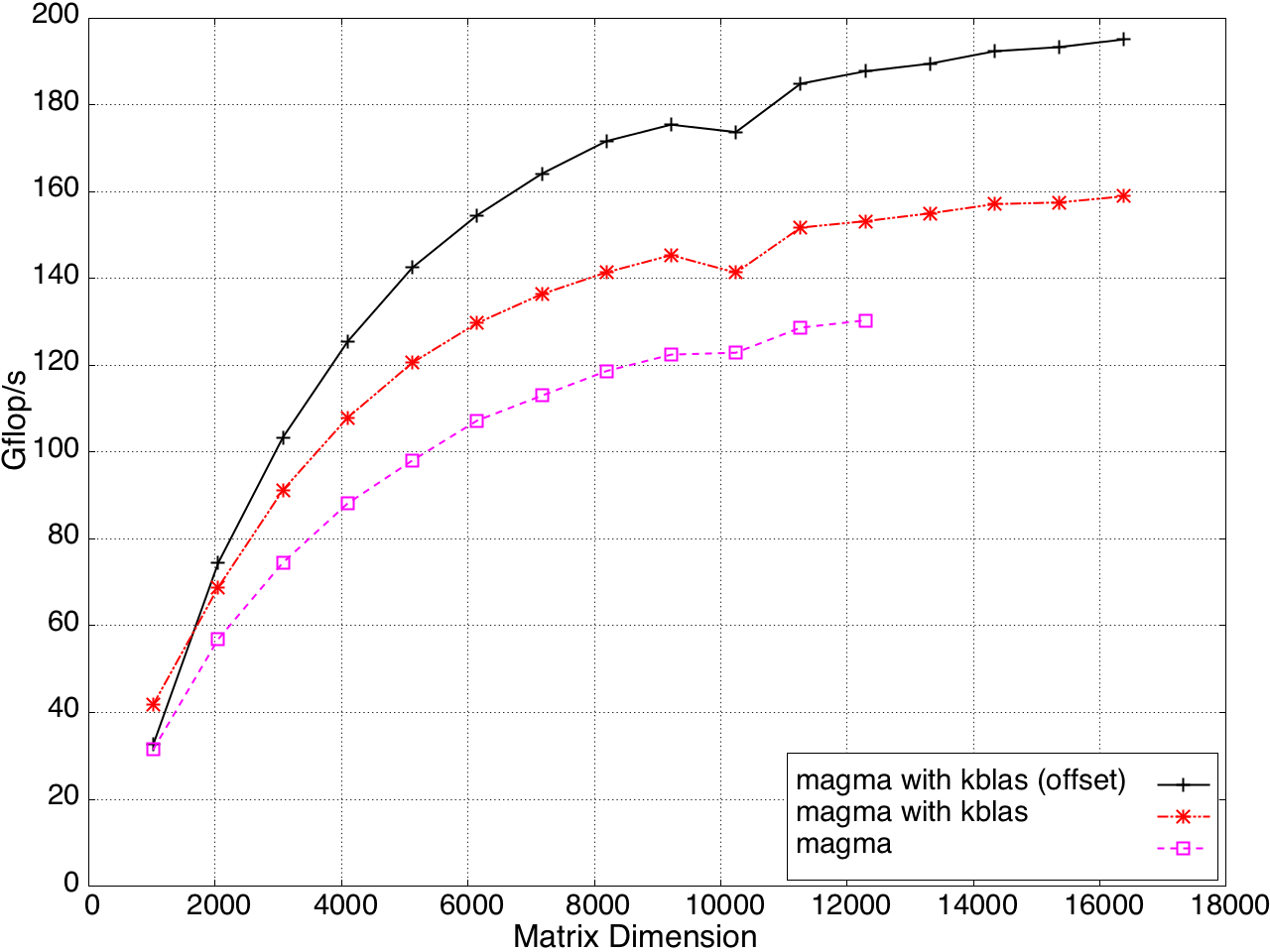}
\label{fig:zhetrd}
}
\caption[]{Tridiagonal Reduction Performance on a K20c GPU, ECC off}
\label{fig:sytrd}
\end{figure}

Figure \ref{fig:sytrd} shows the performance improvement of the MAGMA SYTRD/HETRD algorithm when the KBLAS SYMV/HEMV kernel (both implementations)
is incorporated instead of the original implementation by MAGMABLAS. Similar to the GEBRD algorithm, the new-interface kernels from 
KBLAS are used to give the best possible performance. The improvements are up to $35$\%, $59$\%, $56$\%, and $49$\% for all precisions. We notice 
that the best performance gain comes for relatively small matrices, because KBLAS significantly improves the 
performance of the DSYMV/HEMV kernel for these sizes of matrices against MAGMABLAS, as shown in Figure \ref{fig:symv_offset}.  

\begin{figure}[ht]
\centering
\subfigure[SSYTRD-MGPU]{
\includegraphics[width=0.48\linewidth]{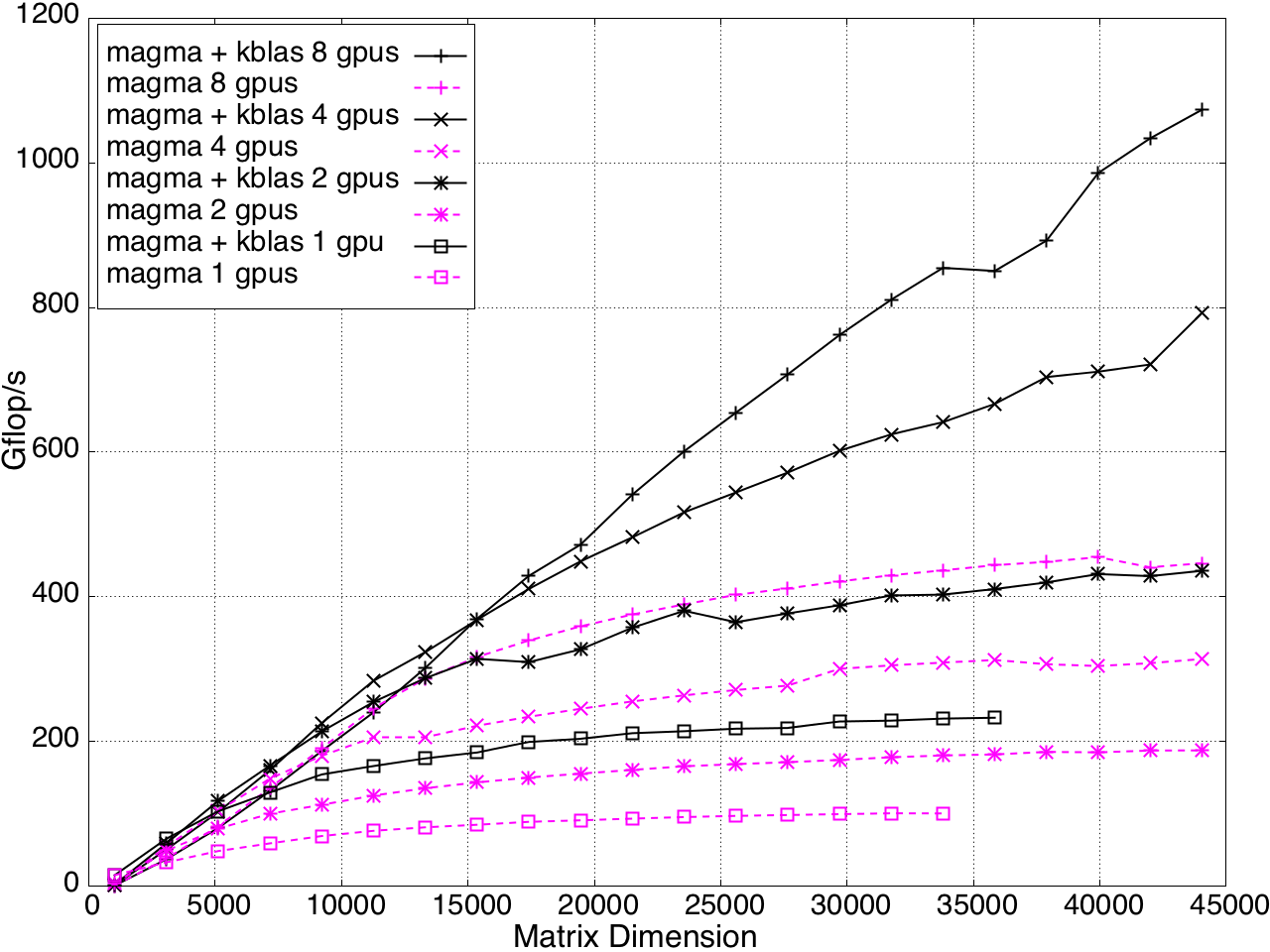}
\label{fig:ssytrd_mgpu}
}
\subfigure[DSYTRD-MGPU]{
\includegraphics[width=0.48\linewidth]{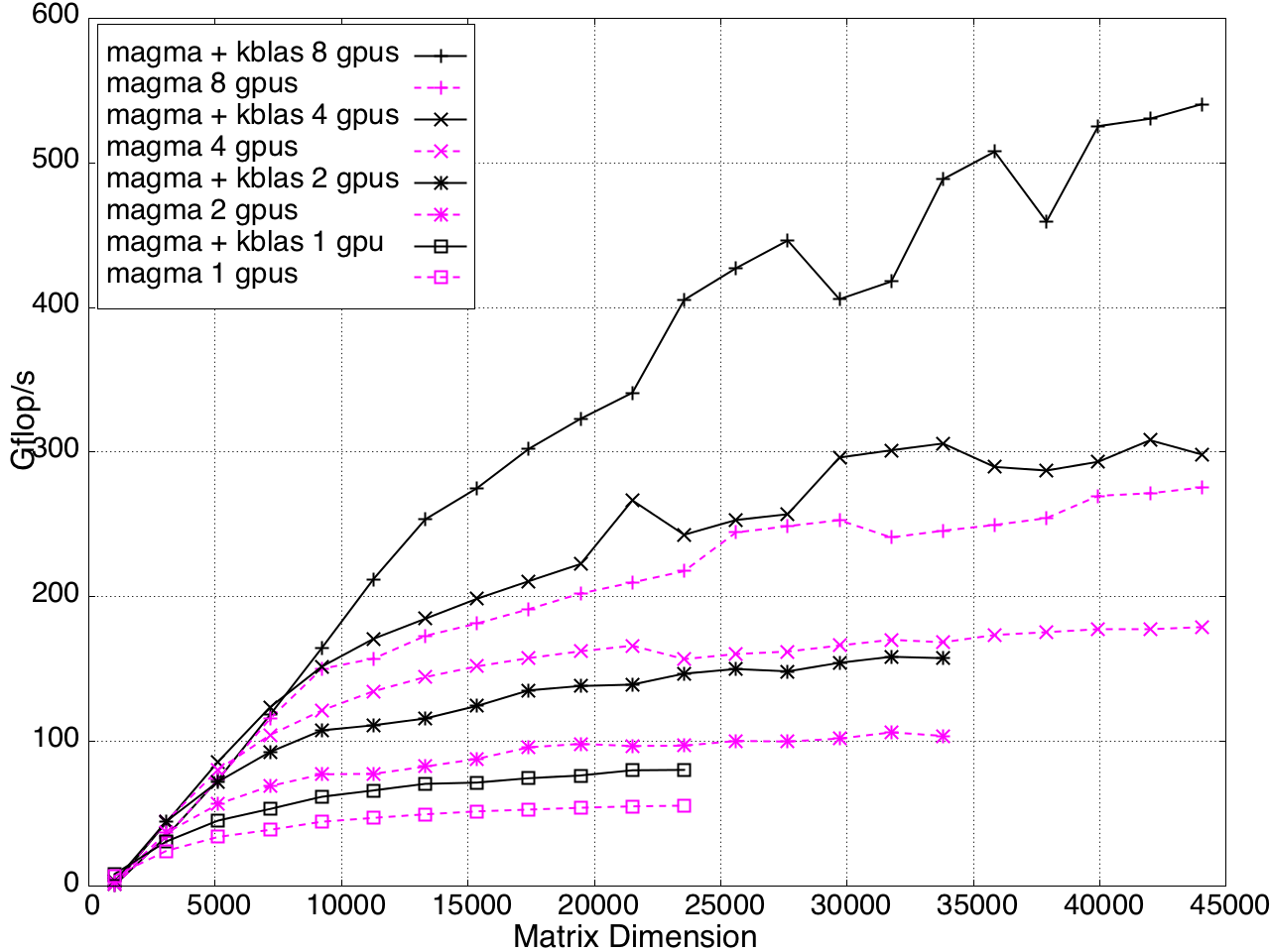}
\label{fig:dsytrd_mgpu}
}
\subfigure[CHETRD-MGPU]{
\includegraphics[width=0.48\linewidth]{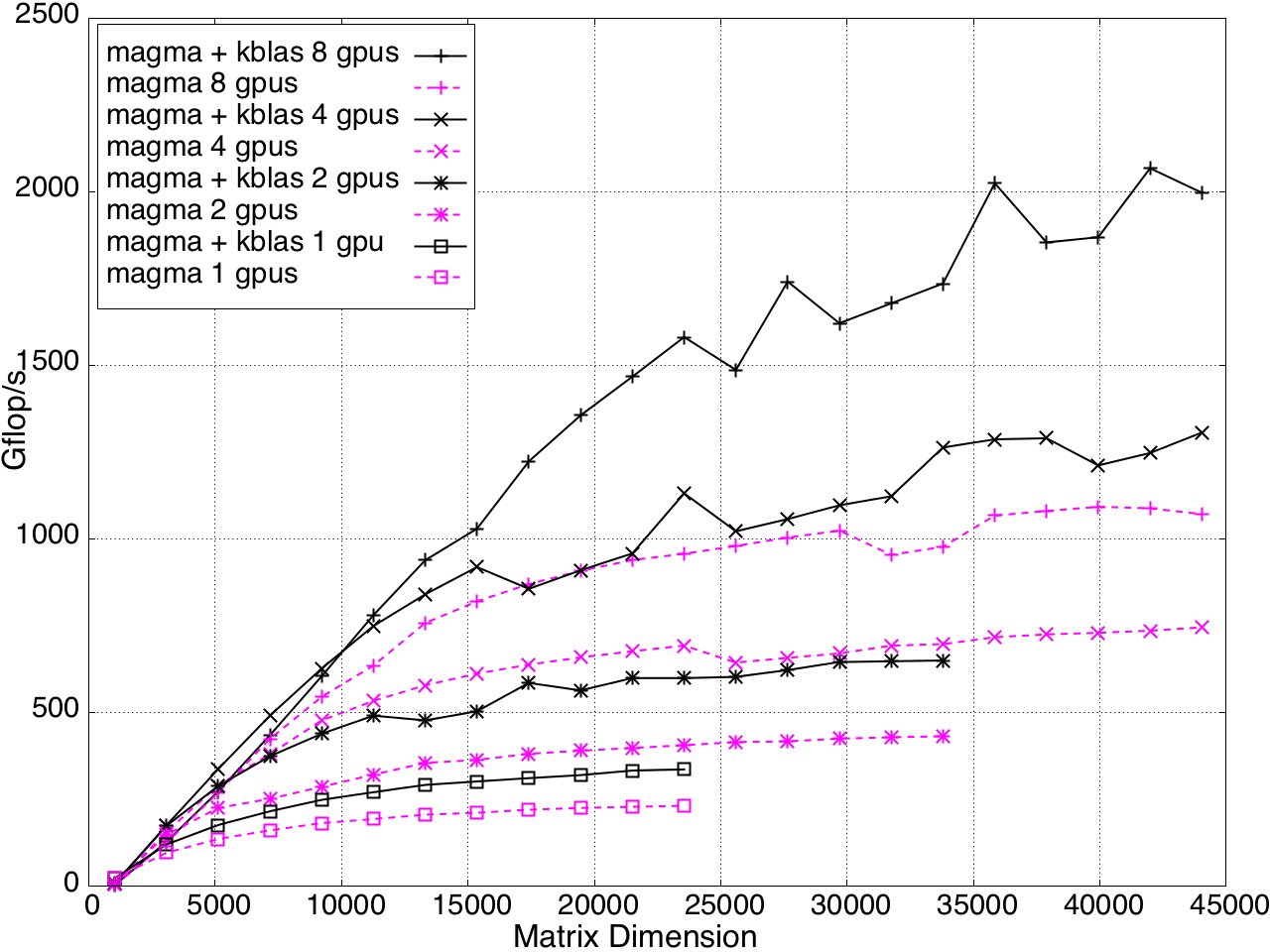}
\label{fig:chetrd_mgpu}
}
\subfigure[ZHETRD-MGPU]{
\includegraphics[width=0.48\linewidth]{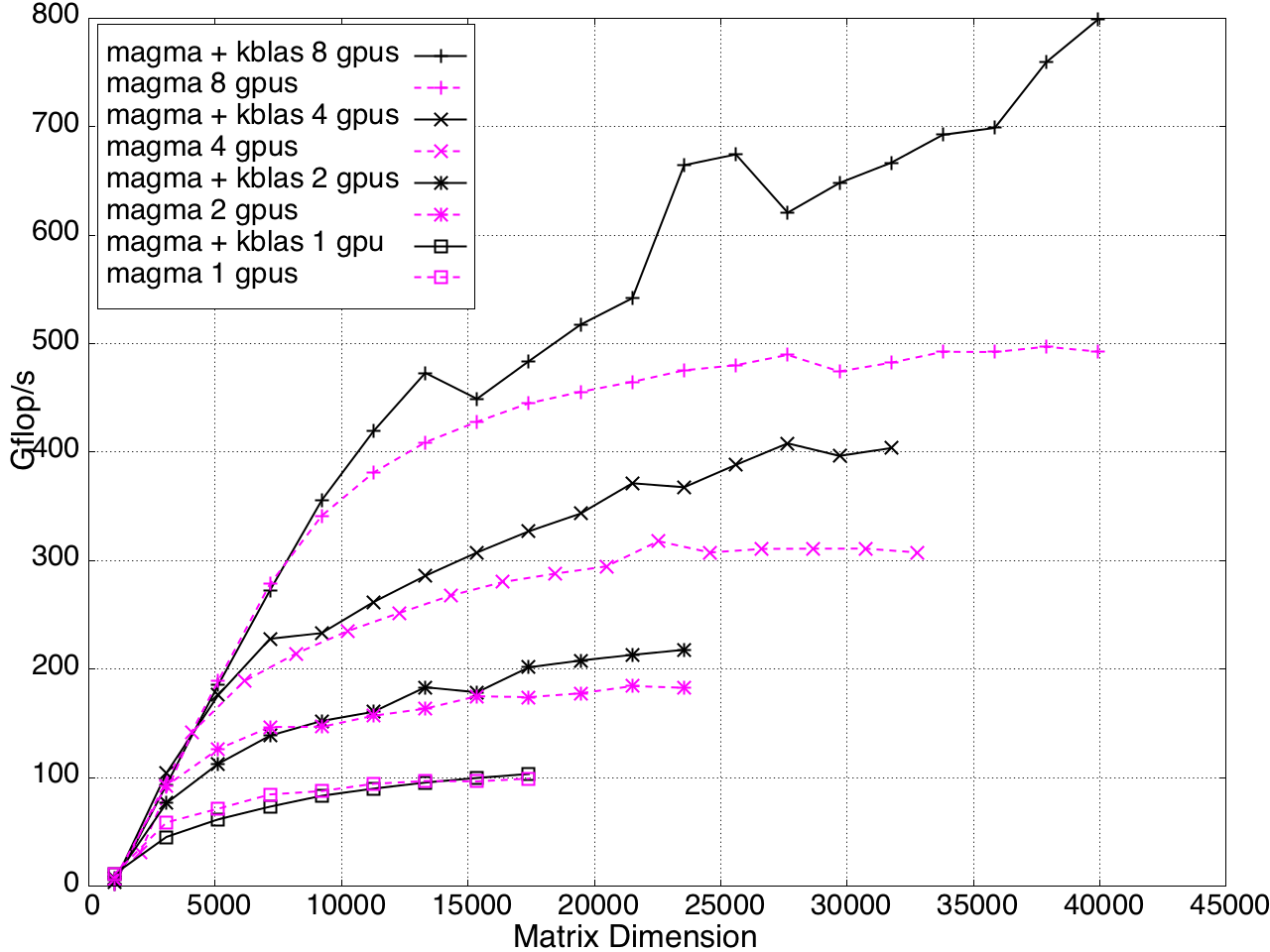}
\label{fig:zhetrd_mgpu}
}
\caption[]{Tridiagonal Reduction Performance on Multi-GPUs, K20c GPU with ECC off}
\label{fig:sytrd_mgpu}
\end{figure}

%-------------------------------------------------------------------------
While MAGMA does not provide a multi-GPU bidiagonal reduction (in which KBLAS GEMV-MGPU kernel can be used), 
%MAGMA also 
it provides an implementation of the SYTRD/HETRD algorithm on multi-GPUs \cite{yamazaki_cpe}. 
Figure \ref{fig:sytrd_mgpu} shows the performance of the this implementation when KBLAS is used 
%as the provider of the SYMV/HEMV kernel on multi-GPUs. 
instead of MAGMABLAS. Using $8$ GPUs, the performance gains are up to $140$\%, $103$\%, $105$\%, and $62$\% 
for all four precisions. We notice that the original MAGMA implementation requires an initialization step for the memory 
workspace each time before calling the SYMV/HEMV kernel. KBLAS, in addition to its better performance, saves such initialization 
time, since it does not need any extra workspace. So, the overall performance gain comes from more optimized kernel + the saving 
in initialization time. The KBLAS-DSYMV kernel on multi-GPU has been used in accelerating a dense symmetric 
eigen solver for very large matrices in a computational astronomy application \cite{aomagma}. 

%------------------------------------------------------------------
%\subsection{Sparse: HLib}
%\label{subsec:hlib}

%% file: summary.tex
%!TEX root = kblas.tex
This paper introduces KBLAS, an open source library that provides optimized kernels for 
dense matrix vector multiplication using NVIDIA GPUs. Through a set of low-level optimizations, 
KBLAS outperform state-of-the-art implementations. Our experiments show that the 
performance is portable across different GPU models and architectures, thanks to the 
tuning parameters KBLAS provides for each kernel. The paper also shows that KBLAS 
can accelerate existing implementations of LAPACK algorithms. 

In the future, KBLAS will continue to provide optimized BLAS kernels, where a room for 
improvement is envisioned. In addition, more data layouts for multi-GPU routines are to be 
supported, for example like the 2D cyclic format used in ScaLAPACK. This step is intended to 
support BLAS operations on distributed systems with multi-GPU accelerated nodes. 
Another direction is to 
provide a sophisticated auto-tuning functionality within KBLAS to facilitate kernel tuning 
on any GPU. This will also open the door for inserting more tuning parameters that can give 
the user more fine grain control on performance. We also plan to use KBLAS building blocks 
in Sparse Matrix Vector Multiplication (SpMV) for sparse matrices that have a substructure 
of dense blocks.

%% file: ack.tex
%!TEX root = kblas.tex
We would like to thank NVIDIA for their hardware donations
and in particular, we would like to give credits to both 
Philippe Vandermersch and Sharan Chetlur from 
NVIDIA for their help and support during the integration 
of the SYMV/HEMV routines into 
cuBLAS library. We also thank the CSCS Swiss National 
Supercomputing Centre for providing access
to their GPU computing platforms. 